\documentclass[11pt]{article}

\usepackage{graphicx,subcaption,float}
\graphicspath{{graphics/}{./graphics/}}
\usepackage[font=small]{caption}
\usepackage[export]{adjustbox}

\usepackage{enumitem}
\setlist[itemize]{itemsep=.5pt, topsep=3pt,leftmargin=*}
\setlist[enumerate]{itemsep=.5pt, topsep=3pt,leftmargin=*}
\usepackage{titlesec}
\titlespacing{\paragraph}{0pt}{1em}{1em} %

\usepackage[dvipsnames]{xcolor}
\definecolor{myOrange}{RGB}{230, 159, 0}
\definecolor{myLightBlue}{RGB}{86, 180, 233}
\definecolor{myGreen}{RGB}{0, 158, 115}
\definecolor{myYellow}{RGB}{240, 228, 66}
\definecolor{myDarkBlue}{RGB}{0, 114, 178}
\definecolor{myRed}{RGB}{213, 94, 0}
\definecolor{myPink}{RGB}{204, 121, 167}

\usepackage{amsmath,amsthm,amssymb,mathtools,thm-restate}

\usepackage[algo2e, linesnumbered, boxruled, vlined]{algorithm2e}
\DontPrintSemicolon
\SetKwInput{Maintain}{Maintain}
\SetKwProg{Procedure}{Procedure}{}{}

\SetArgSty{textnormal}

\usepackage[english]{babel}
\usepackage{fix-cm}
\usepackage[utf8]{inputenc}
\usepackage[T1]{fontenc}
\usepackage{amsfonts}
\usepackage{slantsc}
\usepackage{mlmodern}
\usepackage{csquotes}

\usepackage[protrusion=true,expansion,kerning=true]{microtype}

\usepackage[
    backend=bibtex,    %
    maxcitenames=10,    %
    maxbibnames=10,    %
    maxalphanames=10,    
    maxnames=10,    %
    style=alphabetic,    %
    sorting=anyt,    %
    doi=false,    %
    isbn=false,    %
    eprint=false,    %
    url=false,    %
	backref=true,    %
	backrefstyle=none,    %
        ]{biblatex}

\AtEveryBibitem{
\ifentrytype{inproceedings}{
\clearname{editor}%
\clearfield{valume}%
\clearfield{number}%
\clearfield{series}%
\clearlist{publisher}%
\clearfield{address}%
\clearlist{location}%
\clearfield{month}%
\clearfield{eprint}%
\clearfield{isbn}%
\clearfield{issn}%
}{}
}

\newbibmacro{string+doi}[1]{%
    \iffieldundef{doi}{%
        \iffieldundef{url}{#1}{\href{\thefield{url}}{#1}}
        }{%
        \href{http://dx.doi.org/\thefield{doi}}{#1}
        }
    }

\DeclareFieldFormat*{title}{%
    \usebibmacro{string+doi}{\mkbibemph{#1}}
    }

\renewbibmacro{in:}{%
    \ifentrytype{article}{}{\printtext{\bibstring{in}\intitlepunct}}}

\newtoggle{bbx:datemissing}

\renewbibmacro*{date}{\toggletrue{bbx:datemissing}%
}

\renewbibmacro*{issue+date}{}%

\newbibmacro*{date:print}{%
    \setunit{\addcomma\addspace}    %
    \togglefalse{bbx:datemissing}%
    \printdate}

\renewbibmacro*{chapter+pages}{%
    \printfield{chapter}%
    \setunit{\bibpagespunct}%
    \printfield{pages}%
    \newunit
    \usebibmacro{date:print}%
    \newunit}

\renewbibmacro*{note+pages}{%
    \printfield{note}%
    \setunit{\bibpagespunct}%
    \printfield{pages}%
    \setunit{\addcomma\addspace}%
    \usebibmacro{date}%
    \newunit}

\renewbibmacro*{addendum+pubstate}{%
    \iftoggle{bbx:datemissing}
        {\usebibmacro{date:print}}
        {}%
    \printfield{addendum}%
    \newunit\newblock
    \printfield{pubstate}}

\appto{\bibsetup}{\sloppy}

\DefineBibliographyStrings{english}{%
    page={page},%
    pages={pages},%
    }

\DefineBibliographyStrings{english}{
  backrefpage = {},
  backrefpages = {},
}

\DeclareFieldFormat{bracketswithperiod}{\mkbibbrackets{#1}}

\renewbibmacro*{pageref}{%
  \iflistundef{pageref}
    {}
    {\printtext[bracketswithperiod]{%
       \ifnumgreater{\value{pageref}}{1}
         {\bibstring{backrefpages}}
         {\bibstring{backrefpage}}%
       \printlist[pageref][-\value{listtotal}]{pageref}}%
     }}

\usepackage[
    breaklinks=true,    %
    colorlinks=true,    %
    linkcolor=myDarkBlue,    %
    urlcolor=myRed,    %
    citecolor=myGreen,    %
    bookmarks=true,    %
    bookmarksopenlevel=2,    %
    ]{hyperref}

\usepackage[nameinlink]{cleveref}

\newtheorem{theorem}{Theorem}[section]

\newtheorem{lemma}[theorem]{Lemma}

\newtheorem{remark}[theorem]{Remark}
\newtheorem{definition}[theorem]{Definition}
\newtheorem{claim}[theorem]{Claim}
\newtheorem{observation}[theorem]{Observation}

\newcommand{\overbar}[1]{{\mkern 1.5mu\overline{\mkern-1.5mu#1\mkern-1.5mu}\mkern 1.5mu}}

\newcommand{\eps}{\varepsilon}

\newcommand{\hull}{\operatorname{hull}}

\newcommand{\Tcal}{\mathcal T}
\newcommand{\Mcal}{\mathcal M}
\newcommand{\Pcal}{\mathcal P}
\newcommand{\Hcal}{\mathcal H}

\DeclareMathOperator\poly{poly}
\DeclareMathOperator\dist{dist}
\DeclareMathOperator\Par{Par}
\DeclareMathOperator\ct{count}
\DeclareMathOperator\tail{tail}
\DeclareMathOperator\head{head}

\def\Otil{\widetilde{O}}

\newcommand{\vbar}{\overbar{v}}
\newcommand{\wbar}{\overbar{w}}
\newcommand{\Gbar}{\overbar{G}}
\newcommand{\Rbar}{\overbar{R}}
\newcommand{\pbar}{\overbar{p}}

\def\portal{\textsc{Portals}}
\def\splitportal{\textsc{SplitPortals}}
\def\sliceportal{\textsc{SlicePortals}}
\usepackage{tcolorbox}

\declaretheoremstyle[
    spaceabove=6pt, 
    spacebelow=6pt,
    headfont=\normalfont\bfseries,
    notefont=\mdseries, notebraces={(}{)},
    bodyfont=\normalfont,
    postheadspace=1em,
    numbered=yes,
    shaded=
    {
        bgcolor=myLightBlue!15,
        textwidth=\linewidth,
    },
]{filled}

\declaretheoremstyle[
    spaceabove=6pt, 
    spacebelow=6pt,
    headfont=\normalfont\bfseries,
    notefont=\mdseries, notebraces={(}{)},
    bodyfont=\normalfont,
    postheadspace=1em,
    numbered=yes,
    shaded=
    {
        bgcolor=myLightBlue!0,
        textwidth=.98\linewidth,
        rulecolor=black,
        rulewidth=1pt,
        padding=2mm
    },
]{ruled}
\declaretheorem[style=ruled,name=Procedure]{procedure-ruled}

\crefname{equation}{equation}{equations}
\crefname{subsection}{subsection}{subsections}
\crefname{question}{question}{questions}
\crefname{problem}{problem}{problems}
\crefname{claim}{claim}{claims}
\crefname{algorithm}{algorithm}{algorithms}
\crefname{proof}{proof}{proofs}
\crefname{observation}{observation}{observations}
\crefname{invariant}{invariant}{invariants}

\newcommand{\one}{{\tt Type-1}\xspace}
\newcommand{\two}{{\tt Type-2}\xspace}
\newcommand{\three}{{\tt Type-3}\xspace}
\newcommand{\four}{{\tt Type-4}\xspace}

\usepackage[displaymath, mathlines]{lineno}

\expandafter\let\expandafter\oldequation\csname equation*\endcsname
\expandafter\let\expandafter\endoldequation\csname endequation*\endcsname

\renewenvironment{equation*}{%
  \ifLineNumbers

  \else

  \fi 
  \begin{mathenvironment}}
  {\end{mathenvironment}}

 \usepackage[disable,colorinlistoftodos,prependcaption,textsize=tiny]{todonotes}
\presetkeys{todonotes}{backgroundcolor=Red!10,bordercolor=red}{}
\newcommand{\ali}[1]{\todo[linecolor=red,backgroundcolor=pink!25,bordercolor=red,inline]{\textbf{Ali:~}#1}}
\newcommand{\atodo}[1]{\todo[linecolor=red,backgroundcolor=green!25,bordercolor=red]{\textbf{AF:~}#1}}
\newcommand{\mprobst}[1]{\todo[linecolor=red,backgroundcolor=yellow!25,bordercolor=red]{\textbf{MPG:~}#1}}
\newcommand{\atodoin}[1]{\todo[linecolor=red,backgroundcolor=green!25,bordercolor=red,inline]{\textbf{AF:~}#1}}

\title{Fully Dynamic All-Pairs Min Cuts on Planar Graphs via \\ Planar Pattern Emulators}

\newcommand{\authorbreak}{%
  \end{tabular}%
  \par\vspace{1em}%
  \begin{tabular}[t]{c}%
}

\author{
  Arnold Filtser \\
  \small Bar-Ilan University
  \and
  Gramoz Goranci \\
  \small University of Vienna 
  \and
  Ali Momeni \\
  \small University of Vienna 
  \authorbreak 
  Neel Patel \\
  \small Google Research 
  \and
  Maximilian Probst Gutenberg \\
  \small ETH Zurich 
}

\date{\vspace{1.5em}\today}
 
\begin{document}
\maketitle
\pagenumbering{gobble}
\begin{abstract}
We study the fully-dynamic all-pairs minimum cut problem on planar graphs. Given an $n$-vertex undirected planar graph $G=(V,E)$ undergoing edge insertions and deletions, the goal is to efficiently process these updates while supporting $s$-$t$ minimum cut queries for any pair of vertices $s$ and $t$. We give a $(1+\eps)$-approximate dynamic algorithm that supports edge updates and minimum cut queries in $n^{o(1)}$ time, for any $1/\text{poly}(\log n) < \eps < 1$. For general graphs, previous results either are limited to partially dynamic settings or achieve only logarithmic-factor approximations.

The main conceptual and technical contribution is to show that a suitable adaptation of the planar emulator of Chang, Krauthgamer, and Tan~[STOC'22] satisfies the stronger pattern emulator introduced by Chen and Tan~[ICALP'25]. This leads almost immediately to a nearly linear-time construction of pattern emulators, which, when combined with the recent fully dynamic $(1+\eps)$-approximate all-pairs shortest paths algorithm of Filtser, Goranci, Patel, and Probst Gutenberg~[FOCS'24], yields our result.

\end{abstract}

\newpage
\setcounter{tocdepth}{3}
\addtocontents{toc}{\protect\setcounter{tocdepth}{3}} 
\tableofcontents

\newpage
\pagenumbering{arabic}
\section{Introduction}

\ali{application: global min-cut}

The $s$-$t$ minimum cut is a textbook, fundamental graph problem in combinatorial optimization and algorithm design. It has attracted particular attention for planar graphs, as many real-world applications, including image segmentation~\cite{Boykov:2006aa,Kohli:2010aa} and network reliability~\cite{Provan:1986aa,Ball:2007aa}, naturally arise in such instances. Moreover, the planar setting is especially appealing due to the close relationship between cut problems and shortest-path problems, captured by the elegant mathematical work of \emph{planar duality}. There has been a growing interest in designing efficient algorithms for planar minimum cut, dating back to the seminal work of \citeauthor{Itai:1979aa}~\cite{Itai:1979aa} in the 1970’s, which gave an algorithm with running time $O(n^2 \log n)$. The currently best-known running time, $O(n \log \log n)$, was later achieved by Italiano, Nussbaum, Sankowski, and Wulff-Nilsen~\cite{Italiano:2011aa}.

We study the \emph{all-pairs minimum cuts} on \emph{planar graphs} in the online \emph{fully dynamic} setting. Here, the goal is to preprocess a planar graph $G$ so that we can \emph{efficiently} support the following operations.
\begin{itemize}
\item \underline{Edge Updates:}
    At any point, we may \emph{insert} or \emph{delete} an edge from $G$. 
    Such updates are performed under the constraint that the underlying graph $G$ remains planar and its embedding is not changed.
\item \underline{All-pairs Minimum Cut Queries:}
    We want to be able to answer \emph{queries} about the size of $s$-$t$ minimum cut, for any two given vertices $s$ and $t$, in the current graph $G$.
\end{itemize}

While the runtime complexity of planar $s$-$t$ minimum cut is nearly settled in the static setting, as discussed above, the dynamic setting remains far less understood. 
When the goal is to maintain the minimum cut value \emph{exactly}, Italiano, Nussbaum, Sankowski, and Wulff-Nilsen~\cite{Italiano:2011aa} obtained a dynamic algorithm with $\tilde{O}(n^{2/3})$ update time and $\tilde{O}(n^{2/3})$ query time.\footnote{The notation $\Otil(\cdot)$ hides $\poly(\log n)$ factors.} 
In two weaker models: (1) the incremental model, which allows only edge insertions, and (2) the offline dynamic model, in which the full update sequence is known in advance, the problem admits an algorithm with $\tilde{O}(n^{1/2})$ update and query time~\cite{Das:2022aa}, though only when the terminals $s$ and $t$ are fixed in advance and edge updates are restricted to weight changes. 
This update and query bound is essentially tight up to polylogarithmic factors by a conditional lower bound of Abboud and Dahlgaard~\cite{Abboud:2016aa}.\footnote{More precisely, their lower bound is proved for dynamic bipartite matching on planar graphs, but it can also be adapted to the $s$-$t$ maximum flow problem.}

A possible direction for breaking through the $O(\sqrt{n})$ barrier on update and query time is to pay a small approximation error in the size of the reported minimum cut. We take precisely this approach by addressing the following key open question in the dynamic planar graph algorithms literature: 

\begin{center}
\emph{Is there a fully dynamic algorithm for the planar all-pairs minimum cut problem that supports queries up to a $(1+\varepsilon)$-approximation and can handle updates and queries in \emph{sub-square-root} time?}
\end{center}

In this work, we answer the above question positively, and even show that both update and query times can be made \emph{sub-polynomial}, as formally stated in the theorem below.

\begin{restatable}{theorem}{Intromain} \label{thm:Intromain}
 Let $G$ be an $n$-vertex undirected planar graph undergoing edge insertions and deletions with capacities in $[1, n^c]$ for some constant $c$ at all times. 
 Let $1/\poly(\log n) < \varepsilon < 1$ be an error parameter. 
 Then, there is a deterministic data structure that supports edge updates and queries for the $(1+\varepsilon)$-approximate minimum cut between any pair of vertices $s,t \in V(G)$ in the current graph $G$. 
 The data structure has an initialization time of $\Otil(n/\eps^{O(1)})$ and an amortized update time and query time of $n^{o(1)}$.   
\end{restatable}

\begin{remark}
    We can extend our data structure to support queries for a $(1+\varepsilon)$-approximate $s$-$t$ minimum cut, returning the edges crossing the cut in time proportional to the output size, plus an additive $n^{o(1)}$ overhead.
\end{remark}

One notable strength of the dynamic data structure in \Cref{thm:Intromain} is that it provides a $(1+\varepsilon)$-approximation to the minimum cut value. In general graphs, however, achieving such a strong approximation guarantee in the fully dynamic setting appears to be substantially more difficult, despite significant recent efforts in the literature. More specifically, the best known dynamic algorithms for minimum cuts or maximum flows either (i) apply to the partially dynamic setting\footnote{In the partially dynamic setting, we allow either edge insertions or edge deletions, but not both.} and require $s$ and $t$ to be fixed in advance~\cite{Goranci:2023aa, van2023dynamic, Brand:2024aa, chen2024almost, Brand:2024ab,Goranci:2025aa}; or (ii) work in the fully dynamic setting with a much weaker approximation guarantee, requiring at least a near-logarithmic approximation factor for the returned minimum cut~\cite{Chen:2020aa, Goranci:2021aa, Brand:2024ab, Goranci:2026aa}.

\paragraph*{Our Construction of Static and Dynamic Planar Pattern Emulators.}

The starting point of our construction is the (distance) planar emulator of \citeauthor{Chang:2022aa}~\cite{Chang:2022aa}. 
Given a weighted planar graph $G$ and a subset of terminals $T \subseteq V$, we define an \emph{instance} as a tuple $(G,T)$. 
We call an instance $(H,T)$ a $(1+\varepsilon)$-approximate \emph{planar emulator} of $(G,T)$ if (i) the weighted graph $H$ is a planar graph whose vertex set $V(H)$ contains the terminals $T$ and (i) graph $H$ preserves pair-wise \emph{terminal} distances up to a multiplicative factor of $(1+\varepsilon)$. 
A long line of works considered the construction of such emulators~\cite{Krauthgamer:2014aa,Cheung:2016aa,Filtser:2019aa,Chang:2020aa,Chang:2023aa,Filtser:2024ab}, which culminated in the recent breakthrough by \citeauthor{Chang:2022aa}~\cite{Chang:2022aa} who showed an $\tilde{O}(n/\varepsilon^{O(1)})$ time algorithm for constructing $(1+\varepsilon)$-approximate planar emulators with $|H| \leq \tilde{O}(|T|/\varepsilon^{O(1)})$.

In this paper, we consider a \emph{stronger} variant of planar emulators, called \emph{planar pattern emulators}, studied by Krauthgamer and Rika~\cite{Krauthgamer:2020aa}, and recently revisited by Chen and Tan~\cite{Chen:2025aa}. Informally, such emulators are defined similarly to planar emulators, except that the terminal distances must be preserved with respect to the \emph{faces that contain the terminals} (see \Cref{sec:prelim} for a formal definition). Chen and Tan~\cite{Chen:2025aa} proved that such emulators exist and provided a polynomial-time algorithm to construct them. One of our key technical and conceptual contributions is to show that a simpler and more streamlined construction can be obtained within the CKT framework~\cite{Chang:2022aa}, together with Thorup’s and Klein's $\varepsilon$-covers~\cite{Thorup:2004aa,Klein:2002aa}. Moreover, we show that our algorithm for constructing such emulators can be implemented in near-linear time.\footnote{For a more detailed comparison between the construction of Chen and Tan~\cite{Chen:2025aa} and ours, see \Cref{subsec:technical-static}.}

\begin{restatable}{theorem}{static} \label{thm:static}
    Let $G$ be an $n$-vertex planar graph, $T\subseteq V(G)$ be a terminal set, and $0<\varepsilon <1$ be an error parameter.
    Then, there is an algorithm that constructs a $(1+ \varepsilon)$-approximate planar pattern emulator $(H,T)$ of $(G, T)$ in time $\Otil(n/\varepsilon^{O(1)})$ with size $|H|= \Otil(|T|/\varepsilon^{O(1)})$.
    Moreover, $(H,T)$ is aligned with $(G, T)$.\footnote{Instance alignment is defined in \Cref{def:aligned}; we say that $(H, T)$ is aligned with $(G, T)$ if there exists a one-to-one correspondence between their holes, the terminals on the corresponding holes, and the circular ordering of the terminals on their corresponding holes.}
\end{restatable}

By exploiting the planar duality between cuts and distances, \Cref{thm:static} almost immediately implies a \emph{near-linear}-time algorithm for constructing $(1+\eps)$-approximate \emph{terminal cut sparsifiers} for planar graphs with size $\tilde{O}(|T|/\varepsilon^{O(1)})$ that approximately preserves terminal minimum cuts (see \Cref{thm:static-cut}).

The main technical contribution of our paper is to show that the \emph{versatility} of our static pattern emulator construction extends to the dynamic setting via the recent dynamic all-pairs shortest-paths data structure of Filtser, Goranci, Patel, and Probst Gutenberg~\cite{Filtser:2024aa}. Using cut-distance planar duality once more, we obtain a dynamic data structure that supports edge updates and queries in \emph{sub-polynomial} time and returns $(1+\varepsilon)$-approximations to queried pairwise minimum cuts, thus proving our main result in \Cref{thm:Intromain}. Our dynamic pattern emulator construction is summarized in the theorem below.

\begin{restatable}{theorem}{dynamicpattern} \label{thm:dynamic-pattern}
Let $(G,T)$ be a dynamic instance where $G$ is an $n$-vertex planar graph undergoing edge insertions and deletions with weights $[1, n^{c}]$ for a constant $c$ at all times, and $T\subseteq V(G)$ is a terminal set of at most two vertices at any time.
Let $1 \leq \tau \leq n / \log n$ be a target parameter and $0 < \varepsilon < 1$ be an error parameter. 

Then, there is a deterministic data structure that explicitly maintains a collection $\mathcal P$ of paths in $G$ that induces a $(1+ \varepsilon)$-approximate planar pattern emulator $(\mathcal M[\mathcal P],T)$\footnote{At a high level, one should think of $\mathcal M[\mathcal P]$ as a collection of terminal-to-terminal paths whose contraction, with appropriate reweighting, leads to an pattern emulator of $G$ w.r.t. the terminal set $T$. For a formal definition, see Preliminaries in Section~\ref{sec:prelim}.} of instance $(G, T)$, such that for each batch $B$ of updates in $(G,T)$ (whether edge insertions or deletions in $G$, or terminal additions or removals in $T)$, 
\begin{itemize}
\item \underline{Emulator Size}: 
    At any time, \(|\mathcal P|= \Otil(\sqrt{n\tau}/ \varepsilon^{O(1)})\), and
\item \underline{Emulator Recourse:}  
    The amortized number of paths added or deleted from $\mathcal{P}$ after processing a batch update $B$ is $\Otil(|B|/\varepsilon^{O(1)})$, and  
\item \underline{Update Time:} 
    The amortized update time is $\Otil\Big(n|B|/(\varepsilon^{O(1)} \tau)\Big)$.
\end{itemize}
    The data structure has an initialization time of $\Otil(n/\varepsilon^{O(1)})$.
    Moreover, $(\mathcal M[\mathcal P],T)$ is aligned with $(G,T)$.
\end{restatable}

\subsection{Related Works}

\paragraph{Static Min-cut on Planar and General Graphs.} We begin by reviewing some results for the $s$-$t$ maximum flow problem, which is dual to the $s$-$t$ minimum cut problem~\cite{Ford:1956aa}.
We denote by $m$ the number of edges in the graph, where $m = O(n)$ for planar graphs.

For general graphs, the $\tilde{O}(m)$ time algorithm of \citeauthor{Benczur:1996aa} \cite{Benczur:1996aa} constructs a $(1+\eps)$-approximate (edge) cut sparsifier, which can be used to compute the $s$-$t$ minimum cut value in $\Otil(n^2)$ time.
Alternatively, using the near-linear time algorithm of \citeauthor{Peng:2016aa} \cite{Peng:2016aa} for computing a $(1+\eps)$-approximate $s$-$t$ maximum flow, the $s$-$t$ minimum cut value can be obtained in $\Otil(m)$ time.
Using the exact maximum flow algorithm of \cite{Chen:2022aa}, one can compute the exact value of the $s$-$t$ minimum cut in $m^{1+o(1)}$ time.

For planar graphs, \citeauthor{Itai:1979aa} \cite{Itai:1979aa} obtained an $O(n^2 \log n)$-time algorithm for the $s$-$t$ minimum cut problem.  \citeauthor{Reif:1983aa} \cite{Reif:1983aa} later improved this to a near-linear $O(n \log^2 n)$-time algorithm using minimum weight separating cycles. \citeauthor{Frederickson:1987aa} \cite{Frederickson:1987aa} managed to further improve the running time to $O(n \log n)$. 
More recently, \citeauthor{Italiano:2011aa} \cite{Italiano:2011aa} achieved the state-of-the-art with an $O(n \log \log n)$-time algorithm.

\paragraph{Cut Oracles for Planar Graphs.}
A cut oracle is a data structure that, after preprocessing an input graph, returns the exact weight of the $s$-$t$ minimum cut for any queried vertex pair $s$ and $t$. A compact representation of the pairwise cut structure of a graph is given by the Gomory-Hu tree \cite{Gomory:1961aa}, where the value of the $s$-$t$ minimum cut corresponds to the minimum weight edge on the $s$-$t$ path in the tree.
 \citeauthor{Borradaile:2015aa} \cite{Borradaile:2015aa} showed an algorithm to construct a Gomory-Hu tree for planar graphs in $O(n \log^4 n)$ time and $O(n \log n)$ space, achieving constant query time for all-pairs $s$-$t$ minimum cut.

\paragraph{Static and Dynamic Planar Emulators.}
There is a long line of work on constructing {\it dense distance graphs} (DDGs) for a planar graph $G$.
A dense distance graph is the complete graph on a subset of vertices, where edge weights correspond to shortest-path distances in $G$.
\citeauthor{Fakcharoenphol:2001aa} \cite{Fakcharoenphol:2001aa} show how to construct DDGs in $O(n \log^3 n)$ time and support all-pairs distance queries in $O(\sqrt{n} \log^2 n)$ time.
In the offline dynamic setting, \citeauthor{Das:2022aa} \cite{Das:2022aa} design an algorithm for maintaining dense distance graphs, achieving near-optimal $\tilde{O}(\sqrt{n})$ worst-case update and query time for all-pairs distance queries.

Another line of research focuses on {\it $\alpha$-distance approximating minors} ($\alpha$-DAMs), which are minors of $G$ that preserve the distances between every pair of terminals $s, t \in T$ up to a factor $\alpha$. 
Although the majority of the literature we discuss in this section is concerned with the construction of $(1+\eps)$-approximate planar emulators, the resulting emulator may not necessarily be a minor of $G$. 
\citeauthor{Krauthgamer:2014aa} \cite{Krauthgamer:2014aa} introduced a straightforward construction for an exact DAM of size $O(|T|^4)$, which can be computed in $O(n|T|)$ time. 
Since it is not possible to construct exact DAMs with sub-quadratic size \cite{Krauthgamer:2014aa,Chang:2020aa}, introducing approximation becomes necessary to achieve smaller DAMs. 
In the $(1+\eps)$ stretch regime, \citeauthor{Cheung:2016aa} \cite{Cheung:2016aa} provided a construction with size $\Otil(|T|^2/\eps^2)$. 
More recently, \citeauthor{Chang:2025aa} \cite{Chang:2025aa} achieved a $(1+\eps)$-DAM construction of size $\Otil(|T|/\eps^{O(1)})$ for polynomially bounded weights in near-linear time.

\paragraph{Directed Planar Graphs.}
Graph algorithms for directed planar graphs are generally more challenging, and as a result the literature is comparatively sparse. 
In the static setting, \citeauthor{Borradaile:2009aa} \cite{Borradaile:2009aa} achieved an $O(n \log n)$-time algorithm for the directed $s$-$t$ min-cut problem.

The dynamic setting remains largely unexplored.\mprobst{But all the exact algos work also in directed graphs...?} An exception is the recent work by \citeauthor{Karczmarz:2024aa} \cite{Karczmarz:2024aa} who designed a data structure that, for $r \in [1, n]$, can be constructed in $\Otil(nr)$ preprocessing time and answers {\it exact} $s$-$t$ maximum flow values for queried $s$ and $t$ in $\Otil(n^{3/2}/r^{3/4})$ query time for polynomially bounded weights. 
In the approximate setting, with near-linear preprocessing time, he achieves (1) a static oracle with an $\Otil(n^{3/4})$ query time and (2) a dynamic oracle that supports edge capacity updates in $G$ with an $\Otil(n^{6/7})$ worst-case query time.

\section{Technical Overview}
This section is divided into two parts. First, in \Cref{subsec:technical-static},  we describe our near-linear-time construction of a static terminal cut sparsifier on planar graphs. Building on~\cite{Chen:2025aa}, this construction reduces the problem to computing distance pattern emulators. Then, in \Cref{subsec:technical-dynamic}, we show how to maintain such emulators dynamically using the fully dynamic all-pairs shortest paths data structure of~\cite{Filtser:2024aa}. Together, these yield our main result: a fully dynamic \emph{all-pairs $s$-$t$ min-cut} data structure for planar graphs.

Throughout this section, we use a tuple $(G,T)$, called an instance, where $G$ is an $n$-vertex planar graph and $T \subseteq V(G)$ is a terminal set for which we aim to preserve the cut or distance information (more detailed definitions will be given below).
We let $\mathcal H(G,T)$ denote a minimal set of faces in $G$ that cover all the terminals in $T$, and refer to each face in $\mathcal H(G,T)$ as a hole of $(G,T)$. 
\footnote{In the dynamic setting, we will need to support the addition or deletion of terminal vertices. Note that this in turn enforces changes on $\mathcal{H}(G,T)$, and our goal will be to ensure that the recourse to $\mathcal{H}(G,T)$ is as small as possible. As a result, while $\mathcal{H}(G,T)$ is a minimal set during initialization, this property may no longer hold after processing updates.}

\subsection{Static Terminal Cut Sparsifier} \label{subsec:technical-static}

In this subsection, for any $\eps > 0$, we explain how to construct a $(1+\eps)$-approximate terminal cut sparsifier $(H,T)$ for an instance $(G,T)$.
Specifically, we construct $H$ with the same terminal set as $G$, such that for every subset $\emptyset \subset R \subset T$ of terminals, we have
\[
    \operatorname{mincut}_G(R, T \setminus R) \leq \operatorname{mincut}_H(R, T \setminus R) \leq (1+\eps) \, \operatorname{mincut}_G(R, T \setminus R).
\]

Our algorithm is inspired by the algorithm of \cite[Section 3]{Chen:2025aa} for constructing a $(1+\eps)$-approximate terminal cut sparsifier $(H,T)$ of $(G,T)$.
We first discuss their algorithm, which will also provide the required context on the tools and concepts we use, and then proceed to describe ours.

\subsubsection{The Dual Instance}

The starting point of the discussion is the well-known duality between finding a minimum $s$-$t$ cut in $G$ and a minimum separating cycle in the {\it dual graph} $G^*$ of $G$. 
This cycle is defined as the shortest cycle in $G^*$ that separates two specific dual faces $f_s$ and $f_t$ in $G^*$ (i.e., a cycle that contains exactly one of $f_s$ and $f_t$, but not both), where $f_s$ and $f_t$ are the dual faces in $G^*$ enclosing $s$ and $t$, respectively.
The dual graph $G^*$ is constructed by associating a dual vertex $v_f$ with every face $f$ in $G$, and connecting $v_f$ and $v_{f'}$ if the faces $f$ and $f'$ share an edge in $G$.
The celebrated theorem of Itai and  Shiloach \cite{Itai:1979aa} states that the $s$-$t$ min-cut in $G$ corresponds to the $f_s$-$f_t$ minimum separating cycle in $G^*$.

As is common in the planar graph algorithms literature, our goal is to go one step further and reduce the computation of minimum separating cycles to that of shortest paths. To achieve this, instead of just taking the \emph{standard} dual graph, we first modify the primal graph and then take the standard dual, giving us an \emph{alternative dual instance} $(G_D, T_D)$ of $(G, T)$, where $T_D$ is the set of {\it dual terminal} vertices~\cite{Italiano:2010aa, Chen:2025aa}.
Intuitively, this dual ensures that every minimum separating cycle in $G^*$ is split at a dual terminal vertex; i.e., we split every dual terminal into two dual terminals, so that the path resulting from the split becomes the shortest path between the (split dual terminal) endpoints. 

\begin{figure}[t]
    \centering
    \setkeys{Gin}{width=\linewidth}
    \begin{subfigure}{0.32\textwidth}
        \centering
        \includegraphics[page=1,width=\textwidth]{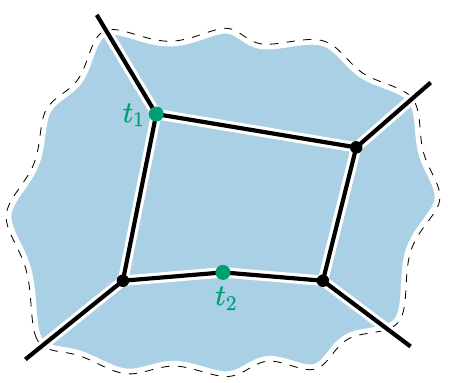}
        \caption{}
    \end{subfigure}
    \hfil
    \begin{subfigure}{0.32\textwidth}
        \centering
        \includegraphics[page=2,width=\textwidth]{dual-instance.pdf}
        \caption{}
    \end{subfigure}

    \vspace{0.5cm}

    \begin{subfigure}{0.32\textwidth}
        \centering
        \includegraphics[page=3,width=\textwidth]{dual-instance.pdf}
        \caption{}
    \end{subfigure}
    \hfil
    \begin{subfigure}{0.32\textwidth}
    \centering
    \includegraphics[page=4,width=\textwidth]{dual-instance.pdf}
    \caption{}
    \label{fig:G-D}
    \end{subfigure}
    \caption{
        An example of computing $(G_D,T_D)$ for a one-hole instance $(G,T)$.
        Each figure depicts a part of $G$ with the hole $h$ being completely shown.
        The faces are colored with a blue background:
        (a) The hole $h$ contains terminals $t_1$ and $t_2$ in $G$.
        (b) The auxiliary vertex $x_h$ (the blue vertex) is placed inside $h$, and it is connected to the terminals $t_1$ and $t_2$ (by blue edges) to form $\Gbar$.
        (c) The standard dual $\Gbar^*$ of $\Gbar$ is obtained by associating a dual vertex (orange squares) with each face of $\Gbar$ and connecting them if the associated faces in $\Gbar$ share an edge.
        Here, $f_{t_1}$ and $f_{t_2}$ are the dual vertices in $\Gbar^*$ associated with the faces containing $t_1$ and $t_2$ in $\Gbar$, respectively.
        (d) The dual $(G_D, T_D)$ is obtained by removing the edges between $f_{t_1}$ and $f_{t_2}$.
    }
    \label{fig:dual}
\end{figure}

The construction proceeds as follows.
For every hole $h \in \mathcal H(G,T)$, we first place an auxiliary vertex $x_h$ inside $h$.
Then, we connect $x_h$ to every terminal on $h$, ensuring that the connections respect the embedding.
As a result, $h$ is divided into faces, each of which is in a one-to-one correspondance with the terminals on $h$.
For a terminal $t$ on $h$, we denote the corresponding face by $f_t$.\atodo{If a terminal lays on several holes in $\mathcal H(G,T)$, you create several corresponding faces?}
Let $\Gbar$ be the graph obtained by adding all the auxiliary vertices and dividing all the holes accordingly.
We then construct the dual graph $\Gbar^*$ of $\Gbar$.
It follows from the construction that each divided face $f_t$ now corresponds to a dual vertex $v_t$ in $\Gbar^*$.
Finally, to obtain the alternate dual graph $G_D$, we remove the dual edge $v_t v_{t'}$ (if any) for every pair of terminals $t, t'$ on $h$.
The dual terminal set $T_D$ consists of all dual vertices $v_t$ for $t \in T$, and there is a one-to-one correspondence between the terminals in $T$ and the dual terminals in $T_D$.\atodo{From uniqueness, I guess that the answer to my previous question is that if a vertex lays on several holes, you pick arbitrary hole and add the edge from the terminal to the auxiliary center only there.}
Moreover, $|\mathcal H(G_D,T_D)| = |\mathcal H(G,T)|$.
Based on the construction, the dual instance of $(G_D,T_D)$ is $(G,T)$ itself, which we refer to the {\it primal instance} of $(G_D,T_D)$.\footnote{Note that the dual instance here is the alternative dual instance and not the normal dual.}
See \Cref{fig:dual} for an illustration.\atodo{The figure is very nice and helpful!}
In \Cref{subsec:dual}, we explain in more detail how $(G_D,T_D)$ is constructed.
A contribution of our work is to show to efficiently maintain $(G_D, T_D)$ under dynamic updates such that each update in $(G, T)$ results in $O(1)$ updates to $(G_D, T_D)$.

\atodoin{I think we should try and say exactly what this construction does and achieves from a higher level. Specifically, as I see it, you have a primal $G$ and dual $\overline{G}$, and then you construct $G_D$ from $\overline{G}$ by taking each $\overline{G}$ vertex $f$ that corresponds to a hole (that covers terminals $t_1,\dots,t_k$) and replacing it by a set of vertices $v_{t_1},\dots,v_{t_k}$ corresponding to the terminals. The edges of $f$ in $\overline{G}$ are split among $v_{t_1},\dots,v_{t_k}$ in $G_D$. As a result, what exact structure did we obtain?}

\subsubsection{Pattern Emulators}

A key observation in \cite{Chen:2025aa} is that for a one-hole instance $(G, T)$, any min-cut between the terminals corresponds to a union of edge-disjoint shortest paths between some dual terminals in $(G_D,T_D)$ (see \Cref{lem:one-hole-case}).
Therefore, for the one-hole instance $(G,T)$, \cite{Chen:2025aa} computes an emulator $(H_D, T_D)$ for $(G_D, T_D)$, which $(1+\eps)$-approximates any shortest path between any pair of terminals in $T_D$, and subsequently shows that the primal instance $(H, T)$ of $(H_D, T_D)$ forms a $(1+\eps)$-approximate terminal cut sparsifier for $(G, T)$.%

\begin{figure}[t]
    \centering
    \setkeys{Gin}{width=\linewidth}
    \begin{subfigure}{0.35\textwidth}
        \centering
        \includegraphics[page=1,width=\textwidth]{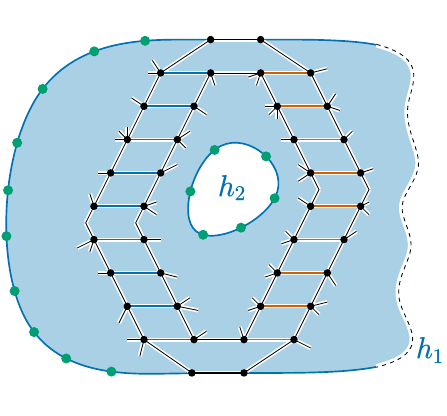}
        \caption{}
        \label{fig:pattern-prelim-G}
    \end{subfigure}
    \hfil
    \begin{subfigure}{0.35\textwidth}
        \centering
        \includegraphics[page=2,width=\textwidth]{pattern-prelim.pdf}
        \caption{}
        \label{fig:pattern-prelim-GD}
    \end{subfigure}
    \caption{
        An example of how the information of minimum cuts $E_1, E_2$ in $(G,T)$ is preserved in $(G_D,T_D)$, where $(G,T)$ consists of two holes $h_1, h_2$ with terminals $T_1, T_2$ lying on them, respectively:
        (a) The interval terminal $T_1' \subset T_1$ is depicted as green terminals on $h_1$, while $T_1 \setminus T_1'$ is on the other side of $h_1$ and is not shown. 
        The minimum cut $E_1$ is the set of blue edges (plus the top and bottom black edges on $h_1$), while the minimum cut $E_2$ is the set of orange edges (plus the top and bottom black edges on $h_1$).
        (b) The path $P_1$ in $(G_D,T_D)$ corresponding to $E_1$ is the blue path (plus the top and bottom gray edges), and the path $P_2$ in $(G_D,T_D)$ corresponding to $E_2$ is the orange path (plus the top and bottom gray edges).
        Here, $h_2'$ is the dual hole in $(G_D,T_D)$ (with dual terminals on it) corresponding to the hole $h_2$ in $(G,T)$.
        Both paths $P_1,P_2$ have the dual terminals $v_t,v_{t'}$ as their endpoint.
        Starting from $v_t$, path $P_1$ has $h_2'$ on its right and path $P_2$ has $h_2'$ on its left.
    }
    \label{fig:pattern-prelim}
\end{figure}

In the multiple-hole case, i.e., $|\mathcal H(G,T)| > 1$, the analysis becomes more involved.
This arises from the fact that when there is more than one hole in $(G,T)$, it is not sufficient to preserve only the shortest paths between the dual terminals in $(G_D,T_D)$. 
We first provide some intuition to explain why preserving only the distance between terminals is not sufficient in this case.
Assume that $(G,T)$ consists of the holes $h_1$ and $h_2$ with terminals $T_1$ and $T_2$ on them, respectively.
Thus, $T = T_1 \cup T_2$.
Let $\emptyset \subset T_1' \subset T_1$ be an interval terminal set on $h_1$.
i.e., if $t_1, \dots, t_k$ is the clockwise ordering of $T_1$ on $h_1$, then $T_1'$ is a subset $t_i, t_{i+1}, \dots, t_j$ of terminals on $h_1$, where $j \geq i$.
We discuss how the minimum cuts $E_1$ and $E_2$ corresponding to $\operatorname{mincut}_G(T_1' \cup T_2, T_1 \setminus T_1')$ and $\operatorname{mincut}_G(T_1', T \setminus T_1')$, respectively, appear in the dual instance $(G_D,T_D)$.
Note that $E_1$ and $E_2$ are cuts that separate $T_1'$ from $T_1 \setminus T_1'$, but $E_1$ separates $T_1'$ along with the entire $T_2$ from the remaining terminals, whereas $E_2$ separates only $T_1'$ from the remaining terminals.
See \Cref{fig:pattern-prelim-G} for an illustration.
Now, the shortest paths $P_1$ and $P_2$ in $(G_D,T_D)$ corresponding to $E_1$ and $E_2$ in $(G,T)$, respectively, are shown in \Cref{fig:pattern-prelim-GD}.
Here, both paths have the same dual terminals $v_t, v_{t'}$ as their endpoints, but exactly one of them, say $P_1$, is the $v_t$-$v_{t'}$ shortest path in $(G_D,T_D)$.\footnote{Throughout this paper, we assume that the shortest path between any pair of terminals is unique. This can be ensured using a standard edge weight perturbation technique.} 
However, to preserve the minimum cut $E_2$, it is also necessary to preserve $P_2$ in $(G_D, T_D)$, which cannot be accomplished solely by maintaining the shortest paths between the dual terminal pairs in $(G_D, T_D)$.
This raises a fundamental question: what should be the criteria for preserving paths in $(G_D,T_D)$ (such as $P_2$) in order to ensure that all min-cuts in $(G,T)$ are preserved?
It turns out that preserving the shortest paths with respect to the holes in $(G_D,T_D)$ is sufficient \cite{Krauthgamer:2020aa}.
In our example, denoting the hole in $(G_D,T_D)$ corresponding to $h_2$ by $h_2'$, path $P_2$ is still the shortest $v_t$-$v_{t'}$ path in $(G_D,T_D)$ that has $h_2'$ on its left (assuming traversal from $v_t$ to $v_{t'}$), while $P_1$ remains the shortest $v_t$-$v_{t'}$ path in $(G_D,T_D)$ with $h_2'$ on its right.
This example also demonstrates why this issue does not arise in the single-hole case: the paths $P_1$ and $P_2$ attempt to ``navigate'' through other holes, whereas in the single-hole case, there are no other holes to consider.
\atodoin{Comment on the paragraph of why just preserving the distances is not sufficient.\\
I think that the description is somewhat verbose. I would move the entire detail specific description to the figure caption. In the text it is better to stick to high level ideas. For example, ``as it turns our, it might be the case that two different min-cuts (separating two different subsets of terminals) will correspond to two different paths, with the exact same end points (see \Cref{fig:pattern-prelim} for an example of such a case).'' Then you continue with ``It turns out...'' }

\atodoin{Comments on \Cref{fig:pattern-prelim}:\\ 1. Don't use blue for edges. It is already overloaded.\\
2. Same for path (don't use blue).\\
3. Add graph names to the figure $G,G_D$.\\
4. It seems that in the face $h'_2$ in $G_D$, you didn't delete the edges between the terminals (as you should according to the reduction).
}

This concept of having a hole on the left or right is what we mean by navigating with respect to the holes in $(G_D,T_D)$, and it is formalized as the pattern $\Phi(P)$ of a path $P$ in $(G_D,T_D)$, defined as follows. 
Let $v_t$ and $v_{t'}$ be the endpoints of $P$, and let $P'$ be the shortest $v_t$-$v_{t'}$ path in $(G_D,T_D)$.
The set $\Phi(P)$ is defined as the set of holes that are inside the closed curve formed by $P \cup P'$.\footnote{The formal definition is more detailed and involves using the dual $G_D^*$ of $G_D$ to precisely formulate what is meant by being inside a closed curve. See \Cref{def:path-pattern} for more details.}
Thus, there are at most $2^h$ distinct patterns in an $h$-hole graph.
If $P$ is the shortest path with pattern $\Phi(P)$, we refer to it as the $\Phi(P)$-respecting $v_t$-$v_{t'}$ shortest path in $(G_D,T_D)$.

To construct the cut sparisifier, the goal is to preserve the shortest paths in $(G_D,T_D)$ for every pattern.
The emulator that, for every pattern $\Phi$, preserves all $\Phi$-respecting shortest paths between the terminals in $(G_D,T_D)$ is called the {\it pattern emulator} of $(G_D,T_D)$.
As mentioned earlier, computing a $(1+\eps)$-approximate pattern emulator $(H_D, T_D)$ of $(G_D, T_D)$ is sufficient for constructing a $(1+\eps)$-approximate terminal cut sparsifier of $(G, T)$ \cite{Krauthgamer:2020aa}.\footnote{In this paper, the pattern emulators and terminal cut sparsifiers are always planar.}
See \Cref{thm:cut-pattern-emulator} for more details.

\subsubsection{Pattern Emulator Construction of Chen and Tan~\cite{Chen:2025aa}}

We now describe how \cite{Chen:2025aa} computes a pattern emulator. Later, we show how to modify this construction so that pattern emulators can be computed in near-linear time. Since, from this point on, we focus exclusively on constructing pattern emulators, we slightly abuse notation and write $(G,T)$ instead of $(G_D,T_D)$ for the dual instance.

\paragraph{Reduction to $O(1)$-hole Instances.}

For an instance $(G,T)$ with an arbitrary number of holes, Chen and Tan~\cite{Chen:2025aa} first partition $E(G)$ into pieces $p_1, \dots, p_k$, referred to as a structured division of $G$.
Intuitively, the structured division allows for the terminals of $G$ to be evenly distributed among the pieces, ensuring that each piece contains $O(1)$ terminals.
Moreover, the boundary vertices of each piece (i.e., vertices shared between the pieces) are guaranteed to lie on $O(1)$ additional holes.
See \Cref{sec:prelim} for the precise definition.
The task of computing a  pattern emulator for $(G, T)$ then translates into the task of computing the  pattern emulator for each piece $(p_i, T_i)$, where $T_i$ is the union of $T \cap V(p_i)$ and the boundary vertices of $p_i$.
By the discussion above, $T_i$ lies on $O(1)$ holes, and thus the primary challenge reduces to computing a  pattern emulator for an instance with $O(1)$ holes.

\paragraph{The $O(1)$-hole Case.}

\begin{figure}[t]
    \centering
    \setkeys{Gin}{width=\linewidth}
    \begin{subfigure}{0.45\textwidth}
        \centering
        \includegraphics[page=2,width=\textwidth]{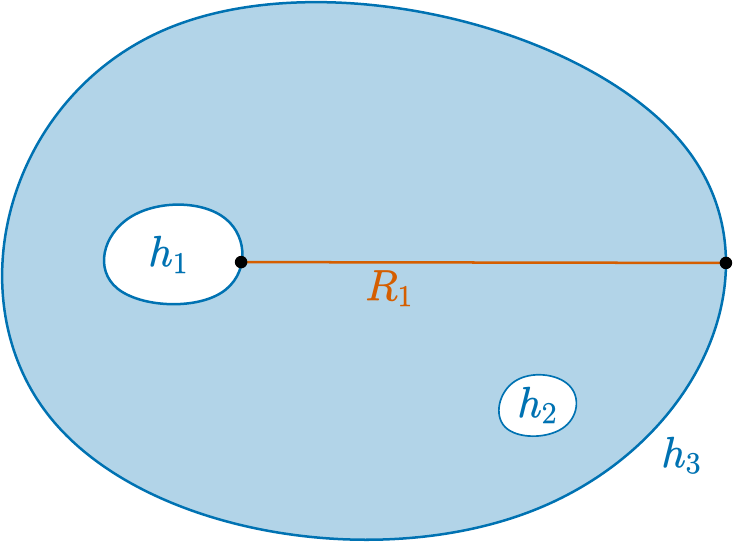}
        \caption{$G = G_1$}
        \label{fig:G1}
    \end{subfigure}
    \hfil
    \begin{subfigure}{0.45\textwidth}
        \centering
        \includegraphics[page=3,width=\textwidth]{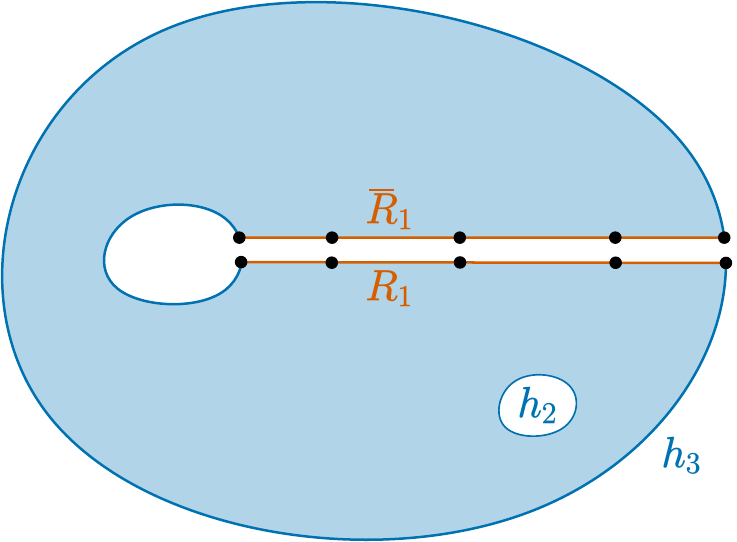}
        \caption{$\Gbar_1$}
        \label{fig:Gbar1}
    \end{subfigure}
    
    \vspace{0.5cm}

    \begin{subfigure}{0.45\textwidth}
        \centering
        \includegraphics[page=4,width=\textwidth]{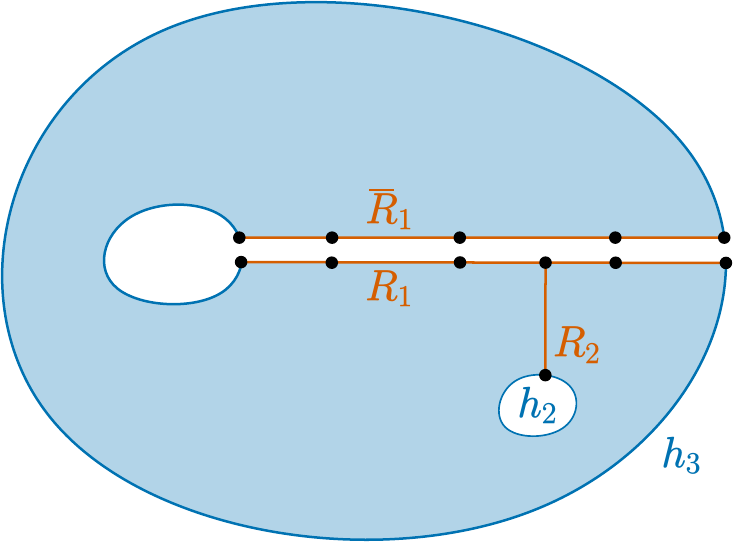}
        \caption{$G_2 = \Gbar_1$}
        \label{fig:G2}
    \end{subfigure}
    \hfil
    \begin{subfigure}{0.45\textwidth}
        \centering
        \includegraphics[page=5,width=\textwidth]{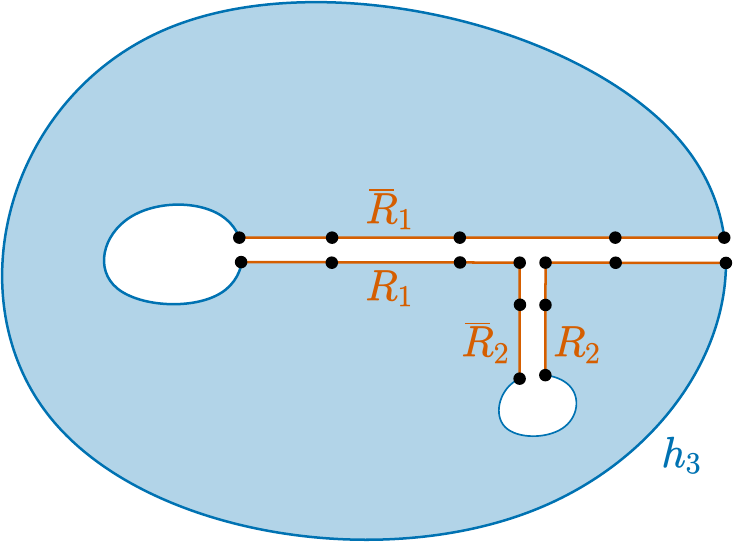}
        \caption{$\Gbar_2 = G_3 = G_3'$}
        \label{fig:Gbar2}
    \end{subfigure}
    \caption[Caption for LOF]{
        An illustration for the slicing procedure used in \Cref{subsub:approx-pattern} for an $h$-hole instance $(G,T)$ with $h=3$ and holes $h_1, h_2, h_3$.
        Starting with $G_1 = G$:
        (a) The path $R_1$ is the shortest path connecting any pair of holes in $G_1$ (e.g., connecting $h_1$ to $h_3$).\protect\footnotemark{}
        (b) The sliced graph $\Gbar_1$ is obtained from $G_1$ by slicing along $R_1$ using \Cref{alg:slice}. The depicted vertices on $R_1$ (and $\Rbar_1$) are the portals (and their copies on $\Rbar_1$) as defined in \Cref{def:portals-all}. Note that, after slicing, $h_1$ is merged into $h_3$.
        (c) The path $R_2$ is a shortest path between any pair of holes in $G_2 = \Gbar_1$.
        (d) The graph $\Gbar_2$ is obtained from $\Gbar_1$ by slicing along $R_2$. Similarly, the portals on $R_2$ and $\Rbar_2$ are depicted. By our definition, $G_3 = \Gbar_2$, and since $h=3$, we have $G_3 = G_3'$.
        Note that $h_2$ is merged into $h_3$.
    }
    \label{fig:slice-definition}
\end{figure}

Assume now that $(G, T)$ is an instance with $h = O(1)$ holes.
To compute a pattern emulator for $(G, T)$, \cite{Chen:2025aa} first employs a similar approach to \cite{Chang:2022aa} to reduce $(G, T)$ into a single-hole instance $(G_h, T_h)$ as follows.
Starting with $G_1 = G$ and $T_1 = T$, they select two arbitrary terminals $t_1, t_2 \in T_1$ lying on different holes in $(G_1,T_1)$ and compute the shortest $t_1$-$t_2$ path $R_1$ in $G_1$.
Next, they slice along $R_1$ to obtain $(G_2,T_2)$; they add a copy $\Rbar_1$ of $R_1$, and for every vertex $v$ on $R_1$, they reassign every edge connected to $v$ on the right side of $R_1$ (with respect to the traversal from $t_1$ to $t_2$) to the vertex $\overbar v$ on $\Rbar_1$ that is the copy of $v$.\footnote{The exact procedure is more involved and requires additional details to precisely define the right-side and left-side of $R_1$. See \Cref{alg:slice} for the detailed slicing procedure.}
This operation merges the holes containing $t_1$ and $t_2$ into a single hole, thus reducing the number of holes in $(G_2,T_2)$ to $h-1$.
This process continues iteratively until a one-hole instance $(G_h,T_h)$ is obtained.
Specifically, for $2 \leq i \leq h-1$, we select two terminals $t_1$ and $t_2$ on different holes of $(G_i,T_i)$ and find the shortest $t_1$-$t_2$ path $R_i$.
The graph $G_i$ is then sliced along $R_i$ to obtain $G_{i+1}$.
See \Cref{fig:slice-definition} for an illustration of the slicing procedure.

\footnotetext{As we will discuss later in this subsection, our slicing procedure uses a different criterion to identify the smallest path $R_i$ between {\it any} holes in $G_i$, rather than defining $R_i$ as in \cite{Chen:2025aa} as the shortest path between any arbitrary terminals lying on different holes of $G_i$. See \Cref{sec:prelim} for more details.}
\atodoin{Comments on \Cref{fig:slice-definition}. Figure is generally good.\\
1. Why (b) is $\bar{G}_1$ while (c) is $G_2=\bar{G}_1$, isn't it suppose to be the same graph?\\
2. I would've also added terminals to (a), and portals to (a), marked differently. Then in (b) this portals should look like terminals, as they got terminal status.
}

To approximately preserve a $\Phi$-respecting $t_1$-$t_2$ shortest path $P$ in $(G,T)$, we next discuss how to expand terminal set $T_i$ based on $T_{i-1}$. To illustrate this, let us first show the intuition behind the expansion of the terminal set $T_2$ in $G_2$, where we distinguish two cases. 
(1) If $P$ does not cross $R_1$, then it remains unchanged in $(G_2,T_2)$. (2) However, if $P$ crosses $R_1$, it is divided into subpaths in $(G_2,T_2)$ (\Cref{fig:slice} illustrates a case where $P$ crosses $R_1$).
It follows that each subpath in $(G_2,T_2)$ is a shortest path between its endpoints with a specific pattern.\footnote{Note that this pattern is different from $\Phi$; the path $P$ no longer exists in $(G_2,T_2)$, and its subpaths, resulting from the slicing of $R_1$, have their patterns defined with respect to the subpath and its endpoints in $(G_2,T_2)$.}
Therefore, considering all the endpoints of these subpaths as terminals in $G_2$ ensures that all the subpaths are preserved.
The major drawback of this approach is that it would require adding $\Omega(n)$ vertices from $T_1 = T$ to $T_2$; different paths may cross $R_1$ at different vertices, and it could be the case that $|R_1| = \Omega(n)$.
To address this issue, \cite{Chen:2025aa} introduces a subset of vertices on $R_1$, called {\it pattern portals}, as additional terminals of $G_2$ (alongside $T_1$), which approximately preserves every $\Phi$-respecting shortest path $P$ crossing $R_1$ for every pattern $\Phi$.

\paragraph{Pattern Portals.}
To explain how pattern portals are selected on $R_1$, we first briefly review the construction of an $\eps$-cover \cite{Thorup:2004aa,Klein:2002aa}, referred to as {\it portals}, which serves as the foundation for the pattern portal construction.
For a terminal $t \in T_1$ not on $R_1$, the portal set of $t$ on $R_1$, denoted by $\portal(R_1, t, \eps)$, is constructed as follows.
First, find the closest vertex $x$ on $R_1$ to $t$.
This vertex induces two maximal subpaths on $R_1$, with $x$ serving as an endpoint.
The algorithm adds $x$ to $\portal(R_1, t, \eps)$ and proceeds as follows for each maximal subpath.
Starting at $y = x$, the algorithm traverses toward the other endpoint of the subpath, selecting the first vertex $z$ (if any) as a portal if $(1+\eps) \dist(t, z) < \dist(t, y) + \dist(y, z)$.
It then sets $y = z$ and repeats the process until reaching the endpoint of the subpath.
This construction ensures that $|\portal(R_1, t, \eps)| = O(1 / \eps)$.
Thus, for the portal set across all terminals, $|\cup_{t \in {T_1}} \portal(R_1, t, \eps)| = O(|T_1| / \eps)$, which is significantly smaller than the $\Omega(n)$ bound that would arise if all vertices on $R_1$ were considered as terminals.
Moreover, computing this portal set when $T_1$ lies on $O(1)$ holes takes $O(n \log |T_1|)$ time \cite[Lemma 2.4]{Chang:2022ab}.
\atodoin{Too long, too many details. This is not new, we don't modify or adapt this construction in any way. There is no reason to go into these details. It will be enough to just state the black box and it's properties. If you prefer, you might first state the black box of \cite{Thorup:2004aa,Klein:2002aa} and then the \cite{Chang:2022aa} adaptation, or just state the \cite{Chang:2022aa} addaptations as is (and only say that it is based on \cite{Thorup:2004aa,Klein:2002aa}).}

The key idea behind computing pattern portals in~\cite{Chen:2025aa} is to compute portals for every pattern.
i.e., for any pattern $\Phi$, the portal set of $t \in T_1$ with respect to $\Phi$ is computed on $R_1$ in a way similar to the discussion above, but the criterion now becomes $(1+\eps) \dist^{\Phi}(t, z) < \dist^{\Phi}(t, y) + \dist^{\Phi}(y, z)$, where $\dist^{\Phi}(x, y)$ is the length of the shortest $\Phi$-respecting $x$-$y$ path.
When $(G,T)$ consists of $h$ holes, it follows that there are $O(2^h |T_1| / \eps)$ pattern portals on $R_1$, which are added to $T_1$ to form $T_2$.
The expansion of $T_{i+1}$ is performed similarly by adding to $T_i$ the pattern portals of $R_i$ computed for each terminal in $T_i$.
It follows that the size of $T_h$ is $O(h^2 2^h |T| / \eps)$, which is $O(|T| / \eps)$ when $h = O(1)$.
In summary, pattern portals prevent a blow-up in the sizes of the terminal sets $T_1,\dots,T_h$ while still ensuring that all patterns are approximately preserved. However, it is unclear how to turn their polynomial-time construction into a near-linear-time one, which is the key step toward obtaining a dynamic pattern emulator.

\begin{figure}[t]
    \centering
    \setkeys{Gin}{width=\linewidth}
    \begin{subfigure}{0.45\textwidth}
        \centering
        \includegraphics[page=6,width=\textwidth]{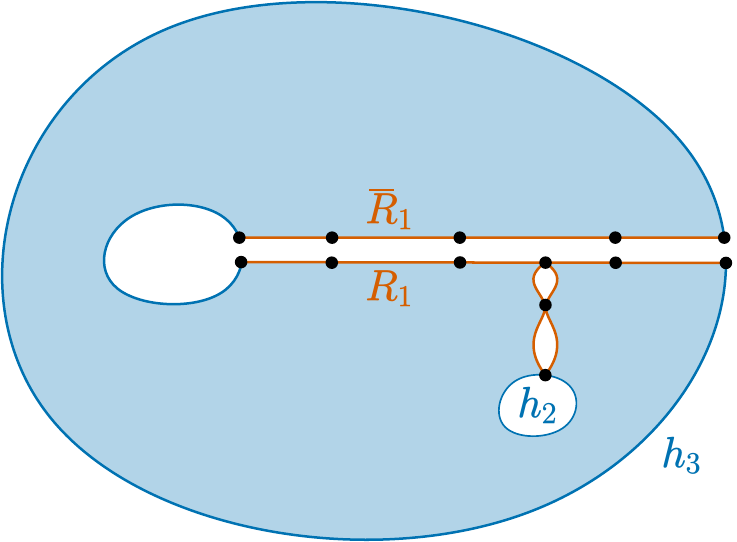}
        \caption{$G_2'$}
        \label{fig:G2prime}
    \end{subfigure}
    \hfil
    \begin{subfigure}{0.45\textwidth}
        \centering
        \includegraphics[page=7,width=\textwidth]{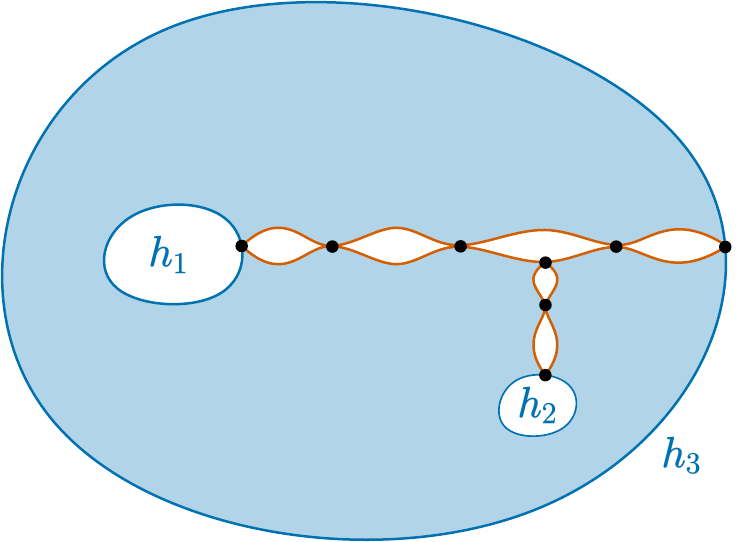}
        \caption{$G_1'$}
        \label{fig:G1prime}
    \end{subfigure}
    \caption{
        An illustration for the gluing notation used in \Cref{subsub:approx-pattern} for the $h$-hole instance $(G_h,T_h)$ obtained from $(G,T)$ in \Cref{fig:slice-definition} with $h=3$ and holes $h_1, h_2, h_3$.
        (a) The glued graph $G_2'$ is obtained from $G_3'$ by gluing back the portals on $R_2$.
        This reestablishes the hole from $G_2$ that was merged to form $G_3$.
        For example, the hole $h_2$ reappears, and, in general, $G_i'$ has the same number of holes as $G_i$.
        (b) The glued graph $G_1'$ is obtained from $G_2'$ by gluing back the portals on $R_1$.
    }
    \label{fig:glue-definition}
\end{figure}

\paragraph{The Bottom-up Procedure.}
After obtaining the instances $(G_1, T_1), \dots, (G_h, T_h)$ by slicing along the path $R_i$ on $G_i$ for $1 \leq i \leq h-1$ in a top-down manner, the  pattern emulator of $(G, T)$ is then computed in a bottom-up fashion. Starting from the single-hole instance $(G_h, T_h)$,~\cite{Chen:2025aa} computes an emulator $(H_h, T_h)$ for $(G_h, T_h)$.
Then, for each $1 \leq i \leq h-1$, the emulator $(H_i, T_i)$ is constructed from $(H_{i+1}, T_{i+1})$ by gluing every portal $z$ on $R_i$ with its copy $\overbar z$ on $\Rbar_i$.
Note that $(H_i, T_i)$ has $T_i$ as its terminal set; during the gluing process, each glued portal on $R_i$ loses its status as a terminal in $(H_i, T_i)$.
It follows that $(H_i, T_i)$ has $h-i+1$ holes, and the instance $(H_1, T_1) = (H, T)$ obtained through this procedure is the  pattern emulator of $(G_1, T_1) = (G, T)$.
See \Cref{fig:glue-definition} for an illustration of the gluing procedure.
\atodoin{It is not clear from the description, the portals lose their terminal status, or do we keep them as terminals? In the second option, you created additional holes... So I guess not? Say so explicitly.}

\subsubsection{Our Contribution}

In the static setting, our main conceptual contribution in is to show that using the standard portals of Thorup-Klein~\cite{Thorup:2004aa,Klein:2002aa} instead of pattern portals is sufficient for preserving all the pattern information.
In other words, we can set $T_{i+1} = T_i \cup \bigcup_{t \in T_i} \portal(R_i, t, \eps'')$, where $\eps'' = \eps / 2^{O(h)}$ is an error parameter, and still obtain a pattern emulator $(H,T)$ of $(G,T)$ by gluing the portals back to their copies in a bottom-up fashion, as described above. Although this does not improve the asymptotic bound on the number of portals used, since $h=O(1)$, it has two important consequences. First, the contruction based on the framework of~\cite{Chang:2022aa} yields a pattern emulator. Second, as a useful byproduct, this construction gives the first near-linear-time implementation of pattern emulators, without relying on additional preprocessing steps such as division or bootstrapping, while simultaneously simplifying the algorithm.

More importantly, we leverage these insights in the dynamic setting to maintain a dynamic pattern emulator based on the dynamic (standard) emulator construction of \citeauthor{Filtser:2024aa} \cite{Filtser:2024aa}, as outlined in \Cref{subsec:technical-dynamic}.

For the remainder of this subsection, we give a high-level technical discussion of where our construction deviates from that of \cite{Chen:2025aa} to show that portals are sufficient instead of pattern portals.
\begin{enumerate}
    \item 
        Our first deviation is in the selection of $R_i$; we choose $R_i$ to be the shortest path between the {\it closest} pair of holes in $(G_i, T_i)$, rather than the shortest path between any pair of holes. 
        This adjustment simplifies our analysis. We define patterns based on the one-hole instance $(G_h,T_h)$, where we assign a {\it unique} path $P_h$ in $(G_h,T_h)$ for every $\Phi$-respecting path $P$ in $(G,T)$.
        The pattern $\Phi$ of $P$ is then defined using both $P$ and $P_h$ (see \Cref{def:pattern-emulator}).
        Slicing along the closest pair of holes allows us to obtain a unique $(G_h,T_h)$, and therefore a unique $P_h$, thus significantly simplifying our pattern analysis for every instance $(G_i, T_i)$.
        See the proof of \Cref{lem:exact-pattern} for an example of how this adjustment simplifies the analysis.
        \atodoin{I don't understand the difference. What is a hole which is not a $(G_i,T_i)$ hole? This is really unclear and I have no idea what you mean here. There should be no holes in $G_i$ which are not $(G_i,T_i)$ holes.}
    \item 
        Another deviation is in our $\eps$-cover procedure.
        We adapt the algorithm of \cite{Thorup:2004aa,Klein:2002aa} for a parameter $\eps'' = \eps/2^{O(h)}$: for the closest vertex $x$ on the path to any terminal $t \in T$, starting from $y = x$, we select the first edge $z'z$ (if any) such that $\dist(x, z) > (1+\eps'') \dist(x, y)$, add $z$ and $z'$ to the portal set, and then repeat the procedure by setting $y = z$ until we reach the other endpoint.
        Moreover, starting from $x$, we choose the edge $z_i'z_i$ (if any) to be the first edge from $x$ satisfying $\dist(x, z_i) > i \eps'' \dist(t,x)$ for $1 \leq i \leq 2/\eps''$ and add $z_i'$ and $z_i$ to the portal set.
        As a result of these changes, our $\eps''$-cover has a size of $O(|T| \log n / \eps)$ instead of $O(|T| / \eps)$, which we can still afford.
        See \Cref{def:portals-all} for the precise definition of $\eps''$.
        
        To determine the value for $\eps''$, we first prove that if $P$ crosses $R$, then $R$ segments $P$ into at most $2^{h+2}$ pieces (see \Cref{clm:subpaths} for more details). 
        Thus, by setting $\eps'' = \eps / 2^{O(h)}$, the distances between the portals become granular enough such that the distance between any two consecutive portals on $R$ is $O(\eps |P|)$ (see \Cref{clm:portals}). 
        This enables us to successfully approximate the length of any $\Phi$-respecting shortest path $P$ in $(G, T)$ by $(1+\eps)|P|$ in $(H, T)$, as it ensures that if $P$ crosses $R_i$, the additional cost incurred when rerouting $P$ through the portals remains sufficiently small relative to $|P|$.
        See \Cref{clm:glued-approximation} for further details.
        \atodoin{Too long too detailed. As I understand, we just use a black box here. We don't need to discuss the details of the black box, only the guarantees it provides, and why they are good enough for our usage.\\
        This second and more important part (of why it is good enough to use classic portals) is missing.}
    \item 
        Another deviation is that we extend the definition of the $\Phi$-respecting shortest path $P$ to allow $P$ to be any path, not necessarily a simple path (as used in \cite{Krauthgamer:2020aa,Chen:2025aa}). 
        As mentioned above, to simplify the proofs, we define the pattern $\Phi$ of a path $P$ in $(G,T)$ using a unique path $P_h$ in $(G_h,T_h)$. 
        Similarly, every path $P_i$ in $(G_i, T_i)$ is also defined using its unique path in $(G_h, T_h)$. 
        The analysis for each such path $P_i$ considers the image of $P_i$ in $(G, T)$, which is obtained by naturally contracting the portals and tracking how $P_i$ evolves. 
        Since the image of $P_i$ may not necessarily be simple in $(G,T)$, generalizing the definition helps us to focus on the image of $P_i$ without requiring us to find another simple path between its endpoints in $(G, T)$. 
        Moreover, this generalization allows us to prove that $R_i$ segments the path $P$ into at most $2^{h+2}$ pieces in \Cref{clm:subpaths}, which we will later use to show that $\eps''$-covers are sufficient to approximate $P$ in the glued instances (see \Cref{clm:glued-approximation}).
        \atodoin{We might want to add something regarding why we need this generalization, while \cite{Chen:2025aa} don't need it.
        I guess it is related to the inductive step when you add a hole? But still, even under \cite{Chen:2025aa} defs it seems that not simples paths are created during the argument... }
\end{enumerate}

In conclusion, we present a method to use $\eps''$-covers instead of requiring pattern portals as in \cite{Chen:2025aa}.
Thus, for a sufficiently fine-grained $\eps''$, employing the \cite{Chang:2022aa} construction (which utilizes $\eps''$-covers to compute portals) also results in pattern emulators.
This shows that the \cite{Chang:2022aa} construction is stronger than previously believed.

\subsection{Dynamic All-Pairs $s$-$t$ Min-cut} \label{subsec:technical-dynamic}

In this subsection, we describe how our main observation from \Cref{subsec:technical-static} (that using $\eps$-covers as portals in the CKT emulator construction \cite{Chang:2022aa} leads to a  pattern emulator) can be combined with the dynamic version of the CKT \cite{Filtser:2024aa} to achieve a fully dynamic all-pairs $s$-$t$ min-cut with subpolynomial  update and query time.
Recall that an update on the instance $(G, T)$ can involve either an edge insertion or deletion in $G$, or the addition or removal of a terminal in $T$.

The high-level intuition is to maintain the dual instance $(G_D, T_D)$ of $(G, D)$ as defined in \Cref{subsec:technical-static}, and subsequently run the dynamic data structure of \cite{Filtser:2024aa} on $(G_D, T_D)$. 
As discussed in \Cref{subsec:technical-static}, the maintained emulator $(H_D, T_D)$ is a  pattern emulator that ensures every cut in $(G,T)$, separating a subset of terminals from the rest, corresponds to a union of shortest paths between certain dual terminals in $T_D$. 
To answer $s$-$t$ queries, we then compute the primal instance $(H, T)$ of $(H_D, T_D)$ and run a static $s$-$t$ min-cut algorithm on $(H,T)$. 
We now elaborate on these steps in more detail.
\subsubsection{Maintaining the Dual Instance}
\begin{figure}[t]
    \centering
    \setkeys{Gin}{width=\linewidth}
    \begin{subfigure}{0.32\textwidth}
        \centering
        \includegraphics[page=4,width=\textwidth]{dual-instance.pdf}
        \label{fig:}
    \end{subfigure}
    \hfil
    \begin{subfigure}{0.32\textwidth}
        \centering
        \includegraphics[page=7,width=\textwidth]{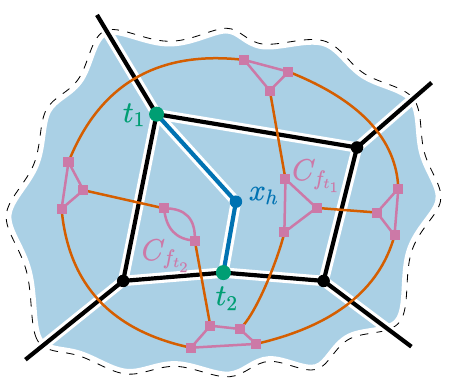}
        \caption{}
    \end{subfigure}
    \caption{
        An example of $(G_D,T_D)$ from \Cref{fig:dual} and the graph $(G_D^*,T_D^*)$ constructed based on it:
        (a) The dual instance $(G_D,T_D)$ of \Cref{fig:G-D}.
        (b) The instance $(G_D^*,T_D^*)$ is obtained from $(G_D,T_D)$ by replacing each dual vertex $f$ with a $\deg_{G_D}(f)$-edge cycle $C_f$ and connecting each vertex of $C_f$ to exactly one of the neighbors of $f$ while respecting the embedding.
    }
    \label{fig:dual-cycle}
\end{figure}

As we run the data structure on the dual instance $(G_D, T_D)$, we need to efficiently maintain $(G_D, T_D)$ after each update in $(G, T)$.
Unfortunately, directly maintaining $(G_D, T_D)$ could result in an $\Omega(n)$ update time. 
For example, consider a face $f$ that is neighboring $\Omega(n)$ other faces in $G$, and let $v_f$ be the dual vertex in $G_D$ corresponding to $f$.
The insertion of an edge in $G$ that splits $f$ can result in a vertex split of $v_f$ in $G_D$, where we need to split $v_f$ into $v_f$ and $v_{f'}$, and reassign almost half of the edges neighboring $v_f$ to $v_{f'}$.
This update takes $\Omega(n)$ time to process.
To overcome this issue, similar to \cite{Italiano:2010aa},\atodo{Shouldn't we cite the conference version and not the arxiv? \href{https://dblp.org/rec/conf/stoc/ItalianoNSW11.html}{DBLP}. more authors, but might be the same paper} we maintain an instance $(G_D^*, T_D^*)$ based on $(G_D, T_D)$ that translates each update in $(G, T)$ to $O(1)$ updates in $(G_D^*, T_D^*)$ (see \Cref{lem:dynamic-instance}).%
The graph $G_D^*$ is constructed as follows.
For every vertex $v_f$ in $G_D$, we replace it with a cycle $C_f$ of $\deg_{G_D}(v_f)$ edges, each edge having a zero length.
Each vertex on $C_f$ is assigned exactly one edge neighboring $v_f$ in $G_D$, such that the constructed graph is planar and respects the embedding. 
See \Cref{fig:dual-cycle} for an illustration.
To form $T_D^*$, if $v_f \in T_D$, we then arbitrarily pick one of the vertices on $C_f$ in $G_D^*$ to be a terminal vertex in $T_D^*$.
It then follows that a single update in $(G, T)$ translates to $O(1)$ edge deletions and insertions between the cycles in $(G_D^*,T_D^*)$ to simulate vertex splits.
See \Cref{subsec:dual} for a more detailed discussion and \Cref{fig:dual-cycle-update} for an update illustration.

\subsubsection{The Dynamic Data Structure}
We then run the data structure of \cite{Filtser:2024aa} on $(G_D^*, T_D^*)$ instead of $(G_D,T_D)$.\footnote{Note that since $(G_D^*, T_D^*)$ contains $0$-weight edges, directly applying the dynamic data structure to $(G_D^*, T_D^*)$ will not compile, as the weights must lie within $[1, n^c]$. This issue can be addressed by replacing the zero weights with $1/\poly(n)$ weights and subsequently rescaling the weights.}
To briefly discuss how the data structure works, for simplicity of notation, we explain how we run the data structure of \cite{Filtser:2024aa} on instance $(G, T)$ instead.

We use terminal updates only when handling min-cut queries; upon an $s$-$t$ min-cut query, we add $s$ and $t$ to the terminal set $T$ and maintain the pattern emulator. 
After answering the query (details will be discussed later), we revert the addition of $s$ and $t$ to $T$. 
Thus, during edge updates, we always have $T = \emptyset$, and in the following discussion, we focus on maintaining a pattern emulator for the instance $(G, \emptyset)$.\footnote{As we will see later, upon receiving an $s,t$ query, our data structure adds $s$ and $t$ to the terminal set and thus maintains a pattern emulator for the instance $(G, \{s,t\})$. The main challenge is therefore to maintain the pattern emulator in subpolynomial time, since the empty graph is trivially a pattern emulator for $(G, \emptyset)$.}

On a high level, to maintain a $(1+\eps)$-approximate pattern emulator $(H,\emptyset)$ of $(G,\emptyset)$,%
we partition $E(G)$ into small enough pieces and then compute a pattern emulator on each piece.\footnote{The pieces must be sufficiently small to ensure that the recourse remains relatively low after each update. To achieve this, we define different types {\tt 1} to {\tt 4} of instances that successively refine one another, ultimately resulting in pieces of sufficiently small size.}
The union of each such piece's pattern emulator forms the pattern emulator $(H, \emptyset)$.

\atodoin{You dive here straight into Type 1-4 jargon. I suggest to start with a high level description. Assume your reader knows nothing about \cite{Filtser:2024aa}. 
Start with the basic idea: hierarchy of pattern emulators of decreasing size. Update them in a cascading fashion. Preform the actual computation to answer the query on the small one.
This is you've set up the first goal: creating an Emulator for empty set of terminals that has small update time.
Next you go into more details of this step. Here you say that one-hole small number of terminals case is easy, called type 1 and is the basis of the recursion. Now tell about the other types. Only what they mean and one key word on the recursive step between them. After you've set up the framework, what we are looking for, and where we are going to, you can start talking about ``to form the pieces'').}
To form these pieces, the data structure starts with what it calls \four instances and reduces their size to obtain \three instances with $O(1)$ holes.
Then, it reduces the number of holes in \three instances to obtain \two instances with one hole.
Finally, the pieces we are looking for, called \one instances, are obtained from \two instances by splitting them to reduce their terminal set.
This procedure for obtaining \one instances naturally forms a tree $\Tcal$, called the {\it decomposition tree}, where each node of $\Tcal$ corresponds to an instance $(R, S)$ of a specific type, and its leaves are \one instances.
We now provide a more detailed explanation of how we compute different types of instances and use them to construct $\Tcal$ in the initialization phase.
\atodoin{The types above are not clear enough. You can try borrow text from FOCS24 paper (e.g. see page 4 \href{https://arnold.filtser.com/papers/FGPP24.pdf}{here}.) 
You should either give high level description with the purpose of each. Or just say that there are different types and then give their description as part of the list below. The current paragraph is hard to read.
}

\paragraph{Initialization.}
For the $n$-vertex graph $G$, target parameter $1 \leq \tau \leq n / \log n$, and error parameter $0<\eps<1$, an instance $(R, S)$:

\begin{itemize}
    \item 
        Is of type \four if $|R| > n / \tau$ and it contains $O(1)$ holes.
        The children of $(R, S)$ in $\Tcal$ are the pieces $(R_1, S_1), \dots, (R_k, S_k)$ obtained by using a division algorithm (see \Cref{thm:r-division})\atodo{What is this mysterious division alg? isn't this just $r$-division, perhaps with additional properties?} that guarantees each instance $(R_i, S_i)$ has $O(1)$ holes with respect to its parent $(R, S)$.
        Each $S_i$ consists of the boundary vertices (i.e., any vertex in $V(R_i)$ that also belongs to $V(R_j)$ for $j \neq i$) and the subset $S \cap V(R_i)$ of $S$ that is present in $R_i$.
        Note that $(G, \emptyset)$ is a \four instance with no holes, which serves as the root of $\Tcal$.
    \item 
        Is of type \three if $|R| = O(n / \tau)$ and it contains $O(1)$ holes.
        The instance $(R, S)$ has only one child $(R', S')$ in $\Tcal$, obtained by slicing along the path $\pi_{uv}$ between the closest holes (as explained in \Cref{subsec:technical-static} and \Cref{alg:slice}) with endpoints $u,v$, where $\pi_{uv}$ is the shortest path between the closest holes in $(R,S)$.
        The terminal set $S'$ consists of $S$ and the portal set $\portal(R, \pi_{uv}, \eps)$ and its copies on $\overbar \pi_{uv}$ as defined in \Cref{subsec:technical-static}.\footnote{More precisely, we use our $\eps'$-cover to construct $\portal(R, \pi_{uv}, \eps')$, where $\eps' = \eps / 2^{O(h)}$ and $h = O(1)$ is the number of holes in the instance $(R, S)$.}
    \item 
        Is of type \two if $|R| = O(n / \tau)$ and it is a one-hole instance.
        The instance $(R, S)$ has two children $(R_1, S_1)$ and $(R_2, S_2)$ in $\Tcal$, which are obtained from $(R, S)$ by splitting it such that $|S_1|, |S_2| \leq \frac{11}{12} |S|$ using a balanced shortest path separator $\pi_{uv}$ (see \Cref{thm:separator}) with endpoints $u,v$.
        Again, $S_1$ and $S_2$ contain the subset of $S$ that is in $R_1$ and $R_2$, respectively, plus the portal set $\splitportal(R, \pi_{uv}, \eps)$.
        Somewhat surprisingly, \cite{Chang:2022aa} shows that one can pick a portal set $\splitportal(R, \pi_{uv}, \eps)$ of size $O(\log n/\eps)$ that satisfies the $O(\eps)$-cover property.\footnote{Throughout this paper, we assume that the edge weights are polynomially bounded; in the terminology of \cite{Chang:2022aa}, this corresponds to the small-spread case.}
        \atodoin{Note on the footnote: are you sure that poly (in n) weights correspond to \cite{Chang:2022aa}, this corresponds to small-spread case? As I recall in \cite{Chang:2022aa} the small spread case is polynomial w.r.t. the number of terminals $k$ and not $n$. So we don't get to be in the small-spread case just by assuming polynomial aspect ratio.\\
        As I understand it, we are not in \cite{Chang:2022aa} small-spread case. Rather, we just don't care and run the algorithm of this case. As a result, the size of the emulator will depend on $n$. \cite{Chang:2022aa} wanted to avoid even polylog dependence on $n$, and this is the reason they have the second complicated case.}
    \item 
        Is of type \one if $|R| = O(n / \tau)$, it is a one-hole instance, and $|S| \leq \log^{10} n$.
        The instance $(R, S)$ is a leaf node in $\Tcal$.
\end{itemize}

The pattern emulator $(H, T)$ of $(G, T)$ is constructed based on $\Tcal$ in a bottom-up fashion: for a \one instance $(R, S)$, we use \cite{Krauthgamer:2014aa} \atodo{Emphasize that this take poly time, but this is fine, as the instances are small.} to construct an {\it exact} pattern emulator $(H, S)$ of $(R, S)$ of size $|H| = \poly(\log n)$ (see \Cref{thm:emulator-exact}).
For any other type of instance $(R, S)$, we define its pattern emulator $(H, S)$ using the pattern emulators of its children in $\Tcal$ as follows.

\begin{itemize}
    \item 
        For a \two instance $(R, S)$, we obtain $(H, S)$ by gluing $(H_1, S_1)$ and $(H_2, S_2)$ on the portals in $\splitportal(R, \pi_{uv}, \eps)$, where $(H_i, S_i)$ is the pattern emulator of the child $(R_i, S_i)$ in $\Tcal$ for $i = 1, 2$.
        i.e., we glue every portal $v \in \splitportal(R, \pi_{uv}, \eps) \cap V(H_1)$ with its copy in $H_2$.
        Using induction, it follows that $|H| = |S| \cdot \poly(\log n) / \eps^{O(1)}$.
        \atodoin{Say something on why it works. Basically there are no holes, so the previous proof just works.}
    \item 
        For a \three instance $(R, S)$, we obtain $(H, S)$ by gluing $(H', S')$ on $\portal(R, \pi_{uv}, \eps)$,\footnote{Recall from \Cref{subsec:technical-static} that in the sliced graph, each portal $v \in \portal(R, \pi_{uv}, \eps)$ has a copy $\overbar{v}$ on the copy $\overbar \pi_{uv}$ of the sliced path.} where $(H', S')$ is the pattern emulator of its child $(R', S')$ in $\Tcal$.
        As we merge vertices, it follows that $|H| = |S| \cdot \poly(\log n) / \eps^{O(1)}$.
        \atodoin{This is the main place where we differ from the previous work. Could you say something on the differences, like now it is patter emulator, but still works and why (if possible to say something easily)?}
    \item 
        For a \four instance $(R, S)$, we set $(H, S)$ as the union $\cup_{i=1}^k (H_i, S_i)$ in $\Tcal$, where $(H_i, S_i)$ is the pattern emulator of its child $(R_i, S_i)$ in $\Tcal$.
        Using the property of the division, it follows that $|H| = \Otil(\sqrt{n \tau} / \eps^{O(1)})$.
\end{itemize}

For more details on the initialization and the size guarantees, see \Cref{subsubsec:initialization}.

\paragraph{Handling the Updates.}

To maintain $(H, \emptyset)$ after each update in $(G, \emptyset)$, we first update the instances in $\Tcal$ by passing the updates in a top-down manner from the root of $\Tcal$ to the affected instances.
After passing the updates, we recompute\atodo{add (from scratch)} the pattern emulator for any affected \one instances and update the pattern emulators of their ancestor instances in $\Tcal$ in a bottom-up fashion accordingly.
i.e., if the pattern emulator of a child of $(R, S)$ is updated, we then update $(H, S)$ by recomputing it using the procedure explained above with the updated pattern emulators of its children.
It follows that after each update in $(G, \emptyset)$, at most $O(\log^2 n)$ instances in $\Tcal$ are affected, and they can be maintained in $\Otil(n / (\eps^{O(1)} \tau))$ amortized update time.
See \Cref{subsubsec:updates} for more details on how the updates are handled and their update guarantees.
\atodoin{The paragraph above should be rewritten. You were given a lot of details before, and now you just jumped over all of them. It is ok if you don't want to go into a lot of details. But there should be at least a clear high level description, and a more detailed description of the differences from the previous FOCS paper.}
\bigskip

Based on the discussion above, the maintained pattern emulator $(H, \emptyset)$ of $(G, \emptyset)$ is of size $\Otil(\sqrt{n \tau} / \eps^{O(1)})$, which is too large for our purpose. 
For example, assume that we want to find a $(1+\eps)$-approximate $s$-$t$ min-cut value. 
We first pass $s, t$ as terminal additions to $(G, \emptyset)$. 
The data structure then maintains the emulator $(H, T)$ of $(G, T)$, where $T = \{s, t\}$, with an amortized update time of $\Otil(\sqrt{n \tau} / \eps^{O(1)})$.
Now, if we want to take the dual $(H_D, T_D)$ of $(H, T)$ and run a static min-cut algorithm on $(H_D, T_D)$, we end up with an $\Otil(\sqrt{n \tau} / \eps^{O(1)})$ query time, which is too expensive for us to afford.
To overcome this issue, we use the data structure in a hierarchical way, called the {\it pattern emulator hierarchy}.

\atodoin{It is not clear what was the purpose of 2.2.2. Especially handling the updates is unclear. It should be to explain what we do, and how it is different from previous work.
I am afraid the answer here is that it is not different. This is fine. So we should explain the main ideas in the prev works, and why surprisingly it still works here.\\
The main idea of the previous work, was that when you have a change, in order to update the emulator, it is enough to add additional portals around the changed area, and don't recompute other stuff. I guess here we should repeat this main idea, and say that it holds also w.r.t. pattern emulators. As I understand this should be the msg here...}

\subsubsection{The Pattern Emulator Hierarchy.}

To achieve subpolynomial update and query time, we form a hierarchy $(H_1, T), \dots, (H_L, T)$ of pattern emulators with $L = \log^{2/3} n$, such that $(H_L, T_L)$ is of size $n^{o(1)}$ and is a $(1 + \eps)$-approximation pattern emulator of $(G, T)$.
Roughly speaking, for $1 \leq i \leq L-1$, the instance $(H_{i+1}, T)$ is a $(1+\eps')$-approximate pattern emulator of $(H_i, T)$, where $\eps' = \eps / (2L)$, and $(H_{i+1}, T)$ is maintained by the dynamic data structure initialized on $(H_{i}, T)$.
Before asking a query, we maintain the hierarchy such that $T= \emptyset$, and for each $s$-$t$ min-cut query, we pass the terminal additions from $(H_{i}, T)$ to $(H_{i+1}, T)$ so that $T = \{s, t\}$ for all pattern emulators while answering the query. 
After answering the query, we revert the terminal additions so that $T = \emptyset$ during edge updates. 

Edge updates also follow this top-down approach; upon an edge update in $(G, \emptyset)$, we first pass the update to the data structure maintaining $(H_1, \emptyset)$, which results in a batch $B_1$ of updates in $(H_1, \emptyset)$. 
By the guarantees of the data structure, $|B_1| = n^{o(1)}$. 
We then send the batch $B_1$ as updates to the data structure maintaining $(H_2, \emptyset)$ to obtain the batch $B_2$ of updates, which will be passed to $(H_3, \emptyset)$, and so on until $(H_L, \emptyset)$ is updated. 
It follows that we can maintain the entire hierarchy in $n^{o(1)}$ amortized update time.
See \Cref{subsec:all-pairs} for more details.
\atodoin{I'll be happy to see here a bit more details. In particular something about why is it so that the batch $B_1$ is so small, and why the sizes of the batches do not explode}

\paragraph{Answering the Queries.}

Recall that to answer an $s$-$t$ min-cut query for an instance $(G,T)$, we maintain the dynamic data structure on the dual instance $(G_D^*, T_D^*)$ of $(G, T)$.
In this case, our pattern emulator hierarchy maintains a pattern emulator $(H_L, T_L)$ of $(G_D^*, T_D^*)$ of size $n^{o(1)}$ in $n^{o(1)}$ amortized update time.
Since $(H_L, T_L)$ is an emulator of $(G_D^*, T_D^*)$, by the discussion we have had so far in this section, its dual instance $(H, T)$ of $(H_L,T_L)$ is a terminal cut sparsifier of $(G, T)$.
As we keep $T = \emptyset$ during the edge updates in $(G, T)$, upon an $s$-$t$ min-cut query, we simply add $s$ and $t$ to $T$ and maintain the pattern emulator $(H_L, T_L)$ in $n^{o(1)}$ amortized update time.
Next, we compute $(H, T)$ and run a static $s$-$t$ min-cut algorithm on it (using \Cref{thm:static-flow}), which takes $n^{o(1)}$ time as the size of $(H_L, T_L)$ is $n^{o(1)}$.
Finally, after answering the $s$-$t$ min-cut query, we revert the addition of $s$ and $t$ to $T$ to restore the state where $T = \emptyset$ for future updates.

\section{Preliminaries} \label{sec:prelim}
\paragraph{Graphs.}
All graphs in this paper are planar, weighted (capacitated), and undirected.
For a graph $G$, we denote its vertex set by $V(G)$ and its edge set by $E(G)$.
For an $n$-vertex graph $G$, all weights (capacities) on $E(G)$ are positive real numbers in the range $[1,n^{c}]$ for some constant $c$.

For vertices $u, v \in V(G)$, we denote by $\dist_G(u, v)$ the shortest distance between $u$ and $v$ in the graph $G$.
When the graph is clear from context, we omit the subscript from $\dist_G$.
We perturb the weights infinitesimally in a standard way so that all shortest paths are unique.

A subgraph $H$ of $G$ has vertex set $V(H)\subseteq V(G)$ and edge set $E(H)\subseteq E(G)$, where every edge in $E(H)$ has both endpoints in $V(H)$.
For a subset $S \subseteq V(G)$, the induced subgraph $G[S]$ is the subgraph of $G$ with vertex set $S$ containing all the edges in $E(G)$ with both endpoints in $S$.
Similarly, for a subset $E \subseteq E(G)$, the subgraph $G[E]$ is the subgraph of $G$ with edge set $E$ and vertex set consisting of the endpoints of $E$.
For an edge $e \in E(G)$, we denote by $G \setminus e$ the subgraph of $G$ defined on the vertex set $V(G)$ with the edge set $E(G) \setminus \{e\}$.

Given two graphs $G_1$ and $G_2$, we define their union $G_1 \cup G_2$ as the graph with vertex set $V(G_1) \cup V(G_2)$ and edge set $E(G_1) \cup E(G_2)$.

\paragraph{Graph Embedding.}
Throughout the paper, every planar graph is assumed to be given together with a fixed plane embedding. We use the combinatorial representation of this embedding, namely its \emph{rotation system}.
Specifically, for each vertex $v\in V(G)$, the combinatorial embedding specifies a cyclic order $e_0, e_1,\dots , e_{k-1}$ of the edges incident to $v$, such that in the planar embedding, every $e_i$ lies between $e_{i-1}$ and $e_{i+1}$ in the cyclic order, with indices modulo $k$.

The embedding is part of the input and is maintained under updates. Edge insertions are allowed only if they preserve planarity.
More precisely, an insertion of an edge $uv$ is assumed to be accompanied by the positions of the new incidences at $u$ and at $v$ in the cyclic orders for $u$ and $v$, and we only allow such insertions when the resulting rotation system is planar.

\paragraph{Faces and Holes.}
The fixed embedding of $G$ divides the plane into regions called {\it faces} of $G$.
The boundary of a face is a closed walk in $G$; 
in particular, an edge may appear more than once on a face boundary.
We say that a face contains a vertex if the vertex lies on its boundary.

Given a set $T \subseteq V(G)$ of terminals, we denote by $\mathcal H(G,T)$ a minimal set of faces of $G$ that covers $T$, i.e., for each $t \in T$, there is at least one face in $\mathcal H(G,T)$ that contains $t$.
We call each face in $\mathcal H(G,T)$ a {\it hole}, and denote the number of holes in $G$ by $h = |\mathcal H(G,T)|$.

\paragraph{Paths and Cycles.}
A path $P$ in $G$ is a sequence of vertices $v_1-v_2-\dots-v_{k+1}$ such that $v_i v_{i+1} \in E(G)$ for $1 \leq i \leq k$.
The length $|P|$ of a path $P$ is defined as the sum of the weights of the edges in $P$.
If $v_i \neq v_j$ for $i \neq j$, then we say $P$ is a {\it simple} path.
We allow paths to repeat vertices unless explicitly called simple.

We denote by $V(P) \coloneq \{v_1, \dots, v_{k+1}\}$ and $E_G(P)$ the multiset of edges present in $P$ within $G$,  
i.e., every edge $v_i v_{i+1}$ for $1 \leq i \leq k$, along with the number of times it is traversed using $P$.
We define $\tail(P) \coloneq v_1$ and $\head(P) \coloneq v_{k+1}$.
For $1 \leq i \leq j \leq k+1$, we denote by $P[v_i, v_j]$ the subpath $v_i-v_{i+1}-\dots-v_{j-1}-v_j$ of $P$. 
We also denote by $\overleftarrow{P}$  the path obtained by traversing $P$ in the reverse direction. 
i.e., we define $\overleftarrow{P} \coloneq v_{k+1} - v_k - \dots - v_1$.

For a path $Q \coloneq v_{k+1} - v_{k+2} - \dots - v_l$, we denote by $P \circ Q$ the path resulting from concatenating $P$ and $Q$. Formally, we define $P \circ Q \coloneq v_1 - v_2 - \dots - v_{k+1} - v_{k+2} - \dots - v_l$.

For a collection of simple paths $\Pcal$ in $G$, we denote by $\Mcal_G[\Pcal]$ the graph where (1) $v \in V(\Mcal_G[\Pcal])$ if and only if $v \in \tail(P)$ or $v \in \head(P)$ for a path $P \in \Pcal$, and (2) for every path $P \in \Pcal$, there is an edge in $\Mcal_G[\Pcal]$ connecting $\tail(P)$ and $\head(P)$ with weight $|P|$.
We drop the subscript $G$ from $\Mcal_G[\Pcal]$ whenever it is clear from the context.

Let $R$ be a simple $u$-$v$ path connecting two holes in $G$, and let $u'$ be the vertex succeeding $u$ on its hole in clockwise order, and $v'$ be the vertex preceding $v$ on its hole in clockwise order.  
Let $e_R^-, e_R^+$ be the edges of $u'-R-v'$ incident to some vertex $w \in V(R)$, and let $e_P^-, e_P^+$ be the edges of $P$ incident to $w$ in an occurrence of $w$ on $P$.  
We say that $P$ crosses $R$ at $w$ if these four incidences are distinct and appear around $w$ in one of the two alternating orders: $e_R^-, e_P^-, e_R^+, e_P^+$ or $e_R^-, e_P^+, e_R^+, e_P^-$, up to cyclic rotation.  
We say that $P$ crosses $R$ if this condition holds at some vertex of $R$.

A closed curve $C$ in $G$ is a sequence of vertices $v_1-v_2-\dots-v_{k+1}$ such that $v_i v_{i+1} \in E(G)$ for $1 \leq i \leq k$ and $v_1 = v_{k+1}$.
We say $C$ is {\it simple} if $v_i$ appears exactly once in $C$ for $2 \leq i \leq k$ and exactly twice for $i = 1, k+1$.
We denote $V(C) \coloneq \{v_1, \dots, v_{k+1}\}$.

\paragraph{Dynamic Instances.}
Given a planar graph $G$, an instance is a tuple $(G,T)$ where $T \subseteq V(G)$ is a set of terminal vertices.
As we allow edge insertions and deletions in $G$, and terminal additions and removals in $T$, we define the following updates on $(G,T)$:
\begin{itemize}
\item Insertion (respectively, deletion) of $uv$ into (respectively, from) $(G,T)$: edge $uv$ is inserted into (respectively, deleted from) $G$.
\item Addition (respectively, removal) of $u$ to (respectively, from) $(G,T)$: vertex $u$ is added to (respectively, removed from) $T$.
\end{itemize}

We define the size of $(G,T)$ as $|(G,T)| = |V(G)| + |E(G)| + |T|$.

Given instances $(G_1,T_1)$ and $(G_2,T_2)$, we define their union $(G_1,T_1) \cup (G_2,T_2)$ to be an instance $(G,T)$ where $G = G_1 \cup G_2$ and $T = T_1 \cup T_2$.

\paragraph{Dual Graph.}
The standard dual $G^*$ of $G$ is simply obtained by associating to every face of $G$ a dual vertex, and connecting the dual vertices by a dual edge if their corresponding faces in $G$ have an edge in common.
We allow loops and parallel edges.
It follows that there is a one-to-one correspondence between the faces in $G$ and the vertices in $G^*$, as well as between the edges in $G$ and the edges in $G^*$.
We denote the outer face of $G$ by $f_\infty$ and its corresponding dual vertex by $v^*$ in $G^*$.

\bigskip

We use the standard dual graph $G^*$ of $G$ in the static setting when defining patterns (later in this section). 
In the dynamic setting, we use {\it dual instances}, which rely on a different definition of the dual graph, as discussed next.

\paragraph{Dual Instance.}
To construct the dual instance $(G_D,T_D)$ of $(G,T)$, we use the dual graph as defined in \cite{Krauthgamer:2020aa,Chen:2025aa} (which differs from the standard dual graph defined above).
Given an instance $(G,T)$, the dual instance $(G_D,T_D)$ of $(G,T)$ is constructed as follows.
For every hole $h \in \mathcal H(G,T)$, we add an auxiliary vertex $x_h$ and connect $x_h$ to every terminal on $h$.
This divides $h$ into faces that are in one-to-one correspondence with the terminals on $h$.
We then take the standard dual of this graph.
Each {\it dual terminal} related to $h$ in this dual graph corresponds to the face to which the terminal was assigned while dividing $h$.
It follows that each terminal on $h$ corresponds to a dual terminal in this dual graph.
We denote $T_D$ by the set of dual terminals.
Finally, to obtain $G_D$, we remove every edge whose two endpoints correspond to auxiliary faces in $G$ formed by connecing an auxiliary vertex $x_h$ to the terminals on $h$.
See \Cref{fig:dual} for an illustration.

A more detailed discussion on the construction of $(G_D, T_D)$ and how updates in $G$ naturally extend to updates in $G_D$ is given in \Cref{subsec:dual}. 
In summary, each edge of $G$ corresponds bijectively to an edge of $G_D$, so an update to the weight of a primal edge induces the corresponding update in $G_D$.

\paragraph{Patterns.}
Before rigorously defining patterns, we provide an intuition similar to the one in \cite{Chen:2025aa}.
For a closed curve $\gamma$ in $G$, we say a vertex $v \notin V(\gamma)$ is inside $\gamma$ if there exists a vertex $u \notin V(\gamma)$ on the outer face of $G$ such that every simple $u$-$v$ path crosses $\gamma$ an odd number of times.\footnote{For this short discussion, we assume that $\gamma$ does not contain all the vertices on the outer face. The exact definition is more involved and requires a path in the dual graph $G^*$ of $G$, as will follow shortly.}

For every hole $h \in \mathcal H(G,T)$, let $z_h$ be an auxiliary vertex inside $h$ connected to an arbitrary vertex of $h$.
For a pair $t_1, t_2 \in T$ of terminals, let $S$ be the shortest $t_1$-$t_2$ path in $G$ and let $P$ be any $t_1$-$t_2$ path in $G$.
The pattern of $P$ in $G$, denoted by $\Phi_G(P)$, is defined as
\[
    \Phi_G(P) \coloneq \{h \in \mathcal H(G,T) \mid \text{$z_h$ is an auxiliary vertex inside the closed curve $S \cup P$}\}.
\]
Intuitively, $\Phi_G(P)$ corresponds to all holes that are inside the closed curve formed by $S \cup P$.
For clarity, we provide the formal definition of a closed curve below.

\begin{restatable}[Closed curve $A \cup B$]{definition}{closedcurve} \label{def:closed-curve}
    Let $G$ be a planar graph, and let $v_1, v_2 \in V(G)$ be two vertices in $G$.
    Let $A$ and $B$ be $v_1$-$v_2$ paths in $G$.
    The closed curve $A \cup B$ is defined as $A \cup B \coloneq A \circ \overleftarrow{B}$.
\end{restatable}

Since our definition of patterns allows $P$ to be any path, we give a precise definition of $\Phi_G(P)$ in the following definitions. 
These definitions will later be used to generalize certain claims from \cite{Krauthgamer:2020aa}, extending them from the case where $P$ is a simple path to the case where $P$ is any path.\footnote{We note that our definition deviates slightly from that in \cite{Krauthgamer:2020aa,Chen:2025aa}, where we allow $P$ to be any path instead of a simple path. Moreover, we define the path $S$ in a graph $G_h$ obtained from $G$, rather than directly in $G$ (see \Cref{def:path-pattern}). We will discuss the importance of this change for our application later.}

We start by defining the parity of a face $f_v$ with respect to a closed curve $M$.
Intuitively, this parity shows whether $f_v$ lies inside $M$ (in which case the parity is one) or outside $M$ (in which case the parity is zero).

\begin{restatable}[Parity of a face]{definition}{parity} \label{def:parity}
    Let $G$ be a planar graph and $f_\infty$ be its outer face.
    Let $G^*$ be the dual of $G$ and $v^*$ be the dual vertex in $G^*$ corresponding to $f_\infty$.
    For a face $f_v$ in $G$, let $v$ be the dual vertex in $G^*$ corresponding to it.
    Let $P^*$ be a simple $v^*$-$v$ path in $G^*$.
    For a closed curve $M$ in $G$, the parity of the face $f_v$ with respect to $M$ is defined as
    \[
        \Par(f_v,M)\coloneq \sum_{e^*\in E_{G^*}(P^*)} \ct\Big(e, E(M)\Big)  \pmod 2,
    \]
    where $E(M)$ is the multiset\footnote{Since an edge $e$ in $M$ could appear multiple times, we define $E(M)$ as a multiset to capture the number of times $e$ appears in $M$.} of edges that are present in $M$, and $\ct(e, E(M))$ is the number of times the edge $e$ corresponding to the dual edge $e^*\in E_{G^*}(P^*)$ appears in $E(M)$.
\end{restatable}

\begin{figure}[t]
    \centering
    \setkeys{Gin}{width=\linewidth}
    \begin{subfigure}{0.45\textwidth}
        \centering
        \includegraphics[page=1,width=\textwidth]{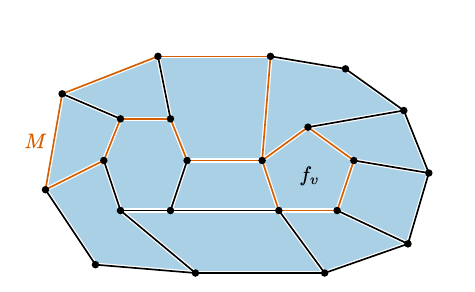}
        \caption{}
    \end{subfigure}
    \hfil
    \begin{subfigure}{0.45\textwidth}
        \centering
        \includegraphics[page=3,width=\textwidth]{pattern.pdf}
        \caption{}
    \end{subfigure}
    \caption{
        An example illustrating \Cref{def:parity}:
        (a) In graph $G$, $M$ is a closed curve (depicted with orange edges). 
        The face $f_v$ is surrounded by some edges of $M$. 
        (b) The dual graph $G^*$ of $G$ is shown in pink. 
        The vertex $v^*$ in $G^*$ corresponds to the outer face of $G$, and the vertex $v$ in $G^*$ corresponds to the face $f_v$ in $G$. 
        The path $P^*$ is a $v^*$-$v$ path in $G^*$ (depicted in green). 
        Each edge in $P^*$ corresponds to an edge in $G$ (depicted in bold). 
        Since all three of these corresponding edges in this example belong to $M$ (the bold orange edges), it follows that $\sum_{e^* \in E_{G^*}(P^*)} \ct\big(e, E(M)\big) = 3$, and so $\Par(f_v, M)=1$.
    }
    \label{fig:pattern}
\end{figure}

See \Cref{fig:pattern} for an illustration of \Cref{def:parity}.

We note that the parity of $f_v$ is independent of the choice of path $P^*$ in \Cref{def:parity}, as summarized in the following claim.

\begin{claim}[\protect{\cite[Claim 4.13]{Krauthgamer:2020aa}}, Rephrased]
    For a fixed face $f_v$ in $G$, let $P_1^*,P_2^*$ be two $v^*$-$v$ paths in $G^*$ as defined in \Cref{def:parity}.
    Then, for every closed curve $M$ in $G$,
    \[
        \sum_{e^*\in E_{G^*}(P_1^*)} \ct\Big(e, E(M)\Big) 
        \equiv
        \sum_{e^*\in E_{G^*}(P_2^*)} \ct\Big(e, E(M)\Big) 
        \pmod 2.
    \]
\end{claim}

\Cref{def:parity} provides a rigorous basis for defining whether a face is inside or outside of a closed curve, formalized below.

\begin{definition} \label{def:in-out}
    Let $G$ be a planar graph, $M$ be a closed curve in $G$, and $f_v$ be a face in $G$.
    Let $\Par(f_v,M)$ denote the parity of $f_v$ with respect to $M$, as defined in \Cref{def:parity}.
    We say that $f_v$ is inside $M$ if $\Par(f_v,M) = 1$, and outside $M$ if $\Par(f_v,M) = 0$.
\end{definition}

We now define the pattern of a closed curve $M$. 
Intuitively, the pattern of $M$ consists of all holes in $\mathcal H(G,T)$ that lie inside $M$.

\begin{restatable}[Pattern of a closed curve]{definition}{curvepattern} \label{def:curve-pattern}
    Let $G$ be a planar graph, $T\subseteq V(G)$ be its terminal set, and $\mathcal H(G,T)$ be the associated hole set.  
    Let $M$ be a closed curve in $G$.  
    Then, the pattern of $M$ in $G$ is defined as  
    \[
        \Phi_G(M)\coloneq \{h\in \mathcal H(G,T)\mid \Par\left(h, M\right)=1 \}.
    \]
\end{restatable}

Finally, we use \Cref{def:curve-pattern} to define the pattern of a $v_1$-$v_2$ path $P$ in $G$ as follows. 
The main intuition behind this definition is to first identify a {\it unique} $v_1$-$v_2$ path $S$ in $G$, and then define the pattern of $P$ as the pattern of the closed curve $P \cup S$ using \Cref{def:curve-pattern}. 
For our later analysis, we define $S$ based on another $v_1$-$v_2$ path $S'$, which is the shortest $v_1$-$v_2$ path in the sliced graph $G_h$ of $G$ (to be defined shortly), where $h = |\mathcal{H}(G,T)|$.\footnote{Throughout the paper, the symbol $h$ is used in two contexts: it represents the total number of holes, $|\mathcal{H}(G,T)|$, and it also denotes a specific hole within $\mathcal{H}(G,T)$. The meaning of $h$ will be clear from the context in which it is used.}
This choice is useful because we use the sliced graph to obtain our emulator; defining $S$ based on $S'$ significantly simplifies our analysis, since $S'$ is preserved by the slicing procedure.

The graph $G_h$ is defined as follows. Starting with $G_1 = G$, for $1 \leq i \leq h-1$, we define $G_{i+1}$ by slicing $G_i$ along the closest pair of holes in $G_i$; we slice along the shortest path between those two holes to obtain $G_{i+1}$.
The slicing procedure along a path $R$ adds a copy $\Rbar$ of $R$ to the graph and reassigns the edges on one side of $R$ to $\Rbar$. 
See \Cref{alg:slice} for a precise description of the slicing procedure. 
The path $S'$ naturally corresponds to a path in $G$: if $S'$ uses a copy $\vbar$ on $\Rbar$, we use $v$ on $R$ in $G$ instead.

\begin{restatable}[Pattern of a path]{definition}{pathpattern} \label{def:path-pattern}
    Let $G$ be a planar graph, $T\subseteq V(G)$ be its terminal set, and $\mathcal H(G,T)$ be the associated hole set.  
    Let $G_h$ be the one-hole graph obtained by repeatedly slicing along the closest pair of holes.\footnote{The slicing procedure in one iteration is detailed in \Cref{alg:slice}.}  
    For a pair $v_1,v_2\in V(G)$, let $P$ be a $v_1$-$v_2$ path and $S$ be the $v_1$-$v_2$ path in $G$ corresponding to the shortest $v_1$-$v_2$ path in $G_h$.  
    Then, the pattern of $P$ in $G$ is defined as  
    \[
        \Phi_G(P)\coloneq \{h\in \mathcal H(G,T)\mid \Par\left(h, P\cup S\right)=1 \}.\footnote{We note that \cite{Krauthgamer:2020aa} only requires $S$ to be a fixed $v_1$-$v_2$ path (in their case, the shortest $v_1$-$v_2$ path in $G$). It is not hard to see that our choice of the $v_1$-$v_2$ path is unique (by perturbing the weights of the copied edges during the slicing procedure) and, as we see later, will greatly simplify the analysis in our case.}
    \]
    Moreover, for $\Phi=\Phi_G(P)$, we say that $P$ is a $\Phi$-respecting $v_1$-$v_2$ path in $G$.
\end{restatable}

We will omit the subscript $G$ from $\Phi_G$ in \Cref{def:curve-pattern,def:path-pattern} when there is no ambiguity.

\paragraph{Pattern Emulator.}
We first discuss the definition of emulators.
Intuitively, an emulator of an instance $(G,T)$ is a planar graph that approximately preserves pairwise distances between vertices in \(T\). 

\begin{definition}[Emulator] \label{def:emulator}
Let $G$ be a planar graph, $T \subseteq V(G)$ be a terminal set, and $0 < \varepsilon < 1$ be an error parameter.
A $(1+ \varepsilon)$-approximate emulator of $(G,T)$ is an instance $(H,T)$ such that for every $s, t \in T$,
\[
    \dist_G(s,t) \leq \dist_{H}(s,t) \leq (1 + \varepsilon) \dist_G(s,t).
\]
\end{definition}

Note that by our convention, $H$ is planar.
Before proceeding, we first define aligned instances, which are used in the following theorems.
Intuitively, aligned instances have similar embedding properties, allowing us to avoid discussing them further in the paper.

\begin{definition}[Aligned Instances] \label{def:aligned}
    Given two instances $(H,T)$ and $(G,T)$ defined on the same terminal set, we say that $(H,T)$ and $(G,T)$ are aligned if 
    there is a bijection between the holes of $(H,T)$ and $(G,T)$ such that corresponding holes contain the same terminals in the same cyclic order.
\end{definition}

We will use the following theorem to obtain an exact emulator (i.e., with zero error) of $(G,T)$.

\begin{restatable}[\cite{Krauthgamer:2014aa}, Rephrased]{theorem}{exactemulator} \label{thm:emulator-exact}
    Let $G$ be an $n$-vertex planar graph and $T \subseteq V(G)$ be a terminal set.
    Then, there is an algorithm that constructs an exact planar emulator $(H,T)$ of $(G,T)$ of size $O(|T|^4)$ in time $O(n |T|)$.
    Moreover, $(H,T)$ is aligned with $(G,T)$.
\end{restatable}

To obtain a $(1+ \varepsilon)$-approximate emulator of $(G, T)$, we use the following theorem.

\begin{theorem}[\cite{Chang:2022aa}, Rephrased]\label{thm:emulator}
    Let $G$ be an $n$-vertex planar graph, $T \subseteq V(G)$ be a terminal set, and $0 < \varepsilon < 1$ be an error parameter.
    Then, there is an algorithm that constructs a $(1+ \varepsilon)$-approximate emulator $(H,T)$ of $(G,T)$ of size $\Otil(|T|/\varepsilon^{O(1)})$ in time $\Otil(n / \varepsilon^{O(1)})$.
    Moreover, $(H,T)$ is aligned with $(G,T)$.
\end{theorem}

We are now ready to define the notion of a pattern emulator.
Intuitively, a pattern emulator maintains the shortest-path distances between terminals for every pattern.

\begin{definition}[Pattern Emulator] \label{def:pattern-emulator}
Let $G$ be a planar graph, $T \subseteq V(G)$ be a terminal set, and $0 < \varepsilon$ be an error parameter.
A $(1+ \varepsilon)$-approximate pattern emulator of $(G,T)$ is an instance $(H,T)$ aligned with $(G,T)$, such that for every pattern $\Phi$ and every $s, t \in T$,
\[
    \dist_G^\Phi(s,t) \leq \dist_{H}^\Phi(s,t) \leq (1 + \varepsilon) \dist_G^\Phi(s,t),
\]
where $\dist_G^\Phi(s,t)$ is the distance of the shortest $\Phi$-respecting $s$-$t$ path in $G$. 
\end{definition}

We now present two properties for pattern emulators: 
(1) transitivity, which states that a pattern emulator of a pattern emulator of $G$ is itself a pattern emulator of $G$, albeit with slightly weaker approximation guarantees, and (2) composability, which states that the union of two pattern emulators of edge-disjoint graphs forms a pattern emulator of the union graph, provided the graphs share \textit{only} terminal vertices in their vertex sets.

\begin{lemma}[Transitivity of Pattern Emulators] \label{lem:transitivity-pattern-emulator}
Let $(H,T)$ be a $(1+ \varepsilon)$-approximate pattern emulator of $(G,T)$, and $(H',T')$ be a $(1 + \varepsilon')$-approximate pattern emulator of $(H, T')$, where $T' \supseteq T$, $T' \subseteq V(H)$, and $T \subseteq V(G)$.
Then, $(H',T)$ is a $(1 + \varepsilon)(1 + \varepsilon')$-approximate pattern emulator of $(G,T)$.
\end{lemma}

\begin{proof}
Suppose that $s, t \in T$ and let $\Phi$ be a pattern.
By assumption, \(s, t \in T'\), and we have
\[
    \dist_{H'}^\Phi(s,t) \leq (1+\varepsilon') \cdot \dist_{H}^\Phi(s,t) \leq (1+\varepsilon)(1+\varepsilon') \cdot \dist_G^\Phi(s,t).
\]
The first inequality holds because $(H',T')$ is a $(1+ \varepsilon')$-approximate pattern emulator of $(H,T')$, and the second inequality holds because \(T \subseteq T'\) and $(H,T)$ is a $(1+ \varepsilon)$-approximate pattern emulator of $(G,T)$. 

For the lower bound, we similarly have
\[
    \dist_{H'}^\Phi(s,t) \geq \dist_H^\Phi(s,t) \geq \dist_G^\Phi(s,t), 
\]   
where the first inequality holds as $(H',T')$ is a $(1+ \varepsilon')$-approximate pattern emulator of $(H,T')$, and the second inequality holds as \(T \subseteq T'\) and $(H,T)$ is a $(1+ \varepsilon)$-approximate pattern emulator of $(G,T)$.
\end{proof}

\begin{lemma}[Composability of Pattern Emulators] \label{lem:composability-pattern-emulators}
Let $(G_1,T_1)$ and $(G_2,T_2)$ be two instances with $V(G_1)\cap V(G_2) \subseteq T_1\cap T_2$. 
If $(H_i,T_i)$ is a $(1+ \varepsilon)$-approximate pattern emulator of $(G_i,T_i)$ for $i = 1, 2$, then $H \coloneq H_1 \cup H_2$ is a $(1 + \varepsilon)$-approximate pattern emulator of $(G,T) \coloneq (G_1 \cup G_2, T_1 \cup T_2)$.
\end{lemma} 

\begin{proof}
Suppose that $s, t \in T$ and let $\Phi$ be a pattern.
Let $P$ be the shortest $\Phi$-respecting $s$-$t$ path in $G$. 
Without loss of generality, suppose that $s \in V(G_1)$.
Then, path $P$ decomposes into subpaths $P_1, P_2, \dots, P_q$, such that $P_1, P_3, \dots$ are fully contained in $G_1$ and $P_2, P_4, \dots$ are fully contained in $G_2$.
Let $s_i, t_i$ be the endpoints of $P_i$, and $\Phi_i$ be the pattern of $P_i$ in the graph containing it (i.e., in $G_1$ or $G_2$).
It then follows that $P_i$ is a $\Phi_i$-respecting $s_i$-$t_i$ shortest path in the graph containing it, since otherwise we could construct a path shorter than $P$ with the same pattern $\Phi$.\footnote{This is formally stated in \Cref{clm:same-pattern} of \Cref{sec:static}.}
We have
\begin{align*}
\dist_H^\Phi(s,t) &= \sum_{i=1}^q \dist_H^{\Phi_i}(s_i,t_i) \\
&= 
\sum_{i=1}^{\lceil q/2 \rceil} \dist_{H_1}^{\Phi_{2i-1}}(s_{2i-1},t_{2i-1}) + \sum_{i=1}^{\lceil q/2 \rceil} \dist_{H_2}^{\Phi_{2i}}(s_{2i},t_{2i}) \\
&\leq
(1+\varepsilon) \sum_{i=1}^{\lceil q/2 \rceil} \dist_{G_1}^{\Phi_{2i-1}}(s_{2i-1},t_{2i-1}) + (1+\varepsilon) \sum_{i=1}^{\lceil q/2 \rceil} \dist_{G_2}^{\Phi_{2i}}(s_{2i} ,t_{2i}) \\
&=
(1+\varepsilon) \sum_{i=1}^q \dist_G^{\Phi_i}(s_i,t_i) \\
&=
(1+\varepsilon) \dist_G^{\Phi}(s,t)~.
\end{align*}

The lower-bound guarantee of the pattern emulator can be obtained similarly. 

It follows that $(H,T)$, where $H=H_1\cup H_2$ and $T=T_1\cup T_2$, is a $(1+\varepsilon)$-approximate pattern emulator of $(G,T)$. \qedhere
\end{proof}

Finally, we use the following theorem to relate pattern emulators to terminal cut sparsifiers.

\begin{restatable}[\protect{\cite[Theorem 16]{Chen:2025aa}} and \protect{\cite[Theorem 4.6]{Krauthgamer:2020aa}}, Rephrased]{theorem}{KR} \label{thm:cut-pattern-emulator}
    Let $(G_D, T_D)$ be the dual instance of $(G, T)$, $0 < \varepsilon < 1$ be an error parameter, and let $(H_D,T_D)$ be an instance aligned with $(G_D,T_D)$ whose primal instance is $(H,T)$.
    If $(H_D, T_D)$ is a $(1+\varepsilon)$-approximate pattern emulator of $(G_D, T_D)$, then $(H, T)$ is a $(1+\varepsilon)$-approximate terminal cut sparsifier of $(G, T)$.
\end{restatable}

\paragraph{Cuts.} 
For an instance $(G,T)$ and a subset $\emptyset \subset R \subset T$ of terminals, we denote by $\operatorname{mincut}_G(R, T \setminus R)$ the value of the minimum cut in $G$ that separates $R$ from $T \setminus R$. 

Using this terminology, we now define terminal cut sparsifiers.

\begin{definition}[Terminal Cut Sparsifier] \label{def:cut-sparsifier}
Let $G$ be a planar graph, $T \subseteq V(G)$ a terminal set, and $0 < \varepsilon < 1$ an error parameter. 
A $(1+ \varepsilon)$-approximate terminal cut sparsifier of $(G,T)$ is an instance $(H,T)$ such that for every subset $\emptyset \subset R \subset T$ of terminals,
\[
    \operatorname{mincut}_G(R, T \setminus R) \leq \operatorname{mincut}_H(R, T \setminus R) \leq (1+\varepsilon) \, \operatorname{mincut}_G(R, T \setminus R).
\]
\end{definition}

\paragraph{Structured Divisions.}
We will use structured divisions, defined below, to compute a terminal cut sparsifier for a graph with an arbitrary number of holes.
Intuitively, we use structured divisions to evenly distribute the terminal vertices across smaller subgraphs.

\begin{restatable}[Structured $r$-Division]{definition}{rdivision} \label{def:r-division}
    Given an $n$-vertex planar graph $G$, a parameter $1 \leq r \leq n$, and a terminal set $T\subseteq V(G)$, a structured $r$-division of $(G,T)$ is a partition of $E(G)$ into $k$ subsets $E_1, \dots, E_k$ such that
    \begin{enumerate}
        \item Each edge-induced subgraph $G[E_i]$ (called a piece) has $O(r)$ vertices, and \label{division-one}
        \item Each piece has $O(\sqrt{r})$ boundary vertices, where a boudary vertex is a vertex shared with another piece, and \label{division-two}
        \item There are $k = O(n/r)$ pieces in the $r$-division, and \label{division-three}
        \item Each piece contains $O(1+|T|/k)$ terminals from $T$.
    \end{enumerate}
\end{restatable}

We note that the above definition is an extension of $r$-division \cite{Klein:2013aa} with the extra property that the terminals in $T$ are evenly distributed among the pieces.
We use the following theorem to construct a structured $r$-division, ensuring that each piece has a constant number of holes.

\begin{restatable}[\protect{\cite[Lemma 5.5]{Chang:2022ab}}, Rephrased]{theorem}{structured} \label{thm:structured-division}
    Given an $n$-vertex planar graph $G$, a parameter $1\le r\le n$, and a terminal set $T\subseteq V(G)$, 
    there is an algorithm that computes a structured $r$-division of $(G,T)$ in $O(n)$ time such that each piece is connected and has $O(1)$ holes.
\end{restatable}

The $r$-division theorem of \cite{Klein:2013aa} is stated below.
Note that \cite{Klein:2013aa} guarantees $O(1)$ holes in the pieces only when $(G, T)$ itself has $O(1)$ holes.

\begin{restatable}[\cite{Klein:2013aa}]{theorem}{structuredrdivision} \label{thm:r-division}
    Given an $n$-vertex planar graph $G$ and a terminal set $T \subseteq V(G)$ on $h$ holes, there is an algorithm that computes an $r$-division of $(G, T)$ satisfying items~\ref{division-one}~to~\ref{division-three} of \Cref{def:r-division} in $O(n)$ time such that, for every $1 \leq i \leq k$, the piece $G[E_i]$ has at most $\frac{3}{4}h + 4$ holes.
\end{restatable}

\paragraph{Separators.}
We use separators to split a one-hole instance $(G, T)$ into two instances with fewer terminals.
The following theorem will be used in \Cref{sec:dynamic}. 
An edge $e$ is called a bridge in $G$ if $G \setminus e$ has more connected components than $G$.

\begin{restatable}[\protect{\cite[Theorem 3.10]{Filtser:2024aa}}, Rephrased]{theorem}{separator} \label{thm:separator}
    Given an $n$-vertex planar graph $G$ and a terminal set $T \subseteq V(G)$ on a single hole, 
    there is an algorithm that finds vertices $u, v$ on the hole of $(G,T)$ in $O(n)$ time such that the two terminal intervals on the hole bounded by $u$ and $v$ contain at most $\frac{11}{12} |T|$ terminals each, and at least one of the following holds.
    \begin{enumerate}
        \item 
            The shortest path $\pi_{uv}$ in $G$ does not contain a bridge.
        \item 
            The edge $uv\in E(G)$ exists and is a bridge.
    \end{enumerate}
\end{restatable}

\paragraph{Portals.}
In our sparsifier constructions, we use the slicing scheme of \cite{Chang:2022aa}, where we pick a shortest path $R$ in $G$ and slice along it to reduce the number of holes in the resulting instance by one (see \Cref{alg:slice} for a detailed explanation of the slicing procedure).
To maintain the distances within a $(1+\varepsilon)$ factor, we pick some vertices on the sliced path $R$, called {\it portals}.
The path $R$ is chosen as the shortest path between any two holes of $G$, and the portals on $R$ form an $\varepsilon$-cover for $T$.
i.e., for every terminal $t \in T$ not on $R$ and every vertex $r \in V(R)$, there is a portal $p$ that satisfies the $\eps$-cover property
\begin{equation} \label{eq:portal}
    \dist(t,r) \leq \dist(t,p) + \dist(p,r) \leq (1+\varepsilon)\dist(t,r).
\end{equation}

Our algorithm for computing an $\eps$-cover is as follows.
For each $t \in T$, the algorithm first finds the shortest path $\Delta_t$ from $t$ to $R$.\footnote{If $t\in V(R)$, we add $t$ as a portal and skip the construction below.}
Let $x$ be the closest vertex to $t$ on $R$.
Then, the algorithm finds the portal set $\portal(R,t,\varepsilon)$ on the two maximal subpaths of $R$ that have $x$ as one endpoint.
The portal set is obtained using the following two procedures.
\begin{enumerate}
    \item \label{itm:portal-close}
        Starting from $x$, the algorithm keeps adding vertices with distance $\eps |\Delta_t|$ from each other until it reaches a vertex $y$ with $\dist(x,y)> 2|\Delta_t|$.
        i.e., starting from $x$, it adds the sequence $z_1,z_2,\dots$ of vertices, where each $z_i$ is the first vertex visited from $x$ with $\dist(x,z_i)> i\eps |\Delta_t|$ and $\dist(x,z_i) \leq 2 |\Delta_t|$.
        Thus, the number of selected portals in the sequence is $O(1/\eps)$.
    \item \label{itm:portal-far}
        Once the algorithm reaches an edge $y'y$ with $\dist(x,y)> 2|\Delta_t|$, it adds $y$ and $y'$ to the portal set.
        Then, starting from $y$, it traverses to the other endpoint until it reaches the first edge $z'z$ (if any) with $\dist(x,z) > (1+\varepsilon) \dist(x,y)$.
        Then, it adds $z$ and $z'$ to the portal set and repeats the procedure for $z$ as $y$.
        i.e., it sets $y = z$ and keeps traversing to the other endpoint until it reaches an edge $w'w$ (if any) such that $\dist(x,w) > (1+\varepsilon) \dist(x,y)$.
        It then adds $w$ and $w'$ to the portal set, and then repeats the procedure for $y = w$.
        The procedure continues until the algorithm reaches the other endpoint of the subpath.
        Since the weights are polynomial in $n$, the number of selected portals is $O(\log n/\eps)$.
\end{enumerate}

The following lemma summarizes the guarantees for computing the portals.
The proof is similar to that of \cite[Lemma 2.4]{Chang:2022ab} and is thus omitted.

\begin{restatable}{lemma}{portals} \label{lem:portals}
    Let $G$ be an $n$-vertex planar graph, $T \subseteq V(G)$ be a terminal set lying on $h$ holes, and $0 < \varepsilon < 1$ be an error parameter.
    Let $R$ be a shortest path in $G$.
    Then, there exists an algorithm that computes the set $\portal(R, T, \varepsilon) \coloneq \cup_{t \in T} \portal(R, t, \varepsilon)$, which lies on $R$ and has size $O(|T| \log n / \varepsilon)$, in time $O(n h \log |T|)$.
\end{restatable}

We will use the following claim to bound the approximation guarantee of our algorithm.

\begin{restatable}{claim}{portaleps} \label{clm:portals}
    Let $G$ be an $n$-vertex planar graph, $T \subseteq V(G)$ be a terminal set, and $0 < \varepsilon < 1$ be an error parameter.
    Let $R$ be a shortest path in $G$ on which an $\varepsilon$-cover $\portal(R, T, \varepsilon)$ is computed as defined in \Cref{lem:portals}.
    Then, for any $t_1$-$t_2$ path $P$ with $t_1, t_2 \in T$, if $P$ crosses $R$, the distance from any crossing vertex on $R$ to its closest portal is $O(\varepsilon |P|)$.
\end{restatable}

\begin{proof}
    Let $x$ be the closest vertex on $R$ to $t_1$, and let $\Delta_{t_1}$ be a shortest path from $t_1$ to $R$, and let $|\Delta_{t_1}|=\dist(t_1,x)$. 
    Suppose that $P$ crosses $R$ on the subpath $R[v_i, v_{i+1}]$, where $v_i$ and $v_{i+1}$ are the $i$-th and $(i+1)$-th portals selected based on item~\ref{itm:portal-far} of the construction. 
    As $P$ crosses $R$, it follows that $|P| \geq |\Delta_{t_1}|$.

    Let $v'$ be a vertex in the subpath $R[v_i, v_{i+1}]$ where $P$ crosses. 
    If $v' = v_i$ or $v' = v_{i+1}$, the claim trivially follows. 
    Otherwise, we assume $v' \neq v_i$ and $v' \neq v_{i+1}$.
    
    We analyze the following two distinct cases.

    \begin{itemize}
        \item \underline{If $\dist(x, v') \leq 2|\Delta_{t_1}|$:}
            By item~\ref{itm:portal-close} of the construction, there exists a portal $z_i$ within a distance of $O(\eps |\Delta_{t_1}|)$ from $v'$. 
            Since $|P| \geq |\Delta_{t_1}|$, the claim follows directly.

        \item \underline{If $\dist(x, v') > 2|\Delta_{t_1}|$:}
            By item~\ref{itm:portal-far} of the construction, we have
            \[
                (1+\eps)^i|\Delta_{t_1}| \leq \dist(x, v') \leq (1+\eps)^{i+1}|\Delta_{t_1}|.
            \]
            Consequently, the distance $d$ between $v'$ and its nearest portal satisfies
            \begin{equation} \label{eqn:vstar}
                d \leq (1+\eps)^{i+1}|\Delta_{t_1}| - (1+\eps)^i|\Delta_{t_1}| = \eps (1+\eps)^i|\Delta_{t_1}| \leq \eps \dist(x, v').
            \end{equation}

            Now, let $P'$ denote the shortest $t_1$-$v'$ path in $G$. 
            Clearly, $|P| \geq |P'|$. 
            Furthermore, using the triangle inequality, we derive
            \begin{equation} \label{eqn:vtriangle}
                |P'| \geq \dist(x, v') - |\Delta_{t_1}| \geq \frac{1}{2} \dist(x, v'),
            \end{equation}
            where the second inequality follows from the assumption $\dist(x, v') > 2 |\Delta_{t_1}|$.

            By combining \Cref{eqn:vstar,eqn:vtriangle}, we obtain
            \[
                d = O(\eps |P'|) = O(\eps |P|),
            \]
            which completes the proof.
        \qedhere
    \end{itemize}
\end{proof}

We will use the following lemma in \Cref{sec:dynamic} to obtain a second set of portals on the shortest path $\pi_{uv}$ as defined in \Cref{thm:separator}.

\begin{restatable}{lemma}{portalsplit} \label{lem:portals-split}
    Let $G$ be an $n$-vertex planar graph, $T \subseteq V(G)$ be a terminal set on a single hole, and $0 < \varepsilon < 1$ be an error parameter.
    Let $\pi_{uv}$ be the separator path obtained from \Cref{thm:separator}.
    Then, there is an algorithm that computes a set of portals $\splitportal(\pi_{uv}, T, \varepsilon)$ on $\pi_{uv}$ of size $O(|T|/\eps)$ that satisfies the $(1+O(\eps))$-cover property of \Cref{eq:portal}.
    Moreover, the portal set can be computed in time $O(n|T|)$.
\end{restatable}

\section{Static Terminal Cut Sparsifier} \label{sec:static}
In this section, we present our algorithm for constructing a $(1+ \varepsilon)$-approximate terminal cut sparsifier $(H,T)$ of $(G,T)$ with near-linear size in near-linear time, whose guarantees are summarized in the following theorem.

\begin{restatable}{theorem}{staticcut} \label{thm:static-cut}
    Let $G$ be an $n$-vertex planar graph, $T\subseteq V(G)$ be a terminal set, and $0<\varepsilon <1$ be an error parameter.
    Then, there is an algorithm that constructs a $(1+ \varepsilon)$-approximate terminal cut sparsifier $(H,T)$ of $(G, T)$ in time $\Otil(n/\varepsilon^{O(1)})$ with size $|H|= \Otil(|T|/\varepsilon^{O(1)})$.
\end{restatable}

We note that \cite{Chen:2025aa} presents a polynomial-time algorithm for constructing a $(1+ \varepsilon)$-approximate terminal cut sparsifier. 
By leveraging standard techniques such as division and bootstrapping, their algorithm can be used to achieve guarantees similar to those in \Cref{thm:static-cut}. 
However, their approach relies on different sets of portals, which makes it unsuitable for use in our dynamic data structure in \Cref{sec:dynamic}, as this would lead to a significant increase in update time.\footnote{More precisely, during the slicing procedure along a shortest path $R$ (see \Cref{alg:slice}), \cite{Chen:2025aa} finds $\varepsilon$-cover portals on $R$ for every pattern in $G$ (see \Cref{sec:prelim} for the definition of a pattern). In the dynamic setting, adding these portals for all patterns upon the addition of a terminal appears challenging in subpolynomial time.} 
Our main technical contribution is showing that employing the same portal set as in \cite{Chen:2022aa} suffices to achieve the same construction. 
This, in turn, allows us to obtain a subpolynomial update time.  

The structure of this section is as follows.  
In \Cref{subsec:reduction}, we explain how obtaining a terminal cut sparsifier as described in \Cref{thm:static-cut} can be reduced to the task of constructing a pattern emulator with guarantees similar to those in \Cref{thm:static}.  
The remainder of the section is then devoted to finding such a pattern emulator.  
In \Cref{subsec:one-hole}, we provide a brief discussion of the special case where $(G, T)$ is a one-hole instance.  
Next, we address the $O(1)$-hole case in \Cref{subsec:constant-hole}.  
Finally, we establish the general case (\Cref{thm:static}) in \Cref{subsec:general}.

\subsection{Reduction to Pattern Emulators} \label{subsec:reduction}

In this subsection, we explain how computing the terminal cut sparsifier of \Cref{thm:static-cut} can be reduced to constructing a pattern emulator with similar guarantees.

The reduction follows the same strategy outlined in \cite{Chen:2025aa} by first using \Cref{thm:cut-pattern-emulator}, restated below.  
Before presenting \Cref{thm:cut-pattern-emulator}, we first revisit the definitions of the parity of a face and the pattern of a path, as these concepts are crucial to the theorem.

\closedcurve*
\parity*
\pathpattern*
\KR*

In other words, \Cref{thm:cut-pattern-emulator} states that obtaining a $(1+\eps)$-approximate terminal cut sparsifier $(H, T)$ of $(G, T)$ can be achieved by first computing the dual instance $(G_D, T_D)$ of $(G, T)$ and then constructing a $(1+\eps)$-approximate pattern emulator $(H_D, T_D)$ of $(G_D, T_D)$.  
If $(H_D, T_D)$ is aligned with $(G_D, T_D)$, we can subsequently conclude that the $(1+\eps)$-approximate terminal cut sparsifier $(H, T)$ of $(G, T)$ corresponds to the dual instance of $(H_D, T_D)$.  
Leveraging this insight, the problem is reduced to proving the following theorem.

\static*

We now formalize the reduction outlined above in the following claim.

\begin{claim} \label{clm:reduction}
    Let $G$ be an $n$-vertex planar graph, $T\subseteq V(G)$ be a terminal set, and $0<\varepsilon <1$ be an error parameter.
    Then, by employing \Cref{thm:static}, we can construct a $(1+ \varepsilon)$-approximate terminal cut sparsifier $(H, T)$ of $(G, T)$ in time $\Otil(n / \varepsilon^{O(1)})$ with size $|H| = \Otil(|T| / \varepsilon^{O(1)})$. 
\end{claim}

\begin{proof}
    We begin by constructing the dual instance $(G_D,T_D)$ of $(G,T)$. 
Recall that the construction of $(G_D,T_D)$ involves adding $O(n)$ auxiliary vertices and $O(n)$ edges to $G$ to form an augmented graph, denoted $G'$, on which we compute the standard dual. 
Since $|G'| = O(n)$, the computation of the standard dual $G'_D$ of $G'$ can be performed in $O(n)$ time. 
    
The dual instance $(G_D,T_D)$ is then derived from $G'_D$ by removing $O(n)$ edges corresponding to certain edges added to $G$ during the formation of $G'$.\footnote{The procedure is detailed in \Cref{sec:prelim}.}
Given that the construction of $G'$ and $(G_D,T_D)$ involves probing $G$ and $G'_D$ exactly once, respectively, it follows that the computation of $(G_D,T_D)$ requires $O(n)$ time.
    
Next, we apply \Cref{thm:static} to $(G_D,T_D)$ to obtain a $(1+\varepsilon)$-approximate pattern emulator $(H_D,T_D)$ of $(G_D,T_D)$ in $\Otil(n/\varepsilon^{O(1)})$ time, with size $|H_D| = \Otil(|T_D|/\varepsilon^{O(1)})$. 
Since $(H_D,T_D)$ is aligned with $(G_D,T_D)$, by \Cref{thm:cut-pattern-emulator}, it follows that the primal $(H,T)$ of $(H_D,T_D)$ serves as a $(1+\varepsilon)$-approximate terminal cut sparsifier for the primal $(G,T)$ of $(G_D,T_D)$.

For the runtime analysis, observe that the computation time is dominated by applying \Cref{thm:static} to constructt $(H_D,T_D)$ in $\Otil(n/\varepsilon^{O(1)})$ time; constructing $(G_D,T_D)$ is only $O(n)$. 
    This follows from the fact that $|H_D| = \Otil(|T_D|/\varepsilon^{O(1)}) = \Otil(n/\varepsilon^{O(1)})$. 
    Additionally, since $|H_D| = \Theta(|H|)$, the size guarantee for $(H,T)$ directly follows from the equality $|T_D| = |T|$ as there is a one-to-one correspondence between the terminals in $T$ and $T_D$ due to the construction of the dual instance.
    
    This concludes the proof.
\end{proof}

Using \Cref{clm:reduction}, for the remainder of this section, we focus on proving \Cref{thm:static} in \Cref{subsec:one-hole,subsec:constant-hole,subsec:general}.

\subsection{The One-Hole Case} \label{subsec:one-hole}

In this subsection, we prove \Cref{thm:static} for the case where $(G,T)$ is a one-hole instance.
i.e., all terminals in $T$ lie on a single hole in $(G,T)$.
The guarantees are summarized in the following lemma.

\begin{lemma} \label{lem:one-hole}
    Let $G$ be an $n$-vertex planar graph, $T\subseteq V(G)$ be a terminal set all lying on a single hole, and $0<\varepsilon <1$ be an error parameter.
    Then, there is an algorithm that constructs a $(1+ \varepsilon)$-approximate pattern emulator $(H,T)$ of $(G, T)$ in time $\Otil(n/\varepsilon^{O(1)})$ with size $|H|= \Otil(|T|/\varepsilon^{O(1)})$.
    Moreover, $(H,T)$ is aligned with $(G, T)$. 
\end{lemma}

\begin{proof}
    Without loss of generality, assume that the hole is the outer face of $G$.
    
    Observe that, for every pair $t_1, t_2 \in T$, any $t_1$-$t_2$ path in $G$ has the trivial pattern. 
    i.e., using the terminology from \Cref{def:path-pattern}, we have $\Phi_G(P) = \emptyset$. 
    This follows from the fact that $f_{\infty}$ is the only hole in $G$, and by the Jordan curve theorem, every $v^*$-$v$ path in the dual graph $G^*$ must cross the dual edges corresponding to the hole edges an even number of times.
    Here, $f_{\infty}$, $v^*$, and $v$ are as defined in \Cref{def:parity}. 
    Note that $v = v^*$ since the hole is the outer face.
    
    Since, for each $t_1$-$t_2$ path $P$, we only need to maintain the shortest $t_1$-$t_2$ path with the pattern $\emptyset$, it follows that we only need to maintain the shortest $t_1$-$t_2$ path for every pair $t_1,t_2\in T$. 
    Consequently, a pattern emulator reduces to a standard emulator, wherein we are required to approximately preserve the shortest paths between any pair in $T$ (see \Cref{def:emulator} for the formal definition of an emulator).
    
    It follows that employing \Cref{thm:emulator} on $(G, T)$ directly results in a (pattern) emulator $(H, T)$ of size $|H| = \Otil(|T| / \varepsilon^{O(1)})$ in $\Otil(n / \varepsilon^{O(1)})$ time. 
    The alignment of $(H, T)$ and $(G, T)$ is a direct consequence of this construction.
\end{proof}

Combined with \Cref{clm:reduction}, \Cref{lem:one-hole} establishes the following lemma.

\begin{restatable}{lemma}{oneholecut} \label{lem:one-hole-cut}
    Let $G$ be an $n$-vertex planar graph, $T \subseteq V(G)$ be a set of terminals all lying on a single hole, and $0<\eps <1$ be an error parameter.
    Then, there is an algorithm that constructs a $(1+ \varepsilon)$-approximate terminal cut sparsifier $(H,T)$ of $(G, T)$ in time $\Otil(n/\varepsilon^{O(1)})$ with size $|H|= \Otil(|T|/\varepsilon^{O(1)})$.
\end{restatable}

Our proof of \Cref{lem:one-hole-cut} is concise and relies solely on the pattern property of $(G,T)$. 
However, it does not provide intuition about how minimum cuts in a one-hole instance $(G,T)$ correspond to shortest paths in its dual instance $(G_D,T_D)$ as described in \Cref{thm:cut-pattern-emulator}.
In contrast, \cite{Chen:2025aa} gives a more direct proof by leveraging the cut property of $(G,T)$. 
We refer the reader to \Cref{sec:one-hole-case} for an outline of their approach, which will also provide valuable intuition about why using the dual instance $(G_D,T_D)$ instead of the standard dual is necessary in \Cref{thm:cut-pattern-emulator}.

\subsection{The $O(1)$-Hole Case} \label{subsec:constant-hole}
We now focus on the case where $(G,T)$ is an $O(1)$-hole instance, and present an algorithm for computing a near-linear size pattern emulator $(H,T)$ of $(G,T)$ in near-linear time.
The main result of this subsection is summarized in the following theorem.

\begin{restatable}{theorem}{constanthole} \label{thm:constant-hole}
    Let $G$ be an $n$-vertex planar graph, $T\subseteq V(G)$ be a terminal set on $O(1)$ holes in $G$, and $0<\varepsilon <1$ be an error parameter.
    Then, there is an algorithm that constructs a $(1+\varepsilon)$-approximate pattern emulator $(H,T)$ of $(G,T)$ in time $\Otil(n/\varepsilon^{O(1)})$ with size $|H|= \widetilde O(|T|/\varepsilon^{O(1)})$.
    Moreover, $(H,T)$ is aligned with $(G,T)$.
\end{restatable}

This subsection is the most technical part of this section.
Before proceeding, we first provide an overview of this subsection below.
\begin{itemize}
    \item 
        In \Cref{subsubsec:auxiliary}, we provide some auxiliary claims that are used in the rest of this subsection.
    \item 
        In \Cref{subsub:exact-pattern}, we discuss how to construct an {\it exact} pattern emulator $(H,T)$ in near-linear time,\footnote{As obtaining such an emulator is not the main goal of this paper, we do not explicitly dicuss the construction of $(H,T)$ and its runtime. Both follow the same analysis as we will see in \Cref{subsub:approx-pattern}.} though the size of $(H,T)$ is not near-linear in $|T|$.
        The main goal of this subsection is to provide the technical material for constructing the $(1+\eps)$-approximation pattern emulator of \Cref{thm:constant-hole} in \Cref{subsub:approx-pattern}.
        We will study how slicing, as the main technique to reduce the $O(1)$-case to the one-hole case of \Cref{subsec:one-hole}, affects the $\Phi$-respecting shortest paths in $G$.
        Most notably, we prove that each such path corresponds to a constant number of paths in the sliced graph (see \Cref{clm:subpaths}), which will play a crucial role in the construction of the $(1+\eps)$-approximation pattern emulator of \Cref{thm:constant-hole} in \Cref{subsub:approx-pattern}.
        We note that the results of this subsection do not depend on the assumption of $(G,T)$ having $O(1)$ holes; this assumption is used in \Cref{subsub:approx-pattern} for obtaining an approximate pattern emulator with a desired size.
    \item 
        In \Cref{subsub:approx-pattern}, we prove \Cref{thm:constant-hole} by using the slicing scheme of \Cref{subsub:exact-pattern}.
        To reduce the size to near-linear, we employ $\eps$-cover portals (as will be defined in \Cref{def:portals-all}) on the sliced paths during the construction of $(H,T)$.
        The main technical part of this subsection is to prove that using these $\eps$-cover portals results in the desired $(1+\eps)$-approximation pattern emulator (see \Cref{clm:glued-approximation,clm:approximation-quality-glued}), and thus the emulator construction of \cite{Chang:2022aa} on the reduced one-hole instance can be used to obtain a $(1+\eps)$-approximation pattern emulator for $(G,T)$.
        We prove \Cref{thm:constant-hole} at the end of this subsection.
\end{itemize}

\subsubsection{Auxiliary Claims} \label{subsubsec:auxiliary}
In this subsection, we prove some claims that will be useful in the rest of the section.

We start by proving a claim about the parity of a face, whose definition is restated below.

\parity*

Intuitively, the following claim states that if $A$, $B$, and $C$ are paths with the same endpoints in $G$, then the holes inside the closed curve $A\cup C$ are the ones that are exactly in {\it one} of $A \cup B$ or $B \cup C$.\footnote{For the case where $A$, $B$, and $C$ are simple paths,  \cite{Krauthgamer:2020aa} proves a similar claim. We generalize their proof to any path $A$, $B$, and $C$.}

\begin{restatable}{claim}{ABC} \label{clm:ABC}
    Let $A$, $B$, and $C$ be three paths with the same endpoints in $G$.
    Then,
    \[
        \Phi_G(A\cup C)= \Phi_G(A\cup B)\bigtriangleup \Phi_G(B\cup C),
    \]
    where $\bigtriangleup$ is the symmetric difference operator.
\end{restatable}
\begin{proof}
    Let $h\in \mathcal H(G,T)$ be a hole in $G$ and $P^*$ be a simple path between the dual vertices in $G^*$ corresponding $f_\infty$ and $h$ in $G$, following the same terminology as in \Cref{def:parity}.
    It is not hard to see that $A\cup C$, $A\cup B$, and $B\cup C$ are closed curves, and that $E(A\cup B)$ is a multiset where the cardinality of $e\in E(A\cup B)$ corresponds to the sum of its cardinalities in $E(A)$ and $E(B)$.
    Thus, we have
    \begin{align*}
        \sum_{e^*\in E_{G^*}(P^*)} \ct\Big(e, E(A\cup C)\Big)
        =&
        \sum_{e^*\in E_{G^*}(P^*)} \ct\Big(e, E(A)\Big) 
        +
        \sum_{e^*\in E_{G^*}(P^*)} \ct\Big(e, E(C)\Big),
        \\
        \sum_{e^*\in E_{G^*}(P^*)} \ct\Big(e, E(A\cup B)\Big)
        =&
        \sum_{e^*\in E_{G^*}(P^*)} \ct\Big(e, E(A)\Big)
        +
        \sum_{e^*\in E_{G^*}(P^*)} \ct\Big(e, E(B)\Big),
        \\
        \sum_{e^*\in E_{G^*}(P^*)} \ct\Big(e, E(B\cup C)\Big)
        =&
        \sum_{e^*\in E_{G^*}(P^*)} \ct\Big(e, E(B)\Big)
        +
        \sum_{e^*\in E_{G^*}(P^*)} \ct\Big(e, E(C)\Big).
    \end{align*}
    By summing all three equations above, we have
    \begin{equation} \label{eq:ABC}
        \Par(h, A\cup C)+ \Par(h, A\cup B)+ \Par(h, B\cup C)
        \equiv
        0
        \pmod 2. 
    \end{equation}
    We now prove the main statement.
    \begin{itemize}
        \item 
            If $h\in \Phi_G(A\cup B)\bigtriangleup \Phi_G(B\cup C)$, then $h$ is inside exactly one of $A\cup B$ or $B\cup C$.
            This implies that exactly one of $\Par(h, A\cup B)$ or $\Par(h, B\cup C)$ equals one.
            By \Cref{eq:ABC}, it follows that $\Par(h, A\cup C) \equiv 1 \pmod 2$, and thus $h\in \Phi_G(A\cup C)$.  
        \item 
            If $h\in \Phi_G(A\cup C)$, then using similar reasoning, it follows that either $h\in \Phi_G(A\cup B)$ or $h\in \Phi_G(B\cup C)$, but $h\notin \Phi_G(A\cup B)\cap \Phi_G(B\cup C)$, and thus $h\in \Phi_G(A\cup B)\bigtriangleup \Phi_G(B\cup C)$.
        \qedhere
    \end{itemize}
\end{proof}

Using the above claim, in the following claim we prove that the closed curve formed by two paths with the same pattern does not contain any hole inside.
Before stating the claim, we restate the definition of the pattern of a path, which is used in the claim.

\pathpattern*

\begin{claim} \label{cor:AC} \label{clm:AC}
    Let $A$ and $C$ be two paths with the same endpoints in $G$.
    Then, $A$ and $C$ have the same pattern if and only if the pattern of the closed curve $A \cup C$ is $\Phi_G(A \cup C) = \emptyset$.
    i.e., $A$ and $C$ have the same pattern if and only if the closed curve $A \cup C$ does not contain any hole of $G$.
\end{claim}
\begin{proof}
    Let $t_1,t_2$ be the endpoints of $A$ and $C$, and let $B$ be the $t_1$-$t_2$ shortest path in $G_h$ as defined in \Cref{def:path-pattern}.
    We prove both directions separately.
    \begin{itemize}
        \item \underline{$\Rightarrow$)}
            Since $A$ and $C$ have the same pattern, by \Cref{def:path-pattern}, $\Phi_G(A \cup B) = \Phi_G(C \cup B)$.
            Using \Cref{clm:ABC}, it immediately follows that $\Phi_G(A \cup C) = \emptyset$.
        \item \underline{$\Leftarrow$)}
            Since $\Phi_G(A \cup C) = \emptyset$, it follows from \Cref{clm:ABC} that $\Phi_G(A \cup B) = \Phi_G(C \cup B)$.
            As $B$ is the $t_1$-$t_2$ shortest path in $G_h$, by \Cref{def:path-pattern}, it follows that $A$ and $C$ have the same pattern.
        \qedhere
    \end{itemize}
\end{proof}

We now prove the following simple claim.

\begin{claim} \label{clm:cross}
    Let $G$ be a planar graph with at least two distinct holes $h_1,h_2$, and let $R$ be a path between $h_1$ and $h_2$.
    Let $C$ be a closed curve in $G$.
    Then, if $R$ does not cross $C$, then both $h_1,h_2$ are either inside $C$ or outside $C$.
\end{claim}

\begin{proof}
    Assume for the sake of contradiction, that $R$ does not cross $C$ and that $h_1$ is inside $C$ while $h_2$ is outside $C$. 
    
    By the definition of the curve $C$, the region inside $C$ and the region outside $C$ are disjoint. 
    Therefore, any path connecting a point inside $C$ to a point outside $C$ must necessarily intersect $C$. 

    Since $R$ connects $h_1$ (inside $C$) to $h_2$ (outside $C$), it must cross $C$ at least once to transition from the interior to the exterior. 
    However, this contradicts the assumption that $R$ does not cross $C$. 
    Hence, the assumption that $h_1$ is inside $C$ while $h_2$ is outside $C$ is incorrect. 
    Consequently, both $h_1$ and $h_2$ must either lie inside $C$ or lie outside $C$.
\end{proof}

We proceed to establish the following claim. 
Intuitively, this claim asserts that replacing a subpath $P'$ of a path $P$ with another subpath $P''$ that shares the same pattern as $P'$ results in a path whose pattern remains identical to that of $P$. 
This claim will play a crucial role in subsequent arguments, particularly when analyzing the behavior of sliced subpaths of $P$ as the graph $G$ undergoes slicing operations.

\begin{claim} \label{clm:same-pattern}
    Let $G$ be a planar graph, and let $P$ be a $t_1$-$t_2$ path in $G$.
    Let $P'$ be a subpath of $P$ with endpoints $p_1$ and $p_2$, such that $P' = P[p_1,p_2]$.
    Let $P''$ be a $p_1$-$p_2$ path in $G$ that has the same pattern as $P'$.
    Then, the path
    \[
        P''' \coloneq P[t_1,p_1] \circ P'' \circ P[p_2,t_2]
    \]
    has the same pattern as $P$.
\end{claim}

\begin{proof}
    Let $S$ be the shortest $t_1$-$t_2$ path in $G_h$, as defined in \Cref{def:path-pattern}.
    Since $P'$ and $P''$ have the same pattern, it follows that $\Phi_G(P' \cup P'') = \emptyset$ by \Cref{clm:AC}.
    By applying \Cref{clm:ABC}, it follows that
    \begin{equation} \label{eqn:clm-same-pattern}
        \Phi_G(P \cup P''') = \Phi_G(P \cup S) \bigtriangleup \Phi_G(P''' \cup S).
    \end{equation}
    Since $P$ and $P'''$ differ only in the subpaths $P'$ and $P''$, we can conclude that
    \[
        \Phi_G(P \cup P''') = \Phi_G(P' \cup P'') = \emptyset.
    \]
    Substituting this into \Cref{eqn:clm-same-pattern}, we obtain $\Phi_G(P \cup S) = \Phi_G(P''' \cup S)$.
    By \Cref{def:path-pattern}, this implies that $P$ and $P'''$ have the same pattern.
\end{proof}

\subsubsection{Exact Maintenance of Patterns} \label{subsub:exact-pattern}

In this subsection, we show how to reduce $(G,T)$ to a one-hole instance such that, for every $t_1,t_2 \in T$ and pattern $\Phi$ in $G$, every $\Phi$-respecting $t_1$-$t_2$ shortest path in $G$ is {\it exactly} maintained in the reduced instance.

The main procedure in the reduction is straightforward: we repeatedly slice along a shortest path connecting a pair of holes (thus reducing the number of holes by one) until we reach the one-hole instance.
At each iteration, we slice along a path $R$, which is the shortest path between any pair of holes in the graph undergoing the slicing.

The slicing along $R$ is performed by adding a copy $\Rbar$ of $R$ to the graph and reassigning the edges incident to vertices of $R$ that lie on one side of $R$, to $\Rbar$.
We denote by $\Gbar$ the resulting sliced graph obtained from $G$.
See \Cref{alg:slice} for more details about the slicing procedure.

\begin{algorithm2e}[t]
    \KwIn{a planar graph $G$ with terminal set $T\subseteq V(G)$ and at least two holes}
    \KwOut{the sliced graph $\overbar G$ with one fewer hole than $G$}
    \BlankLine
    Let $h$ and $h'$ be the two closest holes in $(G,T)$ \;
    Let $R$ be the shortest path between $h$ and $h'$ with $u, v$ as its endpoints \;   
    Let $u'$ be the vertex succeeding $u$ on $h$ in the clockwise order \;
    Let $v'$ be the vertex preceding $v$ on $h'$ in the clockwise order \;
    Create the path \(\overbar{R}\) by duplicating the edges and vertices on $R$ \;
    Let $\overbar{G}$ be a copy of $G$ \;
    Add $\overbar{R}$ to $\overbar{G}$ \;
    \ForEach{vertex $w$ on $R$}{
        Reassign the edges incident to $w$ in $\Gbar$ from the left side of the path $u' - R - v'$ (in the direction from $u'$ to $v'$) to its copy $\wbar$ on $\Rbar$
    }
    \Return{$\overbar G$}
    \caption{Slice($G,T$)}
    \label{alg:slice}
\end{algorithm2e}

\begin{figure}[t]
    \centering
    \setkeys{Gin}{width=\linewidth}
    \begin{subfigure}{0.45\textwidth}
        \centering
        \includegraphics[page=1,width=\textwidth]{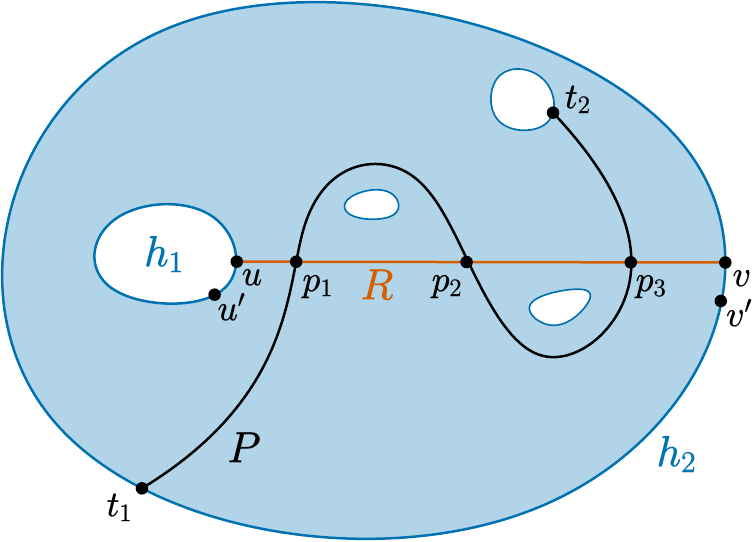}
        \caption{}
        \label{fig:slice-before}
    \end{subfigure}
    \hfil
    \begin{subfigure}{0.45\textwidth}
        \centering
        \includegraphics[page=2,width=\textwidth]{slice.pdf}
        \caption{}
        \label{fig:slice-after}
    \end{subfigure}
    \caption{
        An illustration of the slicing procedure described in \Cref{alg:slice} applied to $G$:
        (a) The holes $h_1$ and $h_2$ (the outer face of $G$) are the closest holes in $G$, and $R$ (the orange path) is the shortest path connecting $h_1$ to $h_2$.
        The vertices $t_1,t_2$ belong to the terminal set $T$, and $P$ (the black path) is the $\Phi$-respecting $t_1$-$t_2$ shortest path in $G$, where $\Phi$ is a pattern.
        The vertices $p_1,p_2,p_3$ are points on $P$ that intersect $R$ during traversal from $t_1$ to $t_2$.
        (b) The instance $\Gbar$ is created by slicing along $R$.
        The vertices $\pbar_1,\pbar_2,\pbar_3$ are the copies of $p_1,p_2,p_3$ on $\Rbar$ (the upper red path), respectively.
    }
    \label{fig:slice}
\end{figure}

The main intuition behind this procedure is that, although the slicing may partition the $\Phi$-respecting $t_1$-$t_2$ shortest path into subpaths, each subpath remains a ``pattern-respecting'' shortest path (defined shortly below) in the sliced graph $\Gbar$.
See \Cref{fig:slice} for an illustration.

Note that the pattern of the subpaths is defined in the sliced graph $\Gbar$, which can differ from the pattern $\Phi$ of $P$ in $G$. 
We use the term ``pattern-respecting'' to refer to the $\Phi'$-respecting shortest path when the exact pattern $\Phi'$ is not known. 
To avoid ambiguity, we formally define this term in the following definition.

\begin{definition}[Pattern-Respecting Shortest Path] \label{def:pattern-respecting}
    Let $(G,T)$ be an instance and let $P$ be a $u$-$v$ path in $G$. 
    We say that $P$ is a pattern-respecting shortest path in $G$ if $P$ is the shortest $u$-$v$ path in $G$ with pattern $\Phi_G(P)$.
\end{definition}

Note that a $\Phi$-respecting shortest path $P$ in $G$ may be split into several subpaths in $\Gbar$ (as depicted in \Cref{fig:slice}). 
Using \Cref{def:pattern-respecting}, we say that each such subpath $P'$ in the sliced graph $\Gbar$ is itself a shortest path between its endpoints with pattern $\Phi_{\Gbar}(P')$: the subpath $P'$ is a pattern-respecting shortest path in $\Gbar$. 
This definition helps reduce the complexity of notation used in our proof later. 

We now prove that the sliced graph $\Gbar$, as described in \Cref{alg:slice}, preserves all the necessary information for maintaining $\Phi$-respecting shortest paths in $G$, as stated in the following lemma.

Intuitively, the lemma asserts that every subpath $P'$ (as defined in the previous paragraphs) is a pattern-respecting shortest path in $\Gbar$. 
For example, using the case in \Cref{fig:slice}, the lemma shows that the segmented subpaths $P[t_1,p_1], P[p_1,p_2], P[p_2,p_3], P[p_3,t_2]$ of $P$, resulting from the slicing along $R$, remain pattern-respecting shortest paths in $\Gbar$, with their respective patterns defined in $\Gbar$. 
e.g., the subpath $P[p_1,p_2]$ in $G$ (with endpoints $p_1$ and $p_2$) is a $\Phi_{\Gbar}(P[\pbar_1,\pbar_2])$-respecting $\pbar_1$-$\pbar_2$ shortest path in $\Gbar$ (with endpoints $\pbar_1$ and $\pbar_2$). 
Therefore, if we preserve all pattern-respecting shortest paths between the endpoints of the subpaths along the sliced path $R$, we can retrieve $P$ using $\Gbar$.

The downside of this approach is that $R$ may contain $O(n)$ vertices, and thus considering all endpoints along $R$ may not reduce the size of the resulting pattern emulator we aim to construct. 
In \Cref{subsub:approx-pattern}, we address this size limitation by selecting $\Otil(|T|/\eps)$ vertices on $R$ instead of all vertices, trading off exactness for a $(1+\eps)$-approximate pattern emulator.

\begin{lemma} \label{lem:exact-pattern}
    Let $G$ be a planar graph with at least two holes, $T\subseteq V(G)$ be a terminal set, and $R$ be the shortest path in $G$ between the closest pair of holes.
    Let $\Gbar$ be the graph resulting from slicing $G$ along $R$ using \Cref{alg:slice}.
    Then, for every $t_1,t_2\in T$ and every pattern $\Phi$, the $\Phi$-respecting $t_1$-$t_2$ shortest path in $G$ is the union of pattern-respecting shortest paths in $\overbar G$ with endpoints in $T\cup V(R)\cup V(\overbar R)$, where $\overbar R$ is the copy of $R$ in $\overbar G$.
\end{lemma}
\begin{proof}
    Let $P$ be the shortest $\Phi$-respecting $t_1$-$t_2$ path in $G$ and $S$ be the shortest $t_1$-$t_2$ path in $G_h$ as defined in \Cref{def:path-pattern}.
    Without loss of generality, suppose that $R$ is the shortest path between the holes $h,h'\in \mathcal H(G,T)$.
    The shortest path in $G_h$ corresponding to $S$ does not cross $R$; after mapping copies back to $G$, $S$ does not cross $R$ as it exists in $G_h$.\footnote{This is where the choice of $S$ in $G_h$, as described in \Cref{def:path-pattern}, simplifies our analysis.}
    Without loss of generality, we assume that the hole $h$ is merged into $h'$ after the slice.
    i.e., the hole $h$ is eliminated in $\Gbar$.
    We consider the following two cases.
    \begin{itemize}
        \item \underline{The path $P$ does not cross $R$:}
            In this case, both $P$ and $S$ exist in $\Gbar$, and the closed curve $S \cup P$ does not cross $R$. 
            We show that $P$ is the shortest $t_1$-$t_2$ path in $\Gbar$ with pattern $\Phi' = \Phi \setminus h$.

            By \Cref{clm:cross}, the holes $h$ and $h'$ are either both contained inside $S \cup P$ or both outside it. 
            To show that $\Phi' = \Phi \setminus h$, we consider the following two distinct cases.

            \begin{itemize}
                \item \underline{Both $h, h'$ are outside $S \cup P$:}
                    By \Cref{def:path-pattern}, we conclude that $h, h' \notin \Phi$. 
                    Furthermore, since $S \cup P$ is not crossed by $R$, it remains unaffected by the slicing along $R$. 
                    Consequently, $\Phi' = \Phi_{\Gbar}(P) = \Phi$. 
                    Given that $h, h' \notin \Phi$, it follows that $\Phi' = \Phi \setminus h$.

                \item \underline{Both $h, h'$ are inside $S \cup P$:}
                    By \Cref{def:path-pattern}, we have $h, h' \in \Phi$. 
                    Since there is one-to-one correspondence between the holes in $\Gbar$ and $\mathcal{H}(G,T) \setminus h$, the only change to $\Phi'$ is the removal of $h$, as $h$ no longer exists in $\Gbar$. 
                    Therefore, it follows that $\Phi' = \Phi_{\Gbar}(P) = \Phi \setminus h$.
            \end{itemize}

            We now proceed to show that $P$ is indeed the shortest $t_1$-$t_2$ path in $\Gbar$. 
            Assume, for the sake of contradiction, that there exists a shorter $t_1$-$t_2$ path $P'$ in $\Gbar$ with pattern $\Phi'$. 
            By \Cref{clm:AC}, there are no holes inside $P \cup P'$. 
            Moreover, since both $P$ and $P'$ exist in $G$, it follows from \Cref{clm:same-pattern} that $P'$ would be a shorter path in $G$ with pattern $\Phi$, which contradicts the assumption that $P$ is the shortest $t_1$-$t_2$ path in $G$ with pattern $\Phi$. 
            Thus, $P$ must be the shortest $t_1$-$t_2$ path in $\Gbar$ with pattern $\Phi'$, completing the proof.

        \item \underline{The path $P$ crosses $R$:}
            Let $p_1,\dots, p_i$ be the sequence of vertices on $P$ (starting from $t_1$) that cross $R$.
            For simplicity of notation, we treat $p_j$s as the vertices on $R$ and their copies on $\Rbar$.\footnote{i.e., we also use $p_j$ to refer to $\pbar_j$, as illustrated in \Cref{fig:slice-before}, in contrast to \Cref{fig:slice-after}. This notation does not affect the subsequent arguments since any subpath $P[p_j, p_{j+1}]$ of $P$ has both of its endpoint edges emanating either from the left of $R$ in $G$ (and consequently from $\Rbar$ in $\Gbar$, meaning the actual endpoints are $\pbar_j, \pbar_{j+1}$) or from the right of $R$ in $G$ (and consequently from $R$ in $\Gbar$, meaning the actual endpoints are $p_j, p_{j+1}$).}
            See \Cref{fig:subpath-exact} for an illustration of the described setting.

            \begin{figure}[t]
                \centering
                \setkeys{Gin}{width=\linewidth}
                \begin{subfigure}{0.65\textwidth}
                    \centering
                    \includegraphics[page=1,width=\textwidth]{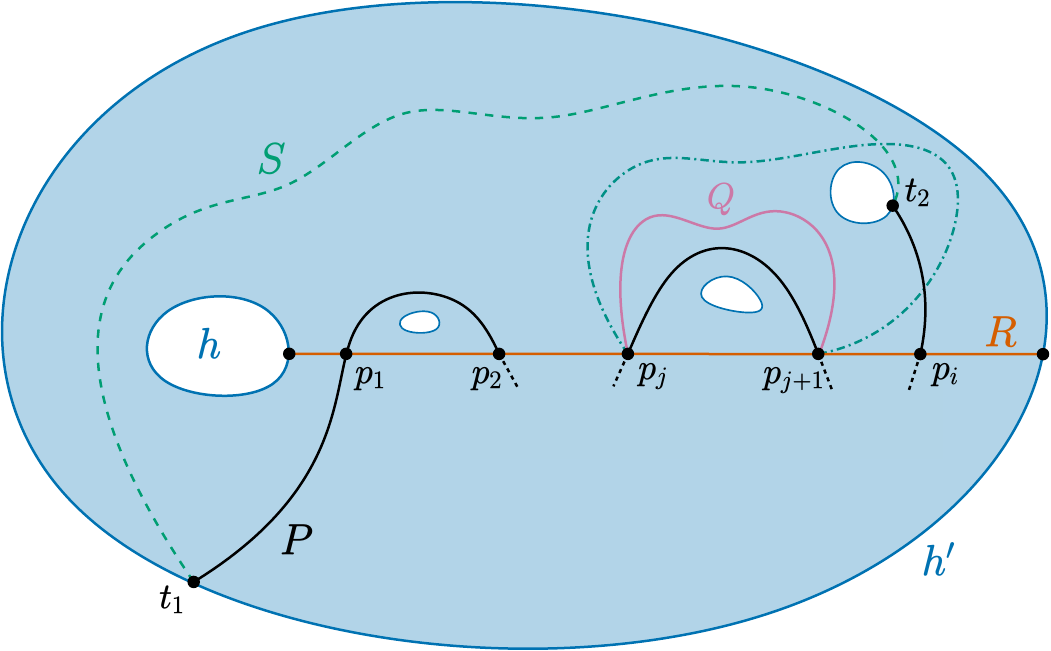}
                \end{subfigure}
                \caption{
                    An illustration of when the path $P$ (the black path) crosses $R$ (the orange path).
                    The path $Q$ (the pink path) is the shortest $p_j$-$p_{j+1}$ path in $\Gbar$ with the same pattern as $P[p_j,p_{j+1}]$.
                    The green paths show the shortest paths in $G_h$: dashed for the $t_1$-$t_2$ path $S[t_1,t_2]$ and dot-dashed for the $p_j$-$p_{j+1}$ path $S[p_j,p_{j+1}]$.
                    Note that the paths in $G_h$ are not crossed by any slicing path (such as $R$).
                }
                \label{fig:subpath-exact}
            \end{figure}

            Clearly, $i\geq 1$ by our assumption.
            If $i \geq 2$, it follows that, for $1 \leq j \leq i-1$, the closed curve $P[p_j, p_{j+1}] \cup R[p_j, p_{j+1}]$ must contain at least one hole; otherwise, since $R[p_j, p_{j+1}]$ is a shorter path than $P[p_j, p_{j+1}]$, replacing $P[p_j, p_{j+1}]$ with $R[p_j, p_{j+1}]$ in $P$ would result in a shorter $t_1$-$t_2$ path in $G$. 
            By \Cref{clm:AC}, this new path would have the same pattern as $P[p_j, p_{j+1}]$. 
            This substitution would ultimately result in a $t_1$-$t_2$ path in $G$ with the same pattern as $P$ by using \Cref{clm:same-pattern}, contradicting the assumption that $P$ is the shortest $t_1$-$t_2$ pattern-respecting path in $G$. 
            Thus, the closed curve $P[p_j, p_{j+1}] \cup R[p_j, p_{j+1}]$ must contain at least one hole.

            We now show that any subpath of $P$ with respect to the sequence $p_1,\dots, p_i$ is a shortest pattern-respecting path in $\overbar G$.
            We discuss the two types of subpaths separately below.
            \begin{itemize} \todo{Remove redundant/repetitive parts.}
                \item \underline{The subpath $P[p_j,p_{j+1}]$ for $1\leq j\leq i-1$:}
                    Let $S[p_j,p_{j+1}]$ be the shortest $p_j$-$p_{j+1}$ path in $G_h$ as defined in \Cref{def:path-pattern}.
                    Suppose that there exists a shorter $p_j$-$p_{j+1}$ path $Q$ in $\overbar G$ with pattern $\Phi_{\overbar G}(P[p_j,p_{j+1}]) \coloneq \Phi_{\overbar G}(R[p_j,p_{j+1}] \cup P[p_j,p_{j+1}])$.\footnote{Recall that the first $\Phi_{\overbar G}(\cdot)$ refers to the pattern of the path $R[p_j, p_{j+1}]$, as defined in \Cref{def:path-pattern}, while the second $\Phi_{\overbar G}(\cdot)$ refers to the pattern of the closed curve $R[p_j, p_{j+1}] \cup P[p_j, p_{j+1}]$, as defined in \Cref{def:curve-pattern}.}
                    Using \Cref{clm:same-pattern}, it follows that the path 
                    \[
                        P' = P[p_1, p_j] \circ Q \circ P[p_{j+1}, p_i]
                    \]
                    has exactly the same pattern as $P[p_1, p_i]$ in $G$.  
                    Specifically, it follows that $\Phi(P') = \Phi(P[p_1, p_i])$, which contradicts the assumption that $P$ is the shortest $t_1$-$t_2$ pattern-respecting path in $G$. 
                    Therefore, $P[p_j, p_{j+1}]$ must be the shortest $p_j$-$p_{j+1}$ path in $\Gbar$.

                    Let $S[t_1, p_1]$ and $S[p_i,t_2]$ be the shortest $t_1$-$p_1$ and $p_i$-$t_2$ paths in $G_h$, respectively.
                    Using the above argument, it follows that for the $t_1$-$t_2$ paths
                    \begin{align*}
                        P'' &= P[t_1,p_1] \circ P' \circ P[p_i,t_2], \text{ and} \\
                        P''' &= S[t_1, p_1] \circ \mathop{\bigcirc}\limits_{j=1}^{i-1} S[p_j,p_{j+1}] \circ S[p_i,t_2],
                    \end{align*}
                    we have
                    \[
                        \Phi_G(P'' \cup P''') = \Phi_G(P \cup P''').
                    \]
                    This follows using the following argument. By using curve decomposition, the only differences between $P'' \cup P'''$ and $P \cup P'''$ are the curves $Q \cup S[p_j,p_{j+1}]$ and $P[p_j,p_{j+1}] \cup S[p_j,p_{j+1}]$, respectively.
                    Since $\Phi_G(Q \cup P[p_j,p_{j+1}]) = \emptyset$, it follows that
                    \[
                        \Phi_G(Q \cup S[p_j,p_{j+1}]) = \Phi_G(P[p_j,p_{j+1}] \cup S[p_j,p_{j+1}]).
                    \]
                    \todo{Do we need more explanation?}

                    Using \Cref{clm:ABC} twice, we obtain  
                    \begin{align*}
                        \Phi_G(S \cup P'') &= \Phi_G(S \cup P''') \bigtriangleup \Phi_G(P'' \cup P''') \\
                                           &= \Phi_G(S \cup P''') \bigtriangleup \Phi_G(P \cup P''') \\
                                           &= \Phi_G(S \cup P),
                    \end{align*}
                    which implies that $P$ and $P''$ have the same pattern in $G$. However, $P''$ has a shorter length than $P$, thereby contradicting that $P$ is the shortest $\Phi$-respecting $t_1$-$t_2$ path.

                \item \underline{The subpaths $P[t_1,p_1]$ and $P[p_i,t_2]$:}
                    Suppose that there exists a shorter $t_1$-$p_1$ path $Q$ in $\overbar G$ with $\Phi_{\overbar G}(Q) = \Phi_{\overbar G}(P[t_1,p_1])$.
                    Using a similar argument as in the previous case, it follows that the path 
                    \[
                        P' \circ Q \circ P[p_1,t_2]
                    \]
                    exists in $G$, and that by \Cref{clm:same-pattern}, $\Phi_G(P' \cup P'') = \Phi_G(P \cup P'')$, where
                    \[
                        P'' = \overbar S[t_1, p_1] \circ P[p_1,p_i] \circ \overbar S[p_i, t_2].
                    \]
                    
                    Again, using \Cref{clm:ABC}, we have 
                    \begin{align*}
                        \Phi_G(S \cup P') &= \Phi_G(S \cup P'') \bigtriangleup \Phi_G(P' \cup P'') \\
                                         &= \Phi_G(S \cup P'') \bigtriangleup \Phi_G(P \cup P'') \\
                                         &= \Phi_G(S \cup P).
                    \end{align*}
                    Therefore, $\Phi_G(P') = \Phi_G(P)$; however, $P'$ is shorter than $P$, thereby contradicting the assumption that $P$ is the shortest path.

                    The argument for the subpath $P[p_i, t_2]$ follows similarly.
                    \qedhere
            \end{itemize}
    \end{itemize}
\end{proof}

We conclude this subsection by establishing an upper bound on the number of segments into which $P$ can be partitioned after slicing along $R$. 
We will use the upper bound later in \Cref{subsub:approx-pattern} to select a sufficient number of portals on $R$ to construct a $(1+\eps)$-approximate pattern emulator.

\begin{figure}[t]
    \centering
    \setkeys{Gin}{width=\linewidth}
    \begin{subfigure}{0.65\textwidth}
        \centering
        \includegraphics[page=1,width=\textwidth]{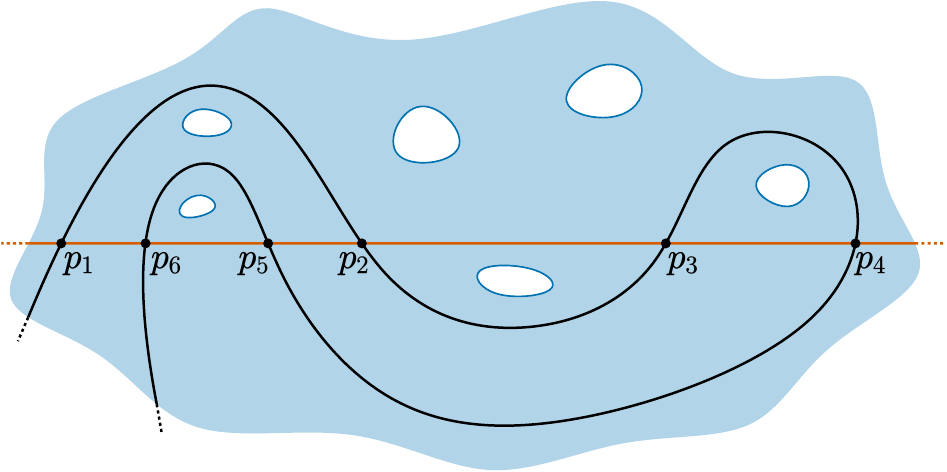}
    \end{subfigure}
    \caption{
        An example of how $P$ can cross $R$.
        The orange and black lines represent subpaths of $R$ and $P$, respectively.
        Note that $P$ does not necessarily partition $R$.
        For example, the subpath $R[p_5,p_6]$ is completely contained within $R[p_1,p_2]$.
        This requires a more detailed reasoning for upper bounding the number of segments of $P$, as the segments of $P$ do not necessarily correspond to a partitioning of $R$.
    }
    \label{fig:segment-number}
\end{figure}

It is important to note that, as $P$ intuitively ``navigates'' around the holes of $G$ to form its pattern, this process can result in a more complex segmentation compared to the scenario depicted in \Cref{fig:subpath-exact}. 
This complexity necessitates a more thorough analysis. 
Refer to \Cref{fig:segment-number} for an illustration of a more complex segmentation.

\begin{claim} \label{clm:subpaths}
    Let $G$ be a planar graph with $h \geq 2$ holes, $T \subseteq V(G)$ be a terminal set, and $R$ be the shortest path in $G$ between any pair of holes.
    Let $P$ be the $\Phi$-respecting $t_1$-$t_2$ shortest path in $G$, for any $t_1,t_2 \in T$ and pattern $\Phi$.
    Then, $R$ divides $P$ into at most $2^{h+1} + 2$ subpaths with both endpoints on $R$.
\end{claim}

\begin{proof}
    Suppose that $R$ partitions $P$ into at least $2^{h+1}+3$ subpaths with both endpoints on $R$.
    This implies that there are at least $2^{h+1}+1$ subpaths with both endpoints on $R$.
    As each subpath can contain some of the holes of $G$, by the pigeonhole principle, there must exist at least three subpaths in $G$ with the same pattern.
    
    Let $P_1$, $P_2$, and $P_3$ be the first, second, and third subpaths (starting from $t_1$) that connect $u_1$ to $v_1$, $u_2$ to $v_2$, and $u_3$ to $v_3$, respectively, with the same pattern.
    Moreover, let $P_1'$ and $P_2'$ denote the subpaths of $P$ between $P_1$ and $P_2$, and between $P_2$ and $P_3$, respectively.
    Let $P[t_1,u_1]$ and $P[v_3,t_2]$ represent the subpaths of $P$ before $P_1$ and after $P_3$, respectively.
    Thus, we can express $P$ as
    \[
        P = P[t_1,u_1] \circ P_1 \circ P_1' \circ P_2 \circ P_2' \circ P_3 \circ P[v_3,t_2].
    \]
    This definition is well-defined as $P_1$, $P_2$, and $P_3$ were chosen as the first, second, and third subpaths with the same pattern, respectively.
    Refer to \Cref{fig:segment-proof} for an illustration.

\begin{figure}[t]
    \centering
    \setkeys{Gin}{width=\linewidth}
    \begin{subfigure}{0.75\textwidth}
        \centering
        \includegraphics[page=,width=\textwidth]{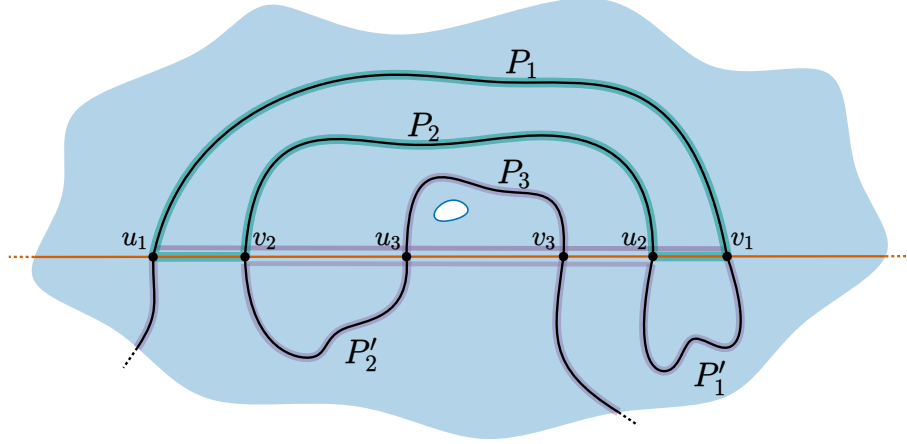}
    \end{subfigure}
    \caption{
        An illustration for the proof of \Cref{clm:subpaths}.
        The subpaths $P_1$, $P_2$, and $P_3$ of $P$ (the black path) have the same pattern.
        Note that $P_1$, $P_2$, and $P_3$ are nested in this example only for clarity of presentation; in the general case, they may not be nested and can cross each other.
        The example does not depict the entire path $P$ but instead focuses only on the sub-part containing $P_1$, $P_2$, $P_3$, and the connecting subpaths $P_1'$ and $P_2'$.
        The path $P'$ is identical to $P$ except in this sub-part, which is depicted in pink. 
        For clarity, its subpaths on $R$ are shown slightly above and below $R$; its difference from $P$ is that it replaces $P_1$ with $R[u_1,v_1]$ and $P_2$ with $R[u_2,v_2]$.
        The closed curve $M''$ is depicted in green.
        Here, $R[u_1, v_1]$ is the smallest subpath of $R$ that covers all subpaths $R[u_1,u_2]$, $R[v_1,v_2]$, $R[u_2,v_1]$, and $R[u_1,v_2]$. 
        Thus, $R[u,v] = R[u_1, v_1]$ in this example.
    }
    \label{fig:segment-proof}
\end{figure}

    We now show that we can form a shorter path by replacing $P_1$ and $P_2$ with shorter paths on $R$ that connect their endpoints.
    Specifically, we show that the path
    \[
        P' = P[t_1,u_1] \circ R[u_1, v_1] \circ P_1' \circ R[u_2, v_2] \circ P_2' \circ P_3 \circ P[v_3,t_2]
    \]
    has the same pattern as $P$ (i.e., $\Phi_{G}(P) = \Phi_{G}(P')$) but is shorter in length.
    
    Following the notation of \Cref{def:parity}, let $P^*$ be a simple $v^*$-$v$ path in the dual graph $G^*$.
    Let $A = \{e \in E(M) \mid e^* \in E_{G^*}(P^*) \}$ be the multiset of edges on the closed curve $M = P \cup S$, where $P$ is the path in $G$ corresponding to the dual path $P^*$, and $S$ is the shortest $t_1$-$t_2$ path in $G_h$.
    By \Cref{def:parity}, $\Par(f_v, M) = |A| \pmod 2$.
    
    Similarly, let $B = \{e \in E(P_1 \cup P_2) \mid e^* \in E_{G^*}(P^*) \}$ be the multiset of edges that are on $P_1$ or $P_2$ and correspond to the edges in $P^*$ in $G$.
    Since $B \subseteq A$, it follows that
    \[
        \Par(f_v, M) = \Big(|A \setminus B| + |B|\Big) \pmod 2.
    \]
    
    We argue that for the closed curve 
    \[
        M' = P' \cup S,
    \]
    it holds that
    \[
        \Par(f_v, M') = \Par(f_v, M),
    \]
    thus proving that $\Phi_{G}(P) = \Phi_{G}(P')$ by \Cref{def:path-pattern} since $f_v$ was arbitrary.
    
    To do so, let
    \[
        M'' = P_1 \cup R[v_1,u_2] \cup P_2 \cup R[v_2,u_1]
    \]
    be the closed curve between $P_1$, $P_2$, and $R$.\footnote{To simplify the presentation, we will use ``$\cup$'' in place of ``$\circ$'' to denote path concatenation as well.}
    Since $R$ is the shortest path connecting the endpoints of $P_1$ and $P_2$, it follows that $\Phi_G(R[v_1, u_2]) = \Phi_G(R[v_2, u_1]) = \emptyset$. 
    Therefore, $P_2$ and $R[v_1, u_2] \cup P_2 \cup R[v_2, u_1]$ have the same pattern. 
    Thus, $R[v_1, u_2] \cup P_2 \cup R[v_2, u_1]$ forms a $v_1$-$u_1$ path with pattern $\Phi_G(P_2) = \Phi_G(P_1)$. 
    It follows from \Cref{clm:AC} that the closed curve $M''$ does not contain any hole. 
    Consequently, the dual path $P^*$, whose endpoint is the dual vertex corresponding to a hole in $G$, cannot end up at a face inside $M''$. 
    That is, the destination face $f_v$ of $P^*$ must lie outside $M''$, as $f_v$ is required to be a hole by \Cref{def:path-pattern}. 
    Therefore, if $P^*$ enters $M''$, it must subsequently exit $M''$ to reach $f_v$. 
    By the parity argument, it thus follows that
    \begin{equation} \label{eq:h-jordan}
        \sum_{e^* \in E_{G^*}(P^*)} \ct\Big(e, E(M'')\Big) \equiv 0 \pmod 2.
    \end{equation}
    
    On the other hand, for the curves $M = P \cup S$ and $M' = P' \cup S$, we have
    \begin{align*}
        M &= M' \cup P_1 \cup P_2 \setminus \Big(R[u_1,v_1] \cup R[u_2,v_2]\Big), \\
          &= M' \cup P_1 \cup P_2 \cup R[u_2,v_1] \cup R[u_1,v_2] \setminus \Big(R[u_1,u_2] \cup R[v_1,v_2] \cup R[u_2,v_1] \cup R[u_1,v_2]\Big), \\
          &= M' \cup M'' \setminus \Big(R[u_1,u_2] \cup R[v_1,v_2] \cup R[u_2,v_1] \cup R[u_1,v_2]\Big),
    \end{align*}
    where the first equality follows from the fact that we replace $P_1$ and $P_2$ with $R[u_1,v_1]$ and $R[u_2,v_2]$, respectively, to form $M'$.
    Let $R[u,v]$ be the minimal subpath of $R$ that covers all subpaths $R[u_1,u_2]$, $R[v_1,v_2]$, $R[u_2,v_1]$, and $R[u_1,v_2]$.  
    It is not hard to see that 
    \begin{multline*}
        \sum_{e^* \in E_{G^*}(P^*)} \ct\bigg(e, E\Big( R[u_1,u_2] \cup R[v_1,v_2] \cup R[u_2,v_1] \cup R[u_1,v_2] \Big) \bigg) \\ \equiv \sum_{e^* \in E_{G^*}(P^*)} \ct\bigg(e, E\Big( R[u,v] \cup R[u,v] \Big) \bigg) \pmod 2 .
    \end{multline*}
    
    Therefore, we obtain 
    \begin{multline} \label{eq:h-parity}
        \sum_{e^* \in E_{G^*}(P^*)} \ct\Big(e, E(M)\Big) =
        \sum_{e^* \in E_{G^*}(P^*)} \ct\Big(e, E(M')\Big) +
        \sum_{e^* \in E_{G^*}(P^*)} \ct\Big(e, E(M'')\Big) \\ 
        - \sum_{e^* \in E_{G^*}(P^*)} \ct\bigg(e, E\Big(R[u,v] \cup R[u,v]\Big)\bigg).
    \end{multline}
    
    Since $R[u,v] \cup R[u,v]$ is a closed curve with no holes inside, by \Cref{clm:AC} and \Cref{def:curve-pattern}, we have
    \begin{equation} \label{eq:h-R}
        \sum_{e^* \in E_{G^*}(P^*)} \ct\bigg(e, E\Big(R[u,v] \cup R[u,v]\Big)\bigg) \equiv 0 \pmod 2.
    \end{equation}
    By combining \Cref{eq:h-jordan,eq:h-parity,eq:h-R}, we obtain 
    \[
        \sum_{e^* \in E_{G^*}(P^*)} \ct\Big(e, E(M)\Big) 
        \equiv
        \sum_{e^* \in E_{G^*}(P^*)} \ct\Big(e, E(M')\Big) 
        \pmod 2,
    \]
    which means that $P'$ is a path shorter than $P$ with the same pattern, thus contradicting the choice of $P$.
    This completes the proof.
\end{proof}

\subsubsection{Approximate Maintenance of Patterns} \label{subsub:approx-pattern}
In this subsection, we explain how we adapt the slicing scheme of \Cref{subsub:exact-pattern} to obtain a pattern emulator for an $O(1)$-hole instance $(G, T)$ with size near-linear in $|T|$.

\constanthole*

Recall that the slicing procedure of \Cref{subsub:exact-pattern} identifies the path $R$ in $G$ as the shortest path between the closest pair of holes, and slices along $R$ to produce $G'$. 
This operation merges the holes connected by $R$ in $G$ into a single hole in $G'$, thereby reducing the number of holes in $G'$ by one.

Generally speaking, to achieve a near-linear size, we select certain vertices on $R$, referred to as {\it portals}. 
To construct an instance aligned with $G$, we subsequently glue $R$ back at these portals. 
This gluing procedure will be applied to a pattern emulator of $G'$ (to be defined later), enabling us to obtain the $(1+\eps)$-approximate pattern emulator of $(G, T)$.

We begin by defining the portal set $\portal(R, T)$ for a shortest path $R$ with respect to a terminal set $T$. 
Before proceeding, we first restate \Cref{lem:portals}, which is referenced in the definition.

\portals*

\begin{restatable}[Portal Set $\sliceportal(R,T,\eps)$]{definition}{portalsall} \label{def:portals-all}
    Let $G$ be a planar graph with $h \geq 2$ holes, $T \subseteq V(G)$ be a terminal set, and $0 <\eps<1$ be an error parameter.
    Let $R$ be the shortest path between the closest pair of holes.
    Then, we define the set $\sliceportal(R,T,\eps)$ of portals on $R$ as
    \[
        \sliceportal(R,T,\eps) \coloneq \portal(R,T,\varepsilon''),
    \]
    where $\portal(\cdot,\cdot,\cdot)$ is from \Cref{lem:portals}, and $\varepsilon'' = \varepsilon' / 2^{h+5}$ and $\varepsilon' = \varepsilon /(2h)$.
\end{restatable}

Intuitively, the choice of $\eps''$ is influenced by two factors:
\begin{enumerate}
    \item 
        By \Cref{clm:subpaths}, for every pattern $\Phi$, $R$ divides each $\Phi$-respecting shortest path into $O(2^h)$ subpaths. 
        Hence, we set $\eps'' = O(\eps'/2^h)$.
    \item 
        As we will observe later, we recursively slice along the shortest path to reduce the $h$-hole instance into a one-hole instance. 
        Since this reduction involves a sequence of $O(h)$ instances, we set $\eps' = O(\eps/h)$.
\end{enumerate}

For convenience, we restate \Cref{clm:portals}, which will be used later in the proofs of this subsection.

\portaleps*

We introduce the notation necessary for our $(1+\varepsilon)$-approximate pattern emulator construction. 
This consists of two parts: first, we define the sliced instance $(G_i, T_i)$ based on $(G,T)$ for $1 \leq i \leq h$ using a top-down approach.  
Next, we use this definition in a bottom-up fashion to define the glued instance $(G_i', T_i')$ based on $(G_h,T_h)$ for $1 \leq i \leq h$.

We now formalize the definition of a sliced instance $(G_i, T_i)$ for $1 \leq i \leq h$. 
Intuitively, this instance is derived from $(G, T)$ by iteratively slicing along the shortest paths between pairs of holes and adding the portals defined in \Cref{def:portals-all} on these shortest paths to the terminal set. 
We present this formally in the following definition.

\begin{restatable}[Sliced Instance $(G_i,T_i)$ Based on $(G,T)$ with Parameter $\eps$]{definition}{defsliced} \label{def:sliced-instance}
    Let $G$ be a planar graph with $h\geq 2$ holes, $T \subseteq V(G)$ be a terminal set, and $0 <\eps<1$ be an error parameter.
    For $1 \leq i \leq h$, we recursively define the instance $(G_i,T_i)$ as follows.
    \begin{enumerate}
        \item 
            We set $(G_1,T_1) \coloneq (G,T)$.
        \item 
            For $2 \leq i \leq h$, we define $(G_i,T_i)$ based on $(G_{i-1},T_{i-1})$ as follows.
            Let $R_{i-1}$ be the shortest path connecting the closest pair of holes in $G_{i-1}$.
            \begin{itemize}
                \item 
                    $G_i$ is the graph obtained from $G_{i-1}$ by slicing along $R_{i-1}$ using \Cref{alg:slice}.
                    i.e., using the terminology of \Cref{alg:slice}, $G_i = \Gbar_{i-1}$.
                \item
                    $T_i$ is obtained from $T_{i-1}$ together with the portals on $R_{i-1}$ (and its copy $\Rbar_{i-1}$) as defined in \Cref{def:portals-all}.
                    i.e., we set 
                    \[
                        T_i \coloneq T_{i-1} \cup \sliceportal(R_{i-1},T_{i-1},\eps) \cup \overbar{\sliceportal}(R_{i-1},T_{i-1},\eps),
                    \]
                    where $\overbar{\sliceportal}(R_{i-1},T_{i-1},\eps)$ is the set of vertices on $\Rbar_{i-1}$ that are copies of $\sliceportal(R_{i-1},T_{i-1},\eps)$.
            \end{itemize}
    \end{enumerate}
\end{restatable}

Refer to \Cref{fig:G1,fig:Gbar1,fig:G2,fig:Gbar2} for an illustration of \Cref{def:sliced-instance}. 
Since each instance $(G_i, T_i)$ has one fewer hole than its predecessor $(G_{i-1}, T_{i-1})$, the following observation is evident.

\begin{observation} \label{obs:sliced-instance}
    For $1 \leq i \leq h$, each sliced instance $(G_i, T_i)$ as defined in \Cref{def:sliced-instance} is a $(h-i+1)$-hole instance.
\end{observation}

We now define the glued instance $(G_i', T_i')$ for $1 \leq i \leq h$. 
Intuitively, $(G_i', T_i')$ is constructed recursively by gluing the shortest path $R_{i}$, as defined in \Cref{def:sliced-instance}, onto its portals in the instance $(G_{i+1}', T_{i+1}')$. 
We formalize this process in the following definition.

\begin{definition}[Glued Instance $(G_i',T_i')$ Based on $(G_h,T_h)$ with Parameter $\eps$] \label{def:glued-instance}
    Let $G$ be a planar graph with $h\geq 2$ holes, $T \subseteq V(G)$ be a terminal set, and $0 <\eps<1$ be an error parameter.
    For $1 \leq i \leq h$, let $(G_i,T_i)$ be the sliced instance as defined in \Cref{def:sliced-instance}.
    For $1 \leq i \leq h$, we recursively define the instance $(G_i',T_i')$ as follows.
    \begin{enumerate}
        \item 
            We set $(G_h',T_h') \coloneq (G_h,T_h)$.
        \item 
            For $1 \leq i \leq h-1$, we define $(G_i',T_i')$ based on $(G_{i+1}',T_{i+1}')$ as follows.
            Let $R_{i}$ be the shortest path connecting the closest pair of holes in $G_{i}$ as defined in \Cref{def:sliced-instance}.
            \begin{itemize}
                \item 
                    $G_i'$ is the graph obtained from $G_{i+1}'$ by gluing the portals of $R_{i}$.
                    i.e., we contract every vertex $v \in \sliceportal(R_i,T_i,\eps)$ with its copy $\vbar \in \overbar{\sliceportal}(R_i,T_i,\eps)$.
                \item
                    $T_i'$ is simply set as $T_i$.
            \end{itemize}
    \end{enumerate}
\end{definition}

Note that the gluing procedure in \Cref{def:glued-instance} is performed on $G_{i+1}'$ and not on $G_{i+1}$. 
However, the shortest path $R_i$ is still available in $G_{i+1}'$ because we only glue the shortest paths $R_{h-1}, \dots, R_{i+1}$ prior to gluing back $R_i$. 
All of these paths were sliced after $R_i$ was sliced in $G_i$. 
Thus, there is no ambiguity in defining $R_i$ within $G_{i+1}'$.

Refer to \Cref{fig:G1prime,fig:G2prime} for an illustration of \Cref{def:glued-instance}. 
Since each instance $(G_i', T_i')$ restores the hole that was removed from $(G_i, T_i)$ during the formation of $(G_{i+1}, T_{i+1})$, the following observation is evident.

\begin{observation} \label{obs:glued-instance}
    For $1 \leq i \leq h$, each glued instance $(G_i',T_i')$ as defined in \Cref{def:glued-instance} is a $(h-i+1)$-hole instance.
\end{observation}

We are now prepared to prove the most technical part of this subsection, which states that each glued instance $(G_i', T_i')$ serves as a good approximation for the corresponding sliced instance $(G_i, T_i)$. 
In particular, $(G_1', T_1')$ approximates $(G_1, T_1) = (G, T)$.

\begin{remark}[Notation]
    Since $T_i' = T_i$ in \Cref{def:glued-instance}, for the remainder of this section, we will use $(G_i', T_i)$ instead of $(G_i', T_i')$ to simplify the notation.
\end{remark}

\begin{claim} \label{clm:glued-approximation} 
    Let $G$ be a planar graph with $h \geq 2$ holes, $T \subseteq V(G)$ be a terminal set, and $0 < \varepsilon < 1$ be an error parameter.
    For each $1 \leq i \leq h$, let $(G_i,T_i)$ and $(G_i',T_i)$ be instances with parameter $\eps$ as defined in \Cref{def:sliced-instance,def:glued-instance}.
    Then, for every $t_1, t_2 \in T_i$ and pattern $\Phi$, the graph $G_i'$ approximates every $\Phi$-respecting $t_1$-$t_2$ shortest path in $G_i$ within a factor of $(1+\varepsilon)$.
\end{claim}

\begin{proof}
    We prove the claim by induction on $I_j \coloneq G_{h-j+1}$ for $1 \leq j \leq h$; we show that, for every $t_1, t_2 \in T_{h-j+1}$ and pattern $\Phi$, the graph $G_{h-j+1}'$, defined in \Cref{def:glued-instance}, approximates every $\Phi$-respecting $t_1$-$t_2$ shortest path in $G_{h-j+1}$ with quality $(1+\varepsilon')^{j-1}$, where $\varepsilon'$ is defined in \Cref{def:portals-all}.
    The claim then follows because $\varepsilon' = \varepsilon /(2h)$, and
    \[
        (1+\varepsilon')^i \leq (1+\varepsilon')^h \leq 1+\varepsilon.
    \]
    
    For the base case $j=1$, the result holds trivially since $G_h' = G_h$ by \Cref{def:glued-instance}.
    
    The remainder of the proof addresses the case $j = i+1$, where $i \geq 1$.  
    We assume that $I_{j-1}' = I_i' \coloneq G_{h-i+1}'$ approximates every $\Phi$-respecting $t_1$-$t_2$ shortest path in $I_i = G_{h-i+1}$ with quality $(1+\varepsilon')^{i-1}$.
    
    Let $P$ be the $\Phi$-respecting $t_1$-$t_2$ shortest path in $I_j = G_{h-i}$ for $t_1, t_2 \in T_{h-i}$.  
    Recall that $G_{h-i+1}$ is obtained from $G_{h-i}$ by slicing along the shortest path $R \coloneq R_{h-i}$ in $G_{h-i}$ (using \Cref{alg:slice}), where $R$ is the shortest path between any two holes of $G_{h-i}$.  
    We denote by $\Rbar$ the copy of $R$ in the sliced graph $G_{h-i+1}$.  
    
    By \Cref{clm:subpaths}, $R$ partitions $P$ into at most $2^{h+1} + 2$ segments. 
    Let $p_1, \dots, p_l$ be the sequence of vertices on $P$ (starting from $t_1$) that cross $R$ while traversing on $P$.  
    Thus, $l \leq 2^{h+1} + 1 \leq 2^{h+2}$.  
    
    If $P$ does not cross $R$, then $P$ exists in $G_{h-i+1}$, so the induction hypothesis directly gives the bound.
    Thus, for the rest of the proof, we assume that $l \geq 1$.

    To construct $G_{h-i}'$, we glue $G_{h-i+1}'$ back along the portals on $R$ and $\Rbar$.  
    Refer to \Cref{fig:subpath-segment-approx} for an illustration of $G_{h-i}'$ and the subpaths of $P$ in it.

    \begin{figure}[t]
        \centering
        \setkeys{Gin}{width=\linewidth}
        \begin{subfigure}{0.7\textwidth}
            \centering
            \includegraphics[page=1,width=\textwidth]{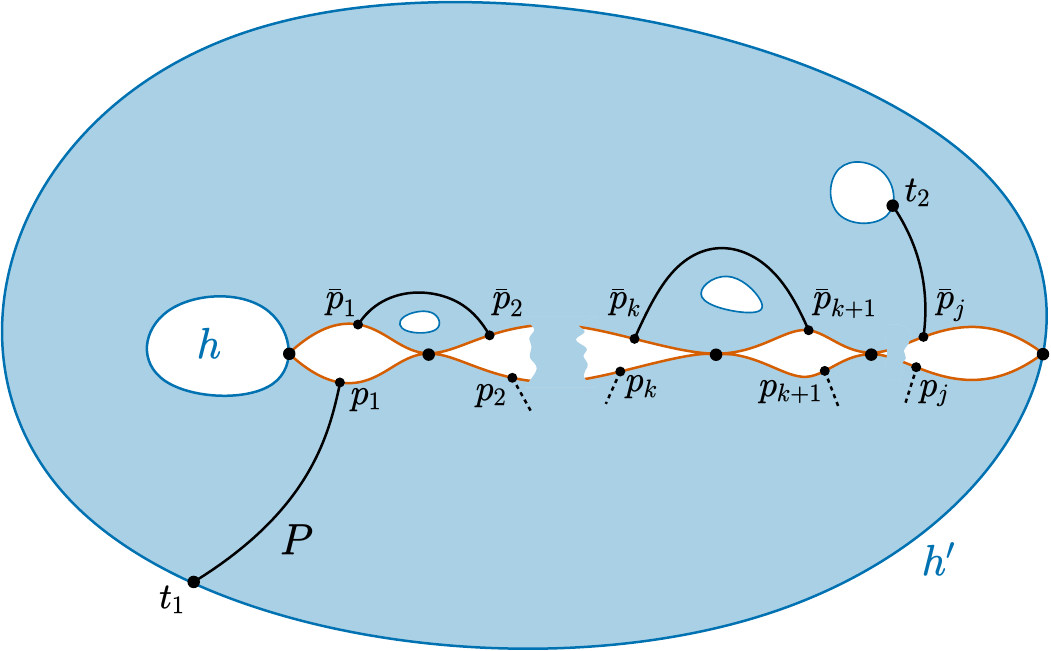}
        \end{subfigure}
        \caption{
            An illustration of $G_{h-i}'$ obtained from $G_{h-i+1}'$, where the (orange) glued lines are $R$ and its copy $\Rbar$ in $G_{h-i+1}$.
            The path $P$ can cross $R$ several times, which are depicted by the vertices $p_1, \dots, p_l$.
            The vertices $\pbar_1, \dots, \pbar_j$ are the corresponding copy vertices on $\Rbar$.
            This results in subpaths of $P$ in $G_{h-i}'$, such as $P[t_1,p_1]$ and $P[\pbar_1,\pbar_2]$ in this example.
            By \Cref{clm:subpaths}, it follows that $l \leq 2^{h+2}$.
        }
        \label{fig:subpath-segment-approx}
    \end{figure}

    To provide an upper bound on the length of the shortest $\Phi$-respecting $t_1$-$t_2$ path in $G_{h-i}'$, we construct a $t_1$-$t_2$ path $P'$ with pattern $\Phi$ and then bound the length of $P'$.  
    The following notation is used in the construction of $P'$.\footnote{Similar to the proof of \Cref{lem:exact-pattern}, we may also use $p_j$ to refer to $\pbar_j$ for inner subpaths of $P$. In these cases, this notation abuse will not affect the subsequent arguments since any subpath $P[p_j, p_{j+1}]$ of $P$ has both of its endpoint edges emanating either from the left of $R$ in $G$ (meaning the actual endpoints are $\pbar_j, \pbar_{j+1}$) or from the right of $R$ in $G$ (meaning the actual endpoints are $p_j, p_{j+1}$).}

    \begin{itemize}
        \item 
            Let $t_1'$ be the portal in $\sliceportal(R,T_{h-i},\eps)$ that is associated with $p_1$ from the terminal $t_1$ (i.e., $t_1'$ is the closest portal to $p_1$ in $\portal(R,\{t_1\}, \eps'')$), and let $S[t_1,t_1']$ be the shortest $t_1$-$t_1'$ path in $G_{h-i+1}$ with the same pattern as $P[t_1,p_1]$.  
            Refer to \Cref{fig:subpath-approx-a} for an illustration.

            \begin{figure}[t]
                \centering
                \setkeys{Gin}{width=\linewidth}
                \begin{subfigure}{0.35\textwidth}
                    \centering
                    \includegraphics[page=2,width=\textwidth]{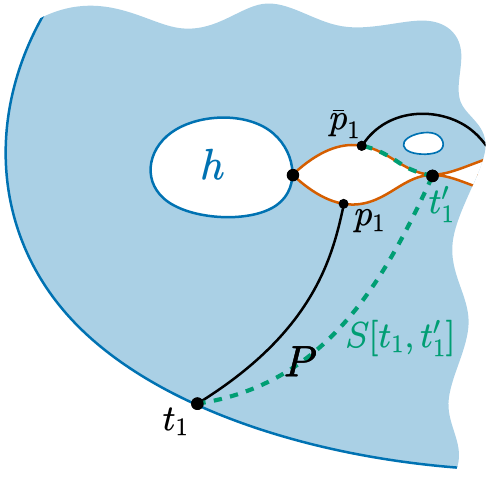}
                    \caption{}
                    \label{fig:subpath-approx-a}
                \end{subfigure}
                \hfil
                \begin{subfigure}{0.33\textwidth}
                    \centering
                    \includegraphics[page=3,width=\textwidth]{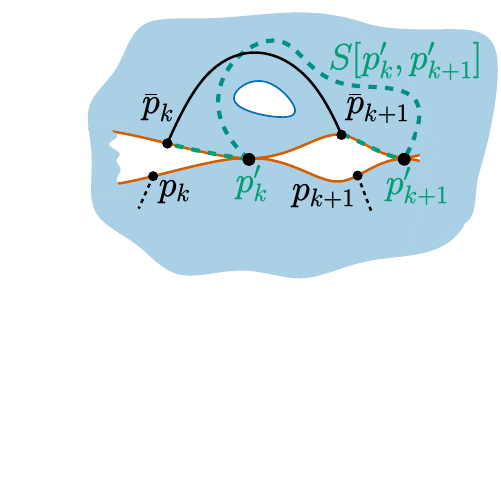}
                    \caption{}
                    \label{fig:subpath-approx-b}
                \end{subfigure}                
                \caption{
                    An illustration of the subpaths of the path $P'$ chosen in $G_{h-i+1}$ for subpaths of $P$ for the case that $P$ enters $R$ and leaves $\Rbar$ (starting from $t_1$).
                    Each subpath of $P'$ has the same pattern as its corresponding subpath of $P$:
                    (a) For the subpath $P[t_1, \pbar_1]$, we select the shortest $t_1$-$t_1'$ path $S[t_1, t_1']$ with the same pattern as $P[t_1, \pbar_1]$, where $t_1'$ is the  portal on $R$ associated with $p_1$ via $t_1$.
                    The subpath $P'[t_1, \pbar_1]$ connecting $t_1$ to $\pbar_1$ is then defined as $S[t_1, t_1'] \circ \Rbar[t_1', \pbar_1]$, represented by the dashed green line.
                    A similar approach is applied to define $P'[t_2, \pbar_j]$.
                    (b) For the subpath $P[\pbar_k, \pbar_{k+1}]$, we first use $\Rbar$ to traverse from $\pbar_k$ to its closest portal $p_k'$ and from $\pbar_{k+1}$ to its closest portal $p_{k+1}'$.
                    Next, we find the shortest $p_k'$-$p_{k+1}'$ path $S[p_k', p_{k+1}']$ that has the same pattern as $P[\pbar_k, \pbar_{k+1}]$.
                    The subpath $P'[\pbar_k, \pbar_{k+1}]$ is then set as $\Rbar[\pbar_k, p_k'] \circ S[p_k', p_{k+1}'] \circ \Rbar[\pbar_{k+1}, p_{k+1}']$, illustrated by the dashed green line.
                }
                \label{fig:subpath-approx}
            \end{figure}

            Since $P[t_1,p_1]$ and $S[t_1,t_1']$ have the same pattern and that $P[t_1,p_1]$ is a shortest pattern-respecing path, it follows that
            \begin{equation}
                \Big|P[t_1,p_1]\Big| \leq \Big|S[t_1,t_1'] \cup R[t_1',p_1]\Big| = \Big|S[t_1,t_1'] \cup R[t_1',\pbar_1]\Big|. %
            \end{equation}

        \item 
            For each $1 \leq k \leq l-1$, let $p_k'$ be the closest portal in $\sliceportal(R,T_{h-i},\eps)$ to $p_k$.  
            Let $S\big[p_k',p_{k+1}'\big]$ denote the shortest $p_k'$-$p_{k+1}'$ path in $G_{h-i+1}$ that has the same pattern as $P\big[p_k,p_{k+1}\big]$ in $G_{h-i+1}$.  
            Similarly, let $S\big[p_k,p_{k+1}\big]$ denote the shortest $p_k$-$p_{k+1}$ path in $G_{h-i+1}$ that has the same pattern as $P\big[p_k,p_{k+1}\big]$ in $G_{h-i+1}$.\footnote{Note that $S[p_k, p_{k+1}]$ may not necessarily be equal to $P[p_k, p_{k+1}]$, as the graphs $G_h', \dots, G_{h-i+2}'$ are glued onto their respective portals prior to addressing $G_{h-i+1}'$. Consequently, the subpath $P\big[p_k, p_{k+1}\big]$ might no longer exist in $G_{h-i+1}'$.}
            Refer to \Cref{fig:subpath-approx-b} for an illustration.

            By using the triangle inequality, we have
            \begin{equation} \label{eqn:triangle}
                \begin{split}
                    \Big|S\big[p_k',p_{k+1}'\big]\Big| \leq& \Big|R\big[p_k,p_k'\big]\cup S\big[p_k,p_{k+1}\big]\cup R\big[p_{k+1},p_{k+1}'\big]\Big| \\
                    =& \Big|R\big[p_k,p_k'\big]\Big| + \Big|S\big[p_k,p_{k+1}\big]\Big| + \Big|R\big[p_{k+1},p_{k+1}'\big]\Big|.
                \end{split}
            \end{equation}

        \item 
            Let $t_2'$ be the portal in $\sliceportal(R,T_{h-i},\eps)$ that is associated with $p_l$ via the terminal $t_2$.  
            Similar to the first item, since $P[t_2,p_l]$ and $S[t_2,t_2']$ have the same pattern and that $P[t_2,p_l]$ is a shortest pattern-respecing path, we have
            \begin{equation}
                \Big|P[t_2,p_l]\Big| \leq \Big|S[t_2,t_2'] \cup R[t_2',p_l]\Big| = \Big|S[t_2,t_2'] \cup R[t_2',\pbar_l]\Big|. %
            \end{equation}
            where $S[t_2,t_2']$ is the shortest $t_2$-$t_2'$ path in $G_{h-i+1}$ with the same pattern as $P[t_2,p_l]$.
    \end{itemize}
    
    Now, we construct the path $P'$ in $G_{h-i}'$ by using the selected portals above in a natural way as follows.
    Note that the shortest paths used to define $P'$ are taken from $G_{h-i+1}'$ (rather than $G_{h-i+1}$); we use the notation $S[\cdot,\cdot]$ for the shortest paths in $G_{h-i+1}$ and the notation $S'[\cdot,\cdot]$ for the shortest paths in $G_{h-i+1}'$ corresponding to one defined in $S[\cdot,\cdot]$.

    \begin{enumerate}
        \item 
            The subpath $P'[t_1,\pbar_1]$ consists of the shortest $t_1$-$t_1'$ path $S'[t_1,t_1']$ in $G_{h-i+1}'$ and the subpath $R\big[t_1',\pbar_1\big]$.  
            Specifically, we define
            \[
                P'\big[t_1,\pbar_1\big] \coloneq S'\big[t_1,t_1'\big] \circ R\big[t_1',\pbar_1\big].
            \]

        \item 
            For $1 \leq k \leq l-1$, we define $P'\big[p_k,p_{k+1}\big]$ and $P'\big[\pbar_k,\pbar_{k+1}\big]$ as
            \begin{align*}
                P'\big[p_k,p_{k+1}\big] & \coloneq R\big[p_k,p_k'\big] \circ S'\big[p_k',p_{k+1}'\big] \circ R\big[p_{k+1},p_{k+1}'\big], \text{ and} \\
                P'\big[\pbar_k,\pbar_{k+1}\big] & \coloneq R\big[\pbar_k,\pbar_k'\big] \circ S'\big[\pbar_k',\pbar_{k+1}'\big] \circ R\big[\pbar_{k+1},\pbar_{k+1}'\big],
            \end{align*}
            where $S'\big[p_k',p_{k+1}'\big]$ (and $S'\big[\pbar_k',\pbar_{k+1}'\big]$) denotes the shortest $p_k'$-$p_{k+1}'$ (and $\pbar_k'$-$\pbar_{k+1}'$) path in $G_{h-i+1}'$ that has the same pattern as $P\big[p_k,p_{k+1}\big]$ (and $P\big[\pbar_k,\pbar_{k+1}\big]$) in $G_{h-i+1}$.  

        \item 
            The subpath $P'\big[t_2,\pbar_l\big]$ is defined similarly as 
            \[
                P'\big[t_2,\pbar_j\big] \coloneq S'\big[t_2,t_2'\big] \circ R\big[t_2',\pbar_l\big].
            \]

    \end{enumerate}

    By the induction hypothesis, the path $S'\big[p_k',p_{k+1}'\big]$ in $G_{h-i+1}'$ approximates $S\big[p_k',p_{k+1}'\big]$ in $G_{h-i+1}$ with quality $(1+\varepsilon')^{i-1}$.  
    Specifically, we have
    \begin{equation} \label{eqn:hypothesis}
        \Big|S'\big[p_k',p_{k+1}'\big]\Big| \leq (1+\varepsilon')^{i-1} \Big|S\big[p_k',p_{k+1}'\big]\Big|,
    \end{equation}
    and
    \begin{equation} \label{eqn:hypothesis-endpoints}
        \Big|S'\big[t_1,t_{1}'\big]\Big| \leq (1+\varepsilon')^{i-1} \Big|S\big[t_1,t_{1}'\big]\Big| 
        \quad \text{and} \quad
        \Big|S'\big[t_2,t_{2}'\big]\Big| \leq (1+\varepsilon')^{i-1} \Big|S\big[t_2,t_{2}'\big]\Big|.
    \end{equation}

    Combining \Cref{eqn:hypothesis,eqn:triangle} yields
    \begin{equation} \label{eqn:middle}
        \Big|S'\big[p_k',p_{k+1}'\big]\Big| \leq \Big(1+\varepsilon'\Big)^{i-1} \Bigg(\Big|R\big[p_k,p_k'\big]\Big| + \Big|S\big[p_k,p_{k+1}\big]\Big| + \Big|R\big[p_{k+1},p_{k+1}'\big]\Big|\Bigg).
    \end{equation}
    
    Using the definition of $P'$ in $G_{h-i}'$, we obtain
    \todo{$P'$ in $G_{h-i}$ or in $G_{h-i}'$?}
    \begin{align*}
        |P'| = & \Big| P'\big[t_1,\pbar_1\big]\Big| + \Big| P'\big[t_2,\pbar_l\big]\Big| + \sum_{k=1}^{l-1} \Big| P'\big[p_k,p_{k+1}\big]\Big| \\
        = & \Big| R\big[p_1,t_1'\big] \cup S'\big[t_1,t_1'\big]\Big| + \Big|R\big[p_j,t_2'\big] \cup S'\big[t_2,t_2'\big]\Big| \\
        & \quad + \sum_{k=1}^{l-1} \Bigg(\Big|R\big[p_k,p_k'\big] \cup S'\big[p_k,p_{k+1}\big] \cup R\big[p_{k+1},p_{k+1}'\big]\Big|\Bigg) \\
        = & \Big|S'\big[t_1,t_1'\big]\Big| + \sum_{k=1}^{l-1} \Big|S'\big[p_k,p_{k+1}\big]\Big| + \Big|S'\big[t_2,t_2'\big]\Big| \\  
        & \quad + \Big|R\big[p_1,t_1'\big]\Big| + \sum_{k=1}^{l-1} \Bigg(\Big|R\big[p_k,p_k'\big]\Big| + \Big|R\big[p_{k+1},p_{k+1}'\big]\Big|\Bigg) + \Big|R\big[p_l,t_2'\big]\Big|.
    \end{align*}
    Using \Cref{eqn:hypothesis-endpoints,eqn:middle} in the previous equality yields
    \begin{align*}
        |P'| \leq & \Big(1+\varepsilon'\Big)^{i-1} \Bigg( \Big|S\big[t_1,t_1'\big]\Big| + \sum_{k=1}^{l-1} \Big|S\big[p_k,p_{k+1}\big]\Big| + \Big|S\big[t_2,t_2'\big]\Big| \Bigg) \\  
        & \quad + \Big|R\big[p_1,t_1'\big]\Big| + \sum_{k=1}^{l-1} \Bigg(\Big|R\big[p_k,p_k'\big]\Big| + \Big|R\big[p_{k+1},p_{k+1}'\big]\Big|\Bigg) + \Big|R\big[p_l,t_2'\big]\Big|.
    \end{align*}
    Now, using the triangle inequality, we obtain 
    \begin{align*}
        |P'| & \leq \Big(1+\varepsilon'\Big)^{i-1} 
            \Bigg(
                \Big|R\big[p_1,t_1'\big]\Big| + \Big|S\big[t_1,p_1\big]\Big| \\
                & \quad\quad\quad\quad\quad\quad + \sum_{k=1}^{l-1} \bigg(\Big|R\big[p_k,p_k'\big]\Big| + \Big|S\big[p_k,p_{k+1}\big]\Big| + \Big|R\big[p_{k+1},p_{k+1}'\big]\Big|\bigg) \\
                & \quad\quad\quad\quad\quad\quad + \Big|R\big[p_j,t_2'\big]\Big| + \Big|S\big[t_2,p_l\big]\Big|
            \Bigg) \\
                    & \quad + \Big|R\big[p_1,t_1'\big]\Big| + \sum_{k=1}^{l-1} \Bigg(\Big|R\big[p_k,p_k'\big]\Big| + \Big|R\big[p_{k+1},p_{k+1}'\big]\Big|\Bigg) + \Big|R\big[p_l,t_2'\big]\Big| \\
                & = \Big(1+\varepsilon'\Big)^{i-1} 
                \Bigg( 
                    \Big|S\big[t_1,p_1\big]\Big| + \sum_{k=1}^{l-1} \Big|S\big[p_k,p_{k+1}\big]\Big| + \Big|S\big[t_2,p_l\big]\Big| 
                \Bigg) \\
                  & \quad + \Big(1+ \big(1+\varepsilon'\big)^{i-1}\Big) 
                  \Bigg( 
                      \Big|R\big[p_1,t_1'\big]\Big| + \sum_{k=1}^{l-1} \bigg(\Big|R\big[p_k,p_k'\big]\Big| + \Big|R\big[p_{k+1},p_{k+1}'\big]\Big|\bigg) + \Big|R\big[p_l,t_2'\big]\Big| 
                  \Bigg) \\
                & = \Big(1+\varepsilon'\Big)^{i-1} |P|\\
                  & \quad + \Big(1+ \big(1+\varepsilon'\big)^{i-1}\Big) 
                  \Bigg( 
                      \Big|R\big[p_1,t_1'\big]\Big| 
                      + \sum_{k=1}^{l-1} \bigg(\Big|R\big[p_k,p_k'\big]\Big| + \Big|R\big[p_{k+1},p_{k+1}'\big]\Big|\bigg) 
                      + \Big|R\big[p_l,t_2'\big]\Big| 
                  \Bigg).
    \end{align*}
    To provide an upper bound for the last line, we use the $\varepsilon''$-cover property of $\sliceportal(R,T_{h-i},\eps)$ by \Cref{clm:portals}.\footnote{Since in $\sliceportal(R, T_{h-i}, \eps)$ of \Cref{def:portals-all}, we compute an $\eps''$-cover on $R$, \Cref{clm:portals} holds for $\eps''$ (instead of $\eps$).}  
    Specifically, for every $1 \leq k \leq l$, we have 
    $\Big|R\big[p_k,p_k'\big]\Big| \leq \varepsilon'' |P|$ as well as $\Big|R\big[p_1,t_1'\big]\Big| , \Big|R\big[p_j,t_2'\big]\Big| \leq \varepsilon'' |P|$.\footnote{We note that the bound in \Cref{clm:portals} is $O(\eps''|P|)$. Using a simple trick, we can absorb the constant into $\eps''$, which we omit here for the sake of clarity in presentation.}
    Since $j \leq 2^{h+2}$, we obtain
    \begin{align*}
        |P'| & \leq \big(1+\varepsilon'\big)^{i-1} |P| + 2^{h+4} \Big(1+ \big(1+\varepsilon'\big)^{i-1}\Big) \varepsilon'' |P| \\
             & = \Bigg( \big(1+\varepsilon'\big)^{i-1} + 2^{h+4} \Big(1+ \big(1+\varepsilon'\big)^{i-1}\Big) \varepsilon'' \Bigg) |P| \\
             & \leq (1+\varepsilon')^i |P|,
    \end{align*}
    where the last inequality follows from $\varepsilon'' = \varepsilon'/2^{h+5}$ and the fact that
    \begin{align*}
        \big(1+\varepsilon'\big)^{i-1} + 2^{h+4} \Big(1+ \big(1+\varepsilon'\big)^{i-1}\Big) \varepsilon'' 
        & = 
        \big(1+\varepsilon'\big)^{i-1} + 2^{h+4} \Big(1+ \big(1+\varepsilon'\big)^{i-1}\Big) \frac{\varepsilon'}{2^{h+5}} \\
        & = \big(1+\varepsilon'\big)^{i-1} + \Big(1+ \big(1+\varepsilon'\big)^{i-1}\Big) \frac{\varepsilon'(1+\varepsilon')^{i-1}}{2(1+\varepsilon')^{i-1}} \\
        & \leq \big(1+\varepsilon'\big)^{i-1} + \Big(1+ \big(1+\varepsilon'\big)^{i-1}\Big) \frac{\varepsilon'(1+\varepsilon')^{i-1}}{1+(1+\varepsilon')^{i-1}} \\
        & = \big(1+\varepsilon'\big)^{i-1} + \Big( \big(1+\varepsilon'\big)^i - \big(1+\varepsilon'\big)^{i-1}\Big) \\
        & = (1+\varepsilon')^i.
    \end{align*}
    As $i = j-1$, the proof is now complete.
\end{proof}

Finally, we discuss the construction of our $(1+\eps)$-approximate pattern emulator $(H, T)$ for $(G, T)$ by leveraging the sliced instances $(G_i, T_i)$ defined earlier. 
The construction of $(H, T)$ is carried out straightforwardly using the following steps.

\begin{enumerate}
    \item 
        We use \Cref{lem:one-hole} from \Cref{subsec:one-hole} to compute a $(1+\eps_H)$-approximate pattern emulator $(H_h, T_h)$ of $(G_h, T_h)$. 
        Note that, by \Cref{obs:sliced-instance}, the instance $(G_h, T_h)$ is a one-hole instance. 
        The choice of $\eps_H$ (discussed shortly afterward) ensures that the resulting pattern emulator $(H, T)$ is a $(1+\eps)$-approximate pattern emulator of $(G,T)$.

    \item 
        For $1 \leq i \leq h-1$, we construct the emulator $(H_i', T_i)$ of $(G_i, T_i)$ in a bottom-up fashion by setting $(H_i', T_i)$ to be the glued instance based on $(H_h, T_h)$ as defined in \Cref{def:glued-instance}. 
        That is, starting with $(H_h', T_h) = (H_h, T_h)$, for each $1 \leq i \leq h-1$, we use $(H_{i+1}', T_{i+1})$ and contract every portal $v\in \sliceportal(R_i,T_i,\eps_H)$ with its corresponding copy $\vbar$. 
        The resulting instance is denoted as the emulator $(H_i', T_i)$.

    \item 
        Let $(H, T)$ be the instance $(H_1', T_1)$. 
        To achieve a $(1+\eps)$-approximation quality, we rely on the guarantee of \Cref{clm:glued-approximation}, combined with the $\eps''$-cover defined in \Cref{def:portals-all}, and the fact that $(H_h', T_h)=(H_h, T_h)$ is a $(1+\eps_H)$-approximate pattern emulator of $(G_h, T_h)$. 
        Specifically, we set $\varepsilon'' = \varepsilon' / 2^{h+5}$ and $\varepsilon' = \eps_H / (2h)$ where $\eps_H = \eps / 4$,.
\end{enumerate}

We now show that $(H, T)$ is indeed a $(1+\varepsilon)$-approximate pattern emulator of $(G, T)$.  
To begin, we prove that $(H_i', T_i)$ $(1+\varepsilon)$-approximates $(G_i', T_i)$.

\begin{claim} \label{clm:approximation-quality-glued}
    Let $G$ be a planar graph with $h \geq 2$ holes, $T \subseteq V(G)$ be a terminal set, and $0 < \varepsilon < 1$ be an error parameter. 
    For each $1 \leq i \leq h$, let $(G_i, T_i)$ be the sliced instance with parameter $\varepsilon$ defined in \Cref{def:sliced-instance}. 
    Let $(H_h, T_h)$ be a $(1+\varepsilon)$-approximate pattern emulator of $(G_h, T_h)$ computed using \Cref{lem:one-hole}. 
    For $1 \leq i \leq h$, let $(H_i', T_i)$ be the glued instance with parameter $\varepsilon$ based on $(H_h, T_h)$, and let $(G_i', T_i)$ be the glued instance with parameter $\varepsilon$ based on $(G_h, T_h)$ as defined in \Cref{def:glued-instance}.

    Then, for every $t_1, t_2 \in T_i$ and pattern $\Phi$, the graph $H_i'$ $(1+\varepsilon)$-approximates every $\Phi$-respecting $t_1$-$t_2$ shortest path in $G_i'$.
\end{claim}

\begin{proof}
    The argument follows from a straightforward adaptation of the proof of \Cref{clm:glued-approximation}. 
    To avoid redundancy, we outline the proof by building upon the reasoning used in \Cref{clm:glued-approximation}.

    The proof proceeds by induction on $I_j \coloneq H_{h-j+1}'$ for $1 \leq j \leq h$, where we show that, for every $t_1, t_2 \in T_{h-j+1}$ and pattern $\Phi$, the graph $H_{h-j+1}'$ approximates every $\Phi$-respecting $t_1$-$t_2$ shortest path in $G_{h-j+1}'$ with approximation quality $(1+\varepsilon')^{j}$. 
    The claim then follows from the observation that $\varepsilon' = \varepsilon/(2h)$ and
    \[
        (1+\varepsilon')^j \leq (1+\varepsilon')^h \leq (1+\varepsilon).
    \]

    The base case follows trivially, as $(H_h', T_h)$ is a $(1+\varepsilon)$-approximate pattern emulator of $(G_h', T_h) = (G_h, T_h)$.

    Assume the claim holds for $j-1$, i.e., $I_{j-1} = I_i = H_{h-i+1}'$ approximates every $\Phi$-respecting $t_1$-$t_2$ shortest path in $G_{h-i+1}'$ with approximation quality $(1+\varepsilon')^{i}$, where $i = j-1$.  
    We now prove the claim for $j = i+1$.

    Let $P$ denote the $\Phi$-respecting $t_1$-$t_2$ shortest path in $G_{h-i}'$ for some $t_1, t_2 \in T_{h-i}$.  
    Recall that $G_{h-i}'$ is obtained from $G_{h-i+1}'$ by gluing the portals along the shortest path $R \coloneq R_{h-i}$ and its copy $\Rbar$ in $G_{h-i}$ (as defined in \Cref{alg:slice}), where $R$ is the shortest path connecting two holes of $G_{h-i}$.  

    We analyze two cases based on whether $P$ crosses $R$:

    \begin{itemize}
        \item \underline{The path $P$ does not cross $R$:}  
            In this case, since $R$ is sliced in $G_{h-i+1}$ and consequently in $G_{h-i+1}'$, the path $P$ exists in $G_{h-i+1}'$.  
            By the inductive hypothesis, $P$ is preserved in $H_{h-i}'$ with approximation quality $(1+\varepsilon')^{i}$.  
            Therefore, the overall quality satisfies $(1+\varepsilon')^{i} \leq (1+\varepsilon')^{i+1}$.

        \item \underline{The path $P$ crosses $R$:}  
            Let $p_1, \dots, p_l$ represent the sequence of vertices on $P$ (starting from $t_1$) that cross $R$ during traversal along $P$.  
            Let $\pbar_1, \dots, \pbar_l'$ and $t_1', t_2'$ denote the closest portals on $R$ corresponding to these intersections, as defined in the proof of \Cref{clm:glued-approximation}.  
            Let $P'$ denote the path in $G_{h-i}$ obtained by replacing the segments of $P$ crossing $R$ with the corresponding portal-based shortcuts as defined in the proof of \Cref{clm:glued-approximation}.

            By the same reasoning in \Cref{clm:glued-approximation}, it follows that $|P'| \leq (1+\varepsilon')|P|$, and that $P'$ has the same pattern $\Phi$ as $P$.  
            Since the approximation quality increases by a multiplicative factor of $(1+\varepsilon')$, the overall quality in $G_{h-i}'$ becomes $(1+\varepsilon')^{i+1}$.
    \end{itemize}

    Hence, the claim holds as $i = j-1$, completing the proof.
\end{proof}

We are now ready to prove that $(H, T)$ is a $(1+\varepsilon)$-approximate pattern emulator of $(G, T)$.

\begin{claim} \label{clm:approximation-quality-G}
    Let $G$ be a planar graph with $h \geq 2$ holes, $T \subseteq V(G)$ be a terminal set, and $0 < \varepsilon < 1$ be an error parameter. 
    For each $1 \leq i \leq h$, let $(G_i, T_i)$ be the sliced instance with parameter $\eps_H \coloneq \eps/4$ defined in \Cref{def:glued-instance}. 
    Let $(H_h, T_h)$ be a $(1+\eps_H)$-approximate pattern emulator of $(G_h, T_h)$ computed using \Cref{lem:one-hole}. 
    For $1 \leq i \leq h$, let $(H_i', T_i)$ be the glued instance based on $(H_h, T_h)$ with parameter $\eps_H$ as defined in \Cref{def:glued-instance}.

    Then, for every $t_1, t_2 \in T_i$ and pattern $\Phi$, the graph $H_i'$ $(1+\eps)$-approximates every $\Phi$-respecting $t_1$-$t_2$ shortest path in $G_i$.
\end{claim}

\begin{proof}
    By \Cref{clm:approximation-quality-glued}, each $H_i'$ is a $(1+\varepsilon_H)$-approximate pattern emulator of $G_i'$, and by \Cref{clm:glued-approximation}, each $G_i'$ is a $(1+\varepsilon_H)$-approximate pattern emulator of $G_i$.  
    Therefore, by the transitivity property of pattern emulators (\Cref{lem:transitivity-pattern-emulator}), $H_i'$ approximates $G_i$ with overall quality
    \(
        (1+\varepsilon_H)(1+\varepsilon_H) \leq (1+\varepsilon),
    \)
    as required.
\end{proof}

We conclude this subsection with the proof of \Cref{thm:constant-hole}.

\constanthole*

\begin{proof}[Proof of \Cref{thm:constant-hole}]
    The $(1+\varepsilon)$ approximation quality directly follows from \Cref{clm:approximation-quality-G}.
    It remains to prove the guarantees for the emulator size, the construction time, and alignment.

    \begin{itemize}
        \item \underline{Emulator Size:}  
            By \Cref{def:sliced-instance}, the sliced instance $(G_i, T_i)$ is constructed by slicing $G_i$ along the path $R_{i-1}$ in $G_{i-1}$.  
            Since $R_{i-1}$ is a shortest path and has a length of $O(n)$, and $h = O(1)$ by assumption, it follows that each sliced graph $G_i$ has size $O(n)$.  
            Additionally, by \Cref{def:portals-all}, we select $\Otil(|T_{i-1}| / \varepsilon^{O(1)})$ portals on $R_i$ to include in $T_i$.  
            Using straightforward induction, it follows that for all $1 \leq i \leq h$, the terminal set $T_i$ has size $\Otil(|T| / \varepsilon^{O(1)})$.  
            Combining these observations, the overall size of each sliced instance $(G_i, T_i)$ is $\Otil(n)$. 
    
            Next, consider the one-hole instance $(H_h, T_h)$.  
            By \Cref{lem:one-hole}, the size of $(H_h, T_h)$ is $\Otil(|T| / \varepsilon^{O(1)})$.  
            Furthermore, $(H_i', T_i)$ is obtained from $(H_{i+1}', T_{i+1})$ by contracting the portals on $R_{i}$ with their corresponding copies on $\Rbar_{i}$.  
            Since the contraction process does not increase the size of the graph, a simple induction shows that for all $1 \leq i \leq h$, the size of $(H_i', T_i)$ is $\Otil(|T| / \varepsilon^{O(1)})$.  
            Hence, the size of the pattern emulator $(H, T)$ satisfies the desired bound.
    
        \item \underline{Construction Time:}  
            By the above reasoning, the size of each graph $G_i$ is $O(n)$.  
            Since each $G_i$ is obtained by slicing along $R_i$, it follows that we can construct $G_i$ in $O(n)$ time.  
    
            For the terminal sets, by \Cref{lem:portals}, the portals for each sliced instance can be computed in $\Otil(n)$ time.  
            Therefore, the total time to construct all $(G_i, T_i)$ instances is $\Otil(n)$.  
    
            Next, consider the one-hole instance $(H_h, T_h)$.  
            Since $(H_h, T_h)$ contains $\Otil(|T| / \varepsilon^{O(1)})$ terminals, by \Cref{lem:one-hole}, we can construct $(H_h, T_h)$ in $\Otil(n / \varepsilon^{O(1)})$ time.  
    
            Finally, constructing $(H_i', T_i)$ for $1 \leq i < h$ involves contracting at most $n$ vertices in $(H_{i+1}', T_{i+1})$ on $R_{i}$ with its corresponding copy $\Rbar_{i}$.  
            Since contraction is a linear-time operation, the total time to construct all $(H_i', T_i)$ instances is also bounded by $\Otil(n / \varepsilon^{O(1)})$.  
            Thus, the total construction time for the emulator $(H, T)$ satisfies the desired bound.
    
        \item \underline{Alignment:}  
            The alignment argument follows directly from the construction process described in \cite{Chang:2022aa}.
            The approach used to build our pattern emulator $(H, T)$ relies on the same foundational building blocks, including slicing, $\eps$-cover portal selection, and contraction.  
            To avoid unnecessary repetition and ensure clarity, we refer the reader to \cite{Chang:2022aa} for a detailed proof of alignment properties.  
            This guarantees that $(H_i', T_i)$ is aligned with $(G_i, T_i)$, and consequently, $(H, T)$ is aligned with $(G, T)$.  
    \qedhere
    \end{itemize}
\end{proof}

\subsection{The General Case: Proof of Theorem~\ref{thm:static}} \label{subsec:general}

In this subsection, we prove \Cref{thm:static}, restated below.

\static*

The approach we use in this subsection is fairly standard, and thus we keep the discussion concise.
Similar arguments can be found for the general case in \cite{Chang:2022aa,Chen:2025aa}.

We first restate \Cref{def:r-division} and \Cref{thm:structured-division}, which are employed in our construction.

\rdivision*
\structured*

To establish \Cref{thm:static}, we first use the following procedure to obtain a $(1+\eps)$-approximate pattern emulator with a smaller size.
\begin{enumerate}
    \item 
        Use \Cref{thm:structured-division} to partition $E(G)$ into pieces $E_1, \dots, E_k$, each containing $O(1)$ holes.
    \item 
        For each piece $G_i \coloneq G[E_i]$, compute a planar pattern emulator $(H_i,T_i)$ of $(G_i,T_i)$ using \Cref{thm:constant-hole}, where $T_i$ contains the boundary vertices of $G_i$ and the terminal vertices $T \cap V(G_i)$ present in $G_i$.
    \item 
        By composability (\Cref{lem:composability-pattern-emulators}), the union $H = \cup_{i=1}^k H_i$ with terminal set restricted to $T$ serves as a planar pattern emulator of $(G,T)$.
\end{enumerate}

We now provide the guarantees for this procedure in the following claim.

\begin{claim} \label{clm:static}
    Let $G$ be an $n$-vertex planar graph, $T \subseteq V(G)$ be a terminal set, and $0< \eps <1$ be an error parameter.
    Then, the procedure above computes a $(1 + \varepsilon)$-approximate pattern emulator $(H,T)$ of $(G,T)$ with size $\Otil(\sqrt{n|T|}/\varepsilon^{O(1)})$ in time $\Otil(n/\varepsilon^{O(1)})$.
\end{claim}

\begin{proof}
    By \Cref{thm:structured-division}, we can construct all the pieces $E_1, \dots, E_k$ in $O(n)$ time such that, for each piece $G_i = G[E_i]$, the size of the boundary set is $|T_i| = O(\sqrt{r})$.
    To ensure $O(1)$-holes for all pieces,\footnote{In this context, the holes cover all the boundary vertices and the terminal vertices $T \cap V(G_i)$ present in $G_i$.} we set $r = \Theta(n/|T|)$, which implies $k = O(|T|)$.

    As the terminals are evenly distributed across the pieces, this results in $O(1)$ terminal vertices in each piece, ensuring that the number of holes remains $O(1)$.
    Consequently, we can use \Cref{thm:constant-hole} for each instance $(G_i,T_i)$ to compute a $(1 + \varepsilon)$-approximate pattern emulator $(H_i,T_i)$ in time $\Otil(|G_i|/\varepsilon^{O(1)}) = \Otil\Big(n/(|T|\varepsilon^{O(1)})\Big)$ with size $\Otil(|T_i|/\varepsilon^{O(1)}) = \Otil\Big(\sqrt{n}/(\sqrt{|T|}\varepsilon^{O(1)})\Big)$.
    Therefore, the total time for constructing $(H,T)$ is
    \[
        \sum_{i=1}^k \Otil\Big(n/(|T|\varepsilon^{O(1)})\Big) = \Otil\Big(n/\varepsilon^{O(1)}\Big),
    \]
    and the size of $(H,T)$ is
    \[
        \sum_{i=1}^k \Otil\Big(\sqrt{n}/(\sqrt{|T|}\varepsilon^{O(1)})\Big) = \Otil(\sqrt{n|T|}/\varepsilon^{O(1)}).
    \qedhere
    \]
\end{proof}

The pattern emulator of \Cref{clm:static} has a significantly larger size than what we are aiming for in \Cref{thm:static}. 
To achieve the desired size and thus prove \Cref{thm:static}, we use \Cref{clm:static} recursively until we obtain the desired size for our emulator. 
We formalize this in the following definition.

\begin{restatable}[$L$-Sequence $(1+\eps)$-Approximate Pattern Emulators of $(G,T)$]{definition}{sequence} \label{def:sequence}
    Let $G$ be an $n$-vertex planar graph, $T \subseteq V(G)$ be a terminal set, $0 < \eps < 1$ be an error parameter, and $L$ be a positive integer. 
    The $L$-sequence $(1+\eps)$-approximate pattern emulators of $(G,T)$ is a sequence $(H_1,T), \dots, (H_L,T)$ of $(1+\eps)$-approximate pattern emulators defined recursively as follows.
    \begin{enumerate}
        \item 
            $(H_1,T)$ is a $(1+\eps)$-approximate pattern emulator of $(G,T)$ obtained via \Cref{clm:static}.
        \item 
            For $1\leq i\leq L-1$, $(H_{i+1},T)$ is a $(1+\varepsilon)$-approximate pattern emulator of $(H_i,T)$ obtained via \Cref{clm:static}.
    \end{enumerate}
\end{restatable}

Using the terminology of \Cref{def:sequence}, our desired pattern emulator $(H,T)$ of $(G,T)$ is then defined as $(H_L,T)$. 
We are now ready to prove \Cref{thm:static}.

\begin{proof}[Proof of \Cref{thm:static}]
    We construct an $L$-sequence of $(1+\eps')$-approximate pattern emulators of $(G,T)$ using \Cref{def:sequence}, with $L = \log\log n$ and $\varepsilon' = \varepsilon / (2L)$. 
    This results in a sequence $(H_1,T), \dots, (H_L,T)$ of $(1+\eps')$-approximate pattern emulators. 
    We analyze the sequence $(H_1,T), \dots, (H_L,T)$ and separately prove the guarantees for $(H,T) \coloneq (H_L,T)$.
    \begin{itemize}
        \item \underline{Construction Time:}
            By \Cref{clm:static} and using simple induction, it follows that $|V(H_i)| = \Otil(n/\varepsilon^{O(1)})$ for $1 \leq i \leq L$.
            The construction time then follows directly from \Cref{clm:static} and the fact that $L = \log\log n$.

        \item \underline{Approximation Quality:}
            Using simple induction, it follows that $(H_i,T)$ approximates $(G,T)$ with quality $(1 + \varepsilon')^i$ for $1 \leq i \leq L$.
            The quality for $(H,T)$ is then given by 
            \[
                (1 + \varepsilon')^L = (1 + \varepsilon/(2L))^L \leq e^{\varepsilon/(2L)L} \leq 1 + \varepsilon.
            \]
        \item \underline{Emulator Size:}
            We show by induction that, for $1 \leq i \leq L$, the size of $H_i$ is
            \[
                \Otil\Bigg(\frac{n^{\frac{1}{2^i}} |T|^{\frac{2^i-1}{2^i}}}{\varepsilon^{c'\big(1+\sum_{j=1}^i \frac{1}{2^j}\big)}}\Bigg),
            \]
            where $c'$ is a sufficiently large constant for the exponent of $\varepsilon$ in \Cref{clm:static}.
            The upper bound on the emulator size then follows as
            \[
                |H_L| =  \Otil\Bigg(\frac{n^{\frac{1}{2^{\log\log n}}} |T|}{\varepsilon^{c'(1+1)}}\Bigg) = \Otil\Big(|T|/\varepsilon^{O(1)}\Big).
            \]
            
            The remainder is dedicated to proving the induction.

            For $i = 1$, the bound on $|H_1|$ trivially follows from \Cref{clm:static}.
            
            For $i = k+1$, assume that 
            \[
                |H_k| =  \Otil\Bigg(\frac{n^{\frac{1}{2^k}} |T|^{\frac{2^k-1}{2^k}}}{\varepsilon^{c'\big(1+\sum_{j=1}^k \frac{1}{2^j}\big)}}\Bigg).                
            \]
            By \Cref{clm:static}, $|H_{k+1}| = \Otil\Big(\sqrt{n_k |T|}/\varepsilon^{c'}\Big)$ where $n_k = |H_k|$.
            Substituting the assumed size of $H_k$ into this bound yields
            \begin{align*}
                \Otil\Bigg(\sqrt{\frac{n^{\frac{1}{2^k}} |T|^{\frac{2^k-1}{2^k}}}{\varepsilon^{c'\big(1+\sum_{j=1}^k \frac{1}{2^j}\big)}}} \cdot \frac{\sqrt{|T|}}{\varepsilon^{c'}}\Bigg) 
                & = 
                \Otil\Bigg(\frac{n^{\frac{1}{2^{k+1}}} |T|^{\frac{1}{2}} |T|^{\frac{-1}{2^{k+1}}}}{\varepsilon^{c'\big(\frac{1}{2}+\sum_{j=1}^k \frac{1}{2^{j+1}}\big)}} \cdot \frac{\sqrt{|T|}}{\varepsilon^{c'}}\Bigg) \\
                & = 
                \Otil\Bigg(\frac{n^{\frac{1}{2^{k+1}}} |T|^{1-\frac{1}{2^{k+1}}}}{\varepsilon^{c'\big(\sum_{j=1}^{k+1} \frac{1}{2^{j}}\big)}} \cdot \frac{1}{\varepsilon^{c'}}\Bigg) \\
                & = 
                \Otil\Bigg(\frac{n^{\frac{1}{2^{k+1}}} |T|^{\frac{2^{k+1} - 1}{2^{k+1}}}}{\varepsilon^{c'\big(1 + \sum_{j=1}^{k+1} \frac{1}{2^{j}}\big)}} \Bigg),
            \end{align*}
            as desired.
        \qedhere
    \end{itemize}
\end{proof}

\ali{Generalize to flow sparsifier? Section 4.4 of KR17.}

\section{Dynamic All-Pairs $s$-$t$ Min-Cut} \label{sec:dynamic}
In this section, we present our fully dynamic algorithm for maintaining all-pairs $s$-$t$ min-cuts with subpolynomial update and query times, as outlined below.

\Intromain*

Our algorithm is derived by dynamizing the static algorithm of \Cref{subsub:approx-pattern}, which computes a $(1+\eps)$-approximate pattern emulator for $(G,T)$. 
The guarantees of this dynamic algorithm are restated below.

\dynamicpattern*

The remainder of this section is organized as follows.
\begin{itemize}
    \item
        In \Cref{subsec:dual}, we describe how to efficiently maintain a dual instance
        $(G_D,T_D)$ associated with $(G,T)$.
        As shown in \Cref{subsec:reduction}, computing a $(1+\eps)$-approximate
        pattern emulator for $(G_D,T_D)$ results in a $(1+\eps)$-approximate terminal cut sparsifier for $(G,T)$.
        Therefore, in order to use a dynamic $(1+\eps)$-approximate pattern
        emulator, it remains to maintain the corresponding dual instance efficiently
        after each update.
        We achieve this by maintaining a dynamic instance $(G_D^*,T_D^*)$ that
        satisfies the same relevant properties as $(G_D,T_D)$.

    \item
        In \Cref{subsec:dynamic-pattern-emulator}, we show how to adapt the
        algorithm from \Cref{sec:static} to the dynamic setting, thereby proving
        \Cref{thm:dynamic-pattern}.
        Our algorithm follows the same framework as that of
        \cite{Filtser:2024aa}; hence, in this subsection we mainly summarize their
        approach and establish the additional guarantees needed for our application.

    \item
        Finally, in \Cref{subsec:all-pairs}, we prove \Cref{thm:Intromain} by
        combining all the ingredients developed throughout the paper.
\end{itemize}

\subsection{Maintaining The Dual Instance} \label{subsec:dual}
In this subsection, we show how an update to the instance $(G,T)$ can be translated
into $O(1)$ updates to an associated dual instance. This is summarized in the
following lemma.

\begin{lemma} \label{lem:dynamic-instance}
    Let $G$ be an $n$-vertex planar graph, $T \subseteq V(G)$ be a
    terminal set, and $(G_D,T_D)$ be the dual instance of $(G,T)$.
    Then, there exists a dynamic instance $(G_D^*,T_D^*)$ with the following
    properties.
    \begin{enumerate}
        \item
            There is a bijection between the terminals in $T_D$ and the terminals
            in $T_D^*$.

        \item
            For every pair of dual terminals $t_1,t_2 \in T_D$, if
            $t_1^*,t_2^* \in T_D^*$ denote their corresponding terminals, then
            \(
                \operatorname{dist}_{G_D}(t_1,t_2)
                =
                \operatorname{dist}_{G_D^*}(t_1^*,t_2^*) .
            \)

        \item
            The size of $(G_D^*,T_D^*)$ is $O(n)$.

        \item
            After each update to $(G,T)$, the instance $(G_D^*,T_D^*)$ can be
            maintained using $O(1)$ elementary updates: terminal additions or removals, and edge insertions or deletions.
    \end{enumerate}
\end{lemma}

We prove \Cref{lem:dynamic-instance} at the end of this subsection.
The construction follows a strategy similar to that of \cite{Italiano:2010aa}: instead of
maintaining the dual instance $(G_D,T_D)$ directly, we maintain a modified
instance $(G_D^*,T_D^*)$ in which every vertex of $G_D$ is replaced by a
zero-length cycle. The construction is described formally in
\Cref{def:dynamic-instance}.

We emphasize that the data structure of \cite{Italiano:2010aa} is designed for
maintaining the standard planar dual $G^*$. In our setting, the dual instance also
depends on the choice of terminals; see \Cref{sec:prelim}. Thus, we need to
extend the same idea to dual instances, while preserving distances between dual
terminals.

Recall that the dual instance $(G_D,T_D)$ is obtained from $(G,T)$ as follows.
First, for every hole $h \in \mathcal{H}(G,T)$, we add an auxiliary vertex $x_h$.
Second, we connect $x_h$ to every terminal on $h$, obtaining the graph
$\overbar{G}$. This subdivides the hole $h$ into faces that are in one-to-one
correspondence with the terminals on $h$; in particular, a terminal $t$ on $h$
corresponds to a face $f_t$ of $\overbar{G}$. Third, we take the standard dual
$\Gbar^*$ of $\overbar{G}$. Finally, we delete from $\Gbar^*$ every dual edge
$f_t f_{t'}$ that corresponds to an auxiliary edge incident to some auxiliary
vertex $x_h$. See \Cref{fig:dual} for an illustration.

\begin{figure}[t]
    \centering
    \setkeys{Gin}{width=\linewidth}
    \begin{subfigure}{0.32\textwidth}
        \centering
        \includegraphics[page=3,width=\textwidth]{dual-instance.pdf}
        \caption{}
    \end{subfigure}
    \hfil
    \begin{subfigure}{0.32\textwidth}
        \centering
        \includegraphics[page=5,width=\textwidth]{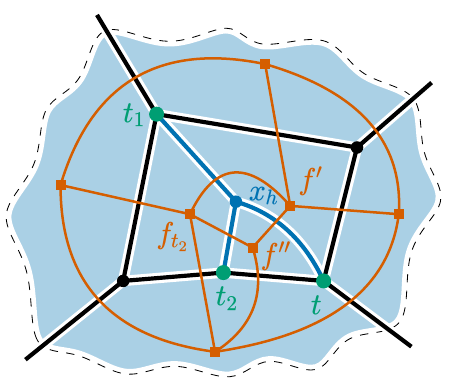}
        \caption{}
    \end{subfigure}

    \vspace{0.5cm}

    \begin{subfigure}{0.32\textwidth}
        \centering
        \includegraphics[page=6,width=\textwidth]{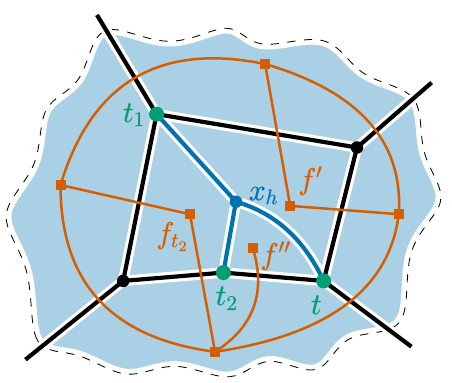}
        \caption{}
        \label{fig:G-D-after}
    \end{subfigure}
    \hfil
    \begin{subfigure}{0.32\textwidth}
        \centering
        \includegraphics[page=4,width=\textwidth]{dual-instance.pdf}
        \caption{}
    \end{subfigure}
    \caption{
        An example based on \Cref{fig:dual}, illustrating how $(G_D,T_D)$ changes
        after adding a terminal $t$ on the hole $h$.
        (a) The standard dual of $\overbar{G}$ before the update.
        (b) The terminal $t$ is added on $h$. In $\overbar{G}$, this corresponds
        to adding the edge $x_h t$, which splits the face $f_{t_1}$ into two faces
        $f'$ and $f''$.
        (c) The updated dual instance $(G_D,T_D)$ is obtained after deleting the
        dual edges corresponding to auxiliary edges.
        (d) The dual instance $(G_D,T_D)$ of \Cref{fig:G-D} before the terminal
        addition, shown for comparison.
    }
    \label{fig:dual-update}
\end{figure}

We next describe the main difficulty in maintaining $(G_D,T_D)$ dynamically.
Suppose that a new terminal $t$ is added on an already existing hole $h$.
Let $t_1$ and $t_2$ be the consecutive terminals immediately before and after $t$
on $h$ in the chosen cyclic order (note that $t_1$ and $t_2$ may coincide). The effect
of this update on the dual instance can be described as follows.
\begin{enumerate}
    \item
        Since the hole $h$ already existed before the update, the auxiliary vertex
        $x_h$ is already present and is connected to the terminals on $h$, including the terminal(s)
        $t_1$ and $t_2$. Let $f_{t_1}$ be the face of $\overbar{G}$ corresponding
        to $t_1$.

    \item
        After $t$ is added as a terminal, the edge $x_h t$ is added to
        $\overbar{G}$. This splits the face $f_{t_1}$ into two faces, say $f'$
        and $f''$. The face $f'$ is bounded by the edges $x_h t_1$ and $x_h t$,
        while $f''$ is bounded by the edges $x_h t$ and $x_h t_2$. We let $f'$
        correspond to the old terminal $t_1$, and let $f''$ correspond to the new
        terminal $t$. See \Cref{fig:dual-update}.

    \item
        Passing to the dual, this face split in $\overbar{G}$ becomes a vertex
        split in the dual graph. All dual edges whose endpoints are different from
        $f_{t_1}$ remain unchanged. Each dual edge incident to $f_{t_1}$ before
        the update becomes incident to exactly one of $f'$ and $f''$ after the
        update. Moreover, because the split is due to a face split in the
        embedded primal graph, the cyclic order of the incident dual edges is
        preserved: the edges assigned to one of the two new vertices form a
        contiguous interval in the embedding of $G_D$.
\end{enumerate}

The obstacle is that the vertex $f_{t_1}$ may have $\Omega(n)$ incident edges.
Thus, explicitly performing the vertex split in $G_D$ may require changing
$\Omega(n)$ edge endpoints, which is too expensive for a dynamic algorithm.
To avoid this, we maintain a modified instance $(G_D^*,T_D^*)$ that preserves all
distances between dual terminals, while allowing such vertex splits and merges to
be implemented by only $O(1)$ elementary updates.

We now define this modified instance.

\begin{definition}[Modified Dynamic instance $(G_D^*,T_D^*)$ based on $(G,T)$]
\label{def:dynamic-instance}
    Let $G$ be a planar graph, $T \subseteq V(G)$ be a terminal set, and $(G_D,T_D)$ be the
    dual instance of $(G,T)$. The dynamic instance $(G_D^*,T_D^*)$ is obtained
    from $(G_D,T_D)$ as follows.
    \begin{itemize}
        \item 
            For every vertex $f \in V(G_D)$, let
            \(
                e_1,e_2,\ldots,e_{\deg_{G_D}(f)}
            \)
            be the edges incident to $f$ in their cyclic order in the embedding of $G_D$.
            We replace $f$ by a cycle $C_f$ with one vertex for each incident edge of $f$.
            The edges touching vertices in $C_f$ are added as follows.
            \begin{itemize}
                \item Consecutive vertices on $C_f$ are connected by edges of length $0$.
                \item Each original edge $e_i$ incident to $f$ is attached to the corresponding vertex of $C_f$, and these attachments preserve the cyclic order of the incident edges around $f$.
            \end{itemize}
            
            \item 
            If $f \in T_D$ is a dual terminal, then we choose an arbitrary vertex of $C_f$ to be the corresponding terminal in $T_D^*$.
    \end{itemize}
\end{definition}

The zero-length cycles ensure that terminal distances are unchanged, as shown in the following claim.

\begin{claim} \label{clm:dynamic-instance}
    Let $(G_D^*,T_D^*)$ be the dynamic instance based on $(G,T)$ as in
    \Cref{def:dynamic-instance}. For every pair of terminals
    $t_1,t_2 \in T_D$, if $t_1^*,t_2^* \in T_D^*$ denote their corresponding
    terminals, then
    \[
        \operatorname{dist}_{G_D}(t_1,t_2)
        =
        \operatorname{dist}_{G_D^*}(t_1^*,t_2^*) .
    \]
\end{claim}

\begin{proof}
    We prove both inequalities.

    First, let $P$ be a $t_1$-$t_2$ path in $G_D$. We construct a corresponding
    $t_1^*$-$t_2^*$ path $P^*$ in $G_D^*$ as follows. Each edge of $P$ is present
    in $G_D^*$ with the same length, except that its endpoints are attached to
    vertices on the corresponding replacement cycles. Whenever $P$ enters and
    leaves a vertex $f$ of $G_D$, we connect the corresponding attachment points
    on $C_f$ by a path along $C_f$. 
    At the first and last vertices, we similarly connect the terminal representative
    to the relevant attachment point along the corresponding zero-length cycle.
    Since all edges of $C_f$ have length $0$,
    this does not increase the length. Thus
    \[
        \operatorname{dist}_{G_D^*}(t_1^*,t_2^*)
        \le
        \operatorname{dist}_{G_D}(t_1,t_2).
    \]

    Conversely, let $P^*$ be a $t_1^*$-$t_2^*$ path in $G_D^*$. Contracting every
    cycle $C_f$ to the original vertex $f$ maps $P^*$ to a $t_1$-$t_2$ walk in
    $G_D$ of the same length, since all contracted edges have length $0$ and all
    original dual edges retain their lengths. Hence
    \[
        \operatorname{dist}_{G_D}(t_1,t_2)
        \le
        \operatorname{dist}_{G_D^*}(t_1^*,t_2^*).
    \qedhere
    \]
\end{proof}

It remains to describe how $(G_D^*,T_D^*)$ is updated after an update to
$(G,T)$.

\paragraph{Edge insertion and terminal addition.}
As discussed above, adding a terminal $t$ on an existing hole $h$ induces a
vertex split in $(G_D,T_D)$, together with the creation of a new dual terminal.
If $t$ is the first terminal on $h$, then the edge $x_h t$ does not split an
existing face of $\overbar{G}$; in this case, $G_D$ remains unchanged, and we
only designate the corresponding vertex as a terminal.

Similarly, inserting an edge in $G$ splits a face of $G$, assuming that $G$
was connected before the insertion. In the dual, this results in a vertex split and the addition of the dual edge
corresponding to the inserted primal edge.

\begin{figure}[t]
    \centering
    \setkeys{Gin}{width=\linewidth}
    \begin{subfigure}{0.32\textwidth}
        \centering
        \includegraphics[page=7,width=\textwidth]{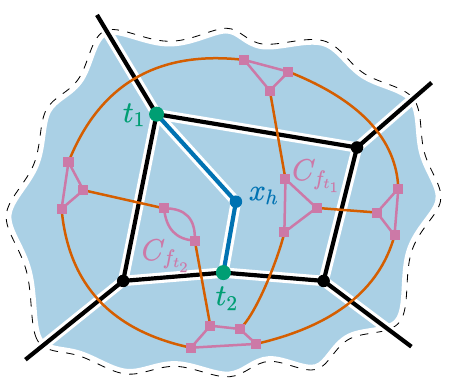}
        \caption{}
    \end{subfigure}
    \hfil
    \begin{subfigure}{0.32\textwidth}
        \centering
        \includegraphics[page=8,width=\textwidth]{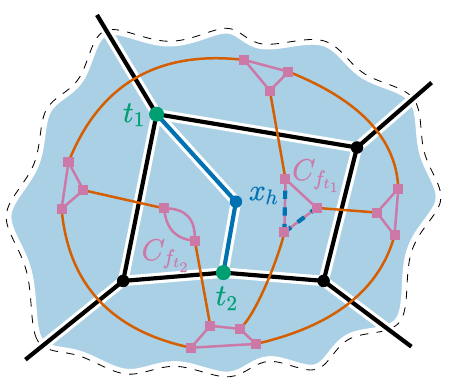}
        \caption{}
    \end{subfigure}

    \vspace{0.5cm}

    \begin{subfigure}{0.32\textwidth}
        \centering
        \includegraphics[page=9,width=\textwidth]{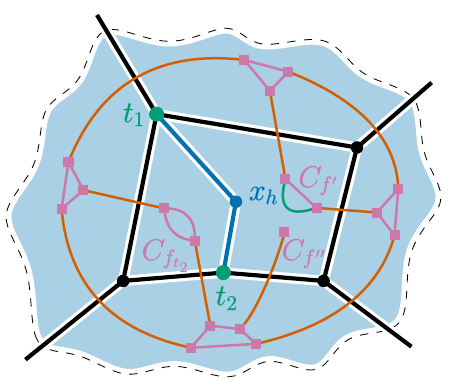}
        \caption{}
    \end{subfigure}
    \hfil
    \begin{subfigure}{0.32\textwidth}
        \centering
        \includegraphics[page=6,width=\textwidth]{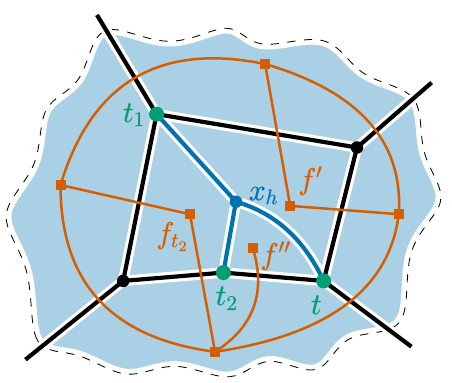}
        \caption{}
    \end{subfigure}
    \caption{
        Updating $(G_D^*,T_D^*)$ after adding the terminal $t$ on the hole $h$, as
        in \Cref{fig:dual-update}.
        (a) The instance $(G_D^*,T_D^*)$ before the update.
        (b) To obtain the cycles $C_{f'}$ and $C_{f''}$ from $C_{f_{t_1}}$, we
        delete two zero-length edges of $C_{f_{t_1}}$ so that the resulting paths
        correspond to the two sides of the split.
        (c) We then add zero-length edges between the endpoints of the resulting
        paths, thereby forming the cycles $C_{f'}$ and $C_{f''}$.
        (d) The corresponding dual instance $(G_D,T_D)$ after the terminal
        addition, shown for comparison.
    }
    \label{fig:dual-cycle-update}
\end{figure}

In both cases, the vertex split respects the embedding: the incident edges
assigned to one side of the split form a contiguous interval in the cyclic order
around the split vertex. We can therefore implement the split in $G_D^*$ as
follows. Suppose the vertex $f$ of $G_D$ is split into vertices $f'$ and $f''$.
The cycle $C_f$ is split into two cycles $C_{f'}$ and $C_{f''}$ by performing the
following operations.
\begin{enumerate}
    \item
        Delete two zero-length edges of $C_f$ so that the two resulting paths
        correspond to the two intervals of incident edges assigned to $f'$ and
        $f''$, respectively.

    \item
        Add zero-length edges between the endpoints of these paths, thereby
        closing them into the cycles $C_{f'}$ and $C_{f''}$.
\end{enumerate}
If the update is an edge insertion in $G$, we also add the corresponding dual
edge between the appropriate vertices of $C_{f'}$ and $C_{f''}$. If the update
creates a new dual terminal, we designate an arbitrary vertex of the corresponding 
cycle as its representative in $T_D^*$. Overall, this requires only $O(1)$
elementary updates. See \Cref{fig:dual-cycle-update} for an illustration.

\paragraph{Edge deletion and terminal removal.}
These operations are the inverses of edge insertion and terminal addition.
Deleting an edge of $G$ merges the two faces incident to that edge, assuming that
$G$ remains connected after the deletion. In the dual, this corresponds to
deleting the dual edge and merging the two incident dual vertices. Similarly,
removing a terminal $t$ from a hole $h$ removes the auxiliary edge $x_h t$ from
$\overbar{G}$, which merges the two faces incident to $x_h t$, if such faces are distinct.

The merge also respects the embedding, and hence it is the inverse of the
embedding-respecting split described above. 
Thus, to merge two cycles $C_{f'}$ and $C_{f''}$ into a single cycle $C_f$, we perform the following operations.
\begin{enumerate}
    \item Delete the zero-length edges that closed the two cycles.
    \item Add zero-length edges that concatenate the two paths into one cycle.
\end{enumerate}
We also remove the corresponding dual edge if the update is an edge deletion in $G$, and update the terminal representatives if the update is a terminal removal.
Again, only $O(1)$ elementary updates are required.

\bigskip

We now prove \Cref{lem:dynamic-instance}.

\begin{proof}[Proof of \Cref{lem:dynamic-instance}]
    We verify the four properties in order.

    \begin{enumerate}
        \item
            By construction, every terminal $t \in T_D$ is represented by exactly
            one designated vertex $t^* \in T_D^*$ on the cycle replacing $t$.
            Hence the correspondence is a bijection.

        \item
            This is exactly \Cref{clm:dynamic-instance}.

        \item
            Each vertex $f \in V(G_D)$ is replaced by a cycle whose size is
            proportional to $\deg_{G_D}(f)$. Therefore,
            \[
                |V(G_D^*)| + |E(G_D^*)|
                =
                O\left(\sum_{f \in V(G_D)} \deg_{G_D}(f)\right)
                =
                O(|E(G_D)|).
            \]
            Since $G_D$ is planar and has size $O(n)$, it follows that
            $(G_D^*,T_D^*)$ has size $O(n)$.

        \item
            By the discussion above, each update to $(G,T)$ induces either no
            change to the underlying dual graph, or one split
            or merge of a dual vertex, together with $O(1)$ changes to dual edges
            and terminal representatives. In $(G_D^*,T_D^*)$, such a split or
            merge is implemented by deleting and inserting $O(1)$ zero-length
            cycle edges, and by performing $O(1)$ additional edge or terminal
            updates. Hence $(G_D^*,T_D^*)$ can be maintained using $O(1)$
            elementary updates after each update to $(G,T)$.
        \qedhere
    \end{enumerate}
\end{proof}

\subsection{Dynamic Pattern Emulator: Proof of Theorem~\ref{thm:dynamic-pattern}} \label{subsec:dynamic-pattern-emulator}
In this subsection, we explain how to dynamically maintain the pattern emulator of \Cref{sec:static} under edge insertions and deletions, and up to two terminal additions or removals to establish \Cref{thm:dynamic-pattern} stated below.

\dynamicpattern*

The dynamic algorithm builds upon the algorithm of \cite{Filtser:2024aa}, and we refer to their proofs where necessary.
We begin by describing the initialization process in \Cref{subsubsec:initialization}.
Next, we outline how the data structure processes updates in \Cref{subsubsec:updates}.
Finally, we show how to adapt the algorithm to achieve the guarantees of \Cref{thm:dynamic-pattern} at the end of \Cref{subsubsec:updates}.
\Cref{thm:dynamic-pattern} is proved at the end of the subsection.

Before moving on to the initialization, we first revisit and outline the components that will be used in this subsection.

\exactemulator*
\structured*
\structuredrdivision*
\separator*
\portals*
\portalsplit*
\portalsall*
\defsliced*

\subsubsection{Initialization} \label{subsubsec:initialization}
The data structure initially partitions $(G,T)$ into sufficiently small pieces and then computes an emulator for each piece. 
The construction of these pieces involves three main steps: reducing the size of the instances, reducing the number of holes, and finally reducing the size of the terminal set to achieve the desired piece size. 
To differentiate these steps, the algorithm classifies the instances formed during this process into four types, as follows.

\begin{itemize}
    \item \underline{\four Instance:} 
        An instance $(R,S)$ is classified as \four if $|R| > n / \tau$ and it contains $O(1)$ holes.
    \item \underline{\three Instance:} 
        An instance $(R,S)$ is classified as \three if $|R| = O(n / \tau)$ and it contains $O(1)$ holes.
    \item \underline{\two Instance:}
        An instance $(R,S)$ is classified as \two if $|R| = O(n / \tau)$ and it is a one-hole instance.
    \item \underline{\one Instance:}
        An instance $(R,S)$ is classified as \one if $|R| = O(n / \tau)$, it is a one-hole instance, and $|S| \leq \log^{10} n$.
\end{itemize}

The main intuition behind this classification is to start with the \four instance $(G,\emptyset)$\footnote{As we will see in \Cref{subsec:all-pairs}, at initialization, graph $G$ does not have any terminals. Thus, $(G, T) = (G, \emptyset)$ represents the instance at initialization. Upon an $s$-$t$ min-cut query, we add $s$ and $t$ to $T$. After answering the query, we revert these additions to restore the state $(G, \emptyset)$.} and sequentially refine it: first obtaining pieces of \three, then using each \three piece to derive \two pieces, and finally using each \two piece to derive \one pieces. 
The \one pieces are the granular pieces that we will use to construct our emulator.

Obtaining \one pieces for $(G,\emptyset)$ naturally induces a tree $\Tcal$, referred to as the {\it decomposition tree}.
Each node in $\Tcal$ corresponds to an instance $(R,S)$.\footnote{We use ``node'' and ``instance'' interchangeably to refer to the node in the decomposition tree $\Tcal$ corresponding to an instance $(R,S)$.}
Similar to \Cref{sec:static}, to approximately preserve distances while constructing the instances, we use a portal set $\portal(R,S)$ for each instance in the tree. 
The portal set is designed to handle the distance changes caused by updates, where portal additions are propagated to the children of $(R,S)$ in $\Tcal$, as will be discussed later in this subsection.

We now describe the construction of the decomposition tree $\Tcal$.

\begin{itemize}
    \item 
        The root of the tree is the \four instance $(G, \emptyset)$.

    \item 
        For every \four instance $(R,S)$, we use \Cref{thm:r-division} to obtain a structured $r$-division of $(R,S)$ with pieces $(R_1,S_1), \dots, (R_k,S_k)$ of $(R,S)$.
        Each piece $(R_i,S_i)$ becomes a child of $(R,S)$ in $\Tcal$. 
        We set the parameter $r$ such that $|R_i| \leq 3/4|R|$ for $1 \leq i \leq k$, and $\sum_{i=1}^k |S_i| = O(\sqrt{|R|})$. 
        Furthermore, by \Cref{thm:r-division}, each instance $(R_i,S_i)$ has $3/4 |\Hcal(R,S)| + 4 \leq 16 = O(1)$ holes.

        For every \four instance $(R,S)$, we set $\portal(R,S) = \left(\bigcup_{i=1}^k S_i\right) \setminus S$. 
        Note that the terminal set $S_i$ is provided by \Cref{thm:r-division} and consists of the boundary vertices of $R_i$ as well as the set $S \cap V(R_i)$, which includes terminals of $(R,S)$ that are present in $(R_i,S_i)$.

    \item 
        For every \three instance $(R,S)$, we reduce the number of holes by one via the slicing procedure (see \Cref{alg:slice}). 
        We select the shortest path $\pi_{uv}$ between the closest holes in $\Hcal(R,S)$, with endpoints $u$ and $v$. 
        After slicing through $\pi_{uv}$, the resulting instance $(R',S')$ is set as the child of $(R,S)$ in $\Tcal$. 
        We denote by $\overbar \pi_{uv}$ the copy of $\pi_{uv}$ used in the slicing procedure. 
        
        For every child $(R',S')$ of a \three instance $(R,S)$, we set $\portal(R',S') = \portal(R,S) \cup \sliceportal(\pi_{uv}, S, \eps) \cup \overbar{\sliceportal}(\pi_{uv}, S, \eps)$, as defined in \Cref{def:portals-all}.\footnote{For simplicity, we adhere to the error parameter $\eps$ throughout this brief summary. Achieving a $(1+\eps)$-approximation for the entire construction is more intricate and requires defining a more granular error parameter, as discussed in \cite{Filtser:2024aa}.}
        Additionally, we set the terminal set $S' = S \cup \sliceportal(\pi_{uv}, S, \eps) \cup \overbar{\sliceportal}(\pi_{uv}, S, \eps)$.

        See \Cref{alg:reduce-hole} for a summary of this procedure.

        \begin{algorithm2e}[]
            \KwIn{an instance $(R, S)$ with at least two holes and error parameter $0 < \varepsilon < 1$}
            \KwOut{an instance $(R', S')$ with one fewer hole than $(R, S)$}
            \BlankLine
            let $h$ and $h'$ be the two closest holes in $(R, S)$\;
            let $\pi_{uv}$ be the shortest path between $h$ and $h'$, where $u \in V(h)$ and $v \in V(h')$\;    
            $\portal(R,S) \gets \portal(R,S) \cup \sliceportal(\pi_{uv}, S, \varepsilon)$\;
            set $R'$ as the graph obtained by slicing ${R}$ along $\pi_{uv}$ using \Cref{alg:slice}\;    
            set $S'$ as the union of $S$, $\sliceportal(\pi_{uv}, S, \eps)$, and their copies $\overbar{\sliceportal(\pi_{uv}, S, \varepsilon)}$ on $\overbar \pi_{uv}$ (as defined in \Cref{def:sliced-instance})\;
            \Return{$(R',S')$}
            \caption{Reduce-Hole($(R,S),\varepsilon$)}
            \label{alg:reduce-hole}
        \end{algorithm2e}

    \item 
        For every \two instance $(R,S)$, we split $(R,S)$ into two instances $(R_1,S_1)$ and $(R_2,S_2)$ such that $|S_i \cap S| \leq 11/12 |S|$ for $i=1,2$. 
        This is accomplished by finding a shortest path $\pi_{uv}$ between the vertices $u$ and $v$ on the single hole in $(R,S)$, using the shortest path separator from \Cref{thm:separator}. 
        Consequently, $(R_1,S_1)$ and $(R_2,S_2)$ still have a single hole, and they are set as the children of $(R,S)$ in $\Tcal$. 

        For every \two instance $(R,S)$, similar to \three instances, we obtain the portal set on the path $\pi_{uv}$.
        We use \Cref{lem:portals-split} to obtain the set $\splitportal(\pi_{uv}, t, \eps)$.
        We set $\portal(R,S) = \cup_{t \in S} \splitportal(\pi_{uv}, t, \eps)$. 
        Since $(R_1,S_1)$ and $(R_2,S_2)$ split $(R,S)$, we set $S_i = \big(S \cap V(R_i)\big) \cup \portal(R,S)$ for $i=1,2$.
        In case $e=uv$ is a bridge edge, then we only add $u$ and $v$ to the portal set as the removal of $\pi_{uv} = e$ disconnects the graph into $(R_1,S_1)$ and $(R_2,S_2)$.

        See \Cref{alg:reduce-terminal} for a summary of this procedure.

        \begin{algorithm2e}[t]
            \KwIn{a one-hole instance $(R,S)$ and error parameter $0 < \varepsilon < 1$}
            \KwOut{two one-hole instances $(R_1,S_1)$ and $(R_2,S_2)$ with $|S_i \cap S| \leq \frac{11}{12}|S|$ for $i=1,2$}
            \BlankLine
            let $u, v$ be the vertices computed by \Cref{thm:separator}, and let $\pi_{uv}$ be the shortest $u$-$v$ path\; 
            \uIf{$\pi_{uv}$ does not contain a bridge edge}{
                $\portal(R,S) \gets \portal(R,S) \cup \splitportal(\pi_{uv}, S, \varepsilon)$\;
                \tcc*[]{the portal set $\splitportal(\cdot,\cdot,\cdot)$ is from \Cref{lem:portals-split}}

                let $R_1, R_2$ be the (non-empty) regions of $R$ enclosed by the (non-simple) cycles formed by $\pi_{uv}$ and its hole    
            }
            \Else{
                $\portal(R,S) \gets \portal(R,S) \cup \{u,v\}$\;
                let $C$ and $C'$ be the connected components of $R \setminus \{e\}$\;
                \tcc*[]{$e=uv$ is a bridge edge by \Cref{thm:separator}}
                let $R_1 \gets R[C] \cup \{e\}$ and $R_2 \gets R[C']$\;
            }
            set $S_i \gets (S \cap V(R_i)) \cup \portal(R,S)$ for $i=1,2$\;
            \Return{$(R_1, S_1)$ and $(R_2, S_2)$}
            \caption{Reduce-Terminal($(R,S),\varepsilon$)}
            \label{alg:reduce-terminal}
        \end{algorithm2e}
\end{itemize}

Using the construction above, it is not hard to see that the type of the instances on any root-to-leaf path in $\Tcal$ monotonically decreases. 
Moreover, the depth of $\Tcal$ is $\Otil(1)$ \cite[Lemma 6.16]{Filtser:2024aa}.

See \Cref{alg:initialize-instance} for a summary of the initialization procedure.

\begin{algorithm2e}[]
    \KwIn{an instance $(R,S)$, target parameter $\tau$, and error parameter $0 < \varepsilon < 1$}
    \KwOut{a decomposition tree $\mathcal T$ rooted at $(R,S)$} 
    \BlankLine
    $\portal(R,S) \gets S$. \tcc*[f]{initialize the portal set}

    \If(\tcc*[f]{handle the degenerate case}){$R$ is not connected}{
        let $(R_p, S_p)$ be the parent of $(R,S)$\;
        remove $(R, S)$ from $\mathcal{T}$\;
        
        \ForEach{connected component $C$ in $R$}{
            add instance $(C, S \cap C)$ as a child of $(R_p,S_p)$ to $\mathcal{T}$\;
            \textsc{Initialize-Instance}($(C, S \cap C), \tau, \varepsilon)$
        }
    \Return
    }
    \uIf{$(R,S)$ is of \texttt{Type-1}}{
        \Return
    }
    \uElseIf{the parent of $(R,S)$ is of \texttt{Type-2} or \texttt{Type-3} and $(R,S)$ is a one-hole instance}{
        let $(R_1,S_1)$ and $(R_2,S_2)$ be the instances returned by \textsc{Reduce-Terminal($(R,S),\varepsilon$)}\;
        add instances $(R_1,S_1)$ and $(R_2,S_2)$ as children of $(R,S)$ to $\mathcal{T}$
    }
    \uElseIf{the parent of $(R,S)$ is of \texttt{Type-3} or \four with $|R|=O(n/\tau)$ and $(R,S)$ is not a one-hole instance}{
        let $(R',S')$ be the instance returned by \textsc{Reduce-Hole($(R,S),\varepsilon$)}\;
        add instance $(R', S')$ as the child of $(R, S)$ to $\mathcal T$ 
    }
    \Else(\tcc*[f]{$(R,S)$ is of \texttt{Type-4}}){
        Use \Cref{thm:r-division} on $(R,S)$ to obtain the subgraphs $(R_1,S_1), \dots, (R_k,S_k)$\;
        
        \ForEach{$i \in [1, k]$}{
            Add instance $(R_i, (S \cup \partial_{{R}} R_i) \cap V(R_i))$ as a child of $(R, S)$ to $\mathcal T$
        }
    }
    
    \ForEach{child $(R', S')$ of $(R, S)$ in $\mathcal{T}$}{
        $\textsc{Initialize-Instance}((R', S'), \tau, \varepsilon)$.
    }
    
    \caption{Initialize-Instance($(R, S), \tau, \varepsilon$)}
    \label{alg:initialize-instance}
\end{algorithm2e}

To initialize $\Tcal$, we simply execute the algorithm on $(G, \emptyset)$, as described in \Cref{alg:initMainAlgo}.

\begin{algorithm2e}[]
    \KwIn{a planar graph $G$, target parameter $\tau$, and error parameter $0 < \varepsilon < 1$}
    \KwOut{a decomposition tree $\mathcal{T}$ with the instance $(G, \emptyset)$ as its root}
    \BlankLine
    Create a decomposition tree $\mathcal{T}$ consisting solely of a root node $({G}, \emptyset)$\;
    $\textsc{Initialize-Instance}(({G}, \emptyset), \tau, \varepsilon)$.
    \caption{Initialize($G, \tau, \eps$)}
    \label{alg:initMainAlgo}
\end{algorithm2e}

We can now construct the emulator for $(G, \emptyset)$ in a bottom-up fashion using $\Tcal$, as follows.

\begin{itemize}
    \item \underline{\one Emulator:}
        For every \one instance $(R, S)$, we compute an exact emulator $(H, S)$ using \Cref{thm:emulator-exact}.
        This results in an emulator of size $|H| = O(|S|^4) = \poly(\log n)$ and takes $O(|V(R)| \cdot |S|) = \Otil(n / \tau)$ time.
    \item \underline{\two Emulator:}
        For every \two instance $(R, S)$ with children $(R_1, S_1)$ and $(R_2, S_2)$, we construct the emulator $(H, S)$ for $(R, S)$ using the emulators $(H_1, S_1)$ and $(H_2, S_2)$ of its children. 
        Recall that these children result from splitting $(R, S)$.
        Thus, $(H, S)$ is defined as the union of $(H_1, S_1)$ and $(H_2, S_2)$, where each portal vertex in $S_1$ on $\pi_{uv}$ is glued to its copy in $S_2$.
        By straightforward induction, it follows that the emulator has a size of $|H| = |S|/\eps^{O(1)} \cdot \poly(\log n)$.
    \item \underline{\three Emulator:}
        For every \three instance $(R, S)$ with child $(R', S')$, we construct the emulator $(H, S)$ for $(R, S)$ using the emulator $(H', S')$ of $(R', S')$. 
        Specifically, $(H, S)$ is obtained from $(H', S')$ by gluing back every portal on $\pi_{uv}$ with its copy on $\overbar \pi_{uv}$.
        Since this process merges some vertices to form $(H, S)$, it follows that $|H| = |S|/\eps^{O(1)} \cdot \poly(\log n)$.
    \item \underline{\four Emulator:}
        For every \four instance $(R, S)$ with children $(R_1, S_1), \dots, (R_k, S_k)$, we define the emulator $(H, S)$ as the union $\bigcup_{i=1}^k (H_i, S_i)$, where $(H_i, S_i)$ is the emulator for $(R_i, S_i)$.
        By the \four property and the guarantees of \Cref{thm:r-division}, it follows that each $H_i$ has size $\Otil(\sqrt{n / \tau})$ and there are $O(\tau)$ such emulators.
        Consequently, $|H| = \Otil(\sqrt{n \tau}/\eps^{O(1)})$.
\end{itemize}

\subsubsection{Handling the Updates} \label{subsubsec:updates}
Updates to $(G, \emptyset)$ consist of edge insertions and deletions in $G$, as well as terminal additions and removals from its terminal set $T$ (recall that at initialization, we have $T = \emptyset$). 
We will explain how the data structure handles each type of update separately.\footnote{It is worth noting that the update procedure described in \cite{Filtser:2024aa} is more intricate; they do not modify the instances in $\Tcal$ directly. Instead, they define the instance $(R^{(t)}, S^{(t)})$ at time $t$ of the update for each node $(R, S)$ in $\Tcal$. For simplicity, we apply the changes directly to $\Tcal$ and refer the reader to \cite{Filtser:2024aa} for a more comprehensive discussion.}

Before proceeding with how updates are handled, we discuss the global rebuilding policy: for every instance $(R, S)$, updates can modify $R$ or its terminal set $S$. 
The data structure recomputes $(R, S)$ if $R$ or $S$ grows significantly relative to their initial size, thereby ensuring that the size of the maintained emulator remains within the desired bounds. 
More precisely,
\begin{itemize}
    \item 
        For a \one instance, whenever an update happens in $(R, S)$, we recompute its emulator.
    
    \item 
        For a \two instance, whenever $R$ undergoes $\widetilde{\Theta} (|S|)$ changes or $\portal(R,S)$ increases by a factor of $\Big(1 + \Theta(1/\log^2 n)\Big)$ compared to its initial size, we reinitialize $(R, S)$.
    
    \item 
        For a \three instance, the reinitialization process is similar to the \two case.
    
    \item 
        For a \four instance, whenever $R$ undergoes $\widetilde{\Theta} (|S|)$ changes, we reinitialize $(R, S)$.
\end{itemize}
This reinitialization removes the subtree of $\Tcal$ rooted at $(R, S)$ and replaces it with a new subtree obtained by executing \Cref{alg:initialize-instance} on $(R, S)$.

\paragraph{Terminal Addition to $(R,S)$.}
Upon the addition of a terminal $t$ to $S$, we propagate this addition naturally to its children as follows.
\begin{itemize}
    \item 
        If $(R,S)$ is a \four instance, then either $t$ is a boundary vertex between its children (and thus already a terminal vertex in them) or it is entirely contained within one child, say $(R_i, S_i)$. 
        In the latter case, we add $t$ to $S_i$.
    \item 
        If $(R,S)$ is a \three instance, we add $t$ to $S$ and include $\sliceportal(\pi_{uv}, t, \eps)$ in $\portal(R,S)$. 
        We then propagate the addition of $t$ to the child $(R', S')$ of $(R, S)$.
    \item 
        If $(R,S)$ is a \two instance, we add $t$ to $S$ and include $\splitportal(\pi_{uv}, t, \eps)$ in $\portal(R,S)$. 
        We then pass the addition of $t$ to every child of $(R, S)$ that contains $t$.
    \item 
        If $(R,S)$ is a \one instance and $t$ is not already in $S$, we add $t$ to $S$ and then recompute the emulator $(H, S)$ of $(R, S)$ using \Cref{thm:emulator-exact}.
\end{itemize}
This procedure affects $\Otil(1)$ instances in $\Tcal$ \cite[Claims 6.30 to 6.32]{Filtser:2024aa}.

Intuitively, the propagation of $t$ through $\Tcal$ ensures that the emulator preserves the distances between the newly added terminal $t$ and any other terminal. 
This is achieved by adding sufficient portals on the sliced paths of \two and \three instances, and recomputing the emulator for the \one instance containing $t$.

\paragraph{Terminal Removal from $(R,S)$.}
Upon the removal of a terminal $t$ from $(R, S)$, we simply remove $t$ from $S$ and propagate the deletion of $t$ to the children of $(R, S)$ that contain it. 
Note that this update does not change $\portal(R, S)$; however, once the number of portals increases significantly relative to its initial size, the rebuilding policy triggers a recomputation of $(R, S)$, thereby removing stale portals in a single rebuilding step.

\paragraph{Edge Deletion from $(R,S)$.}
The deletion of an edge $e$ from $(R,S)$ is propagated from $(R,S)$ to its children. 
Note that if $(R,S)$ is a \three or \two instance, the edge $e$ may have two copies in the children of $(R,S)$ if it was part of the split path $\pi_{uv}$. 
In such cases, we remove all copies of $e$ from the children and propagate these deletions to deeper levels of $\Tcal$. 

If $e$ is part of a hole, then its removal updates the hole to include the edges of the other face containing $e$ prior to its deletion.

This process affects $\Otil(1)$ instances in $\Tcal$ \cite[Lemma 6.22]{Filtser:2024aa}.

\paragraph{Edge Insertion into $(R,S)$.}
Upon the insertion of an edge $e = xy$ into $(R,S)$, the data structure adds $e$ to $R$ if the endpoints of $e$ lie on a face that is not a hole in $(R,S)$. 
Otherwise, the data structure treats the endpoints $x$ and $y$ as terminals and processes their addition to $(R,S)$, as discussed earlier. 
This transition from edge insertion to terminal addition ensures that if $e$ is not added directly to $(R,S)$, the distance information from $x$ and $y$ to other terminals is preserved, and directly adding $e$ to the emulator results in the updated emulator.\footnote{In the case of terminal addition, the insertion of $e$ is applied at higher levels of $\Tcal$. Note that $e$ is always inserted into the root $(G, \emptyset)$ since $G$ has no holes. The insertion propagates through $\Tcal$ until reaching the first instance $(R, S)$ that inserts $e$ while none of its children can insert $e$. In \cite{Filtser:2024aa} terms, this instance $(R, S)$ is called a {\it minimal valid} instance, and the edge $e$ is referred to as a {\it valid} update for $(R, S)$ if it can be inserted into $(R, S)$.}

Similar to edge deletion, this process affects $\Otil(1)$ instances in $\Tcal$ \cite[Lemma 6.22]{Filtser:2024aa}.

In the following theorem, the guarantees of the data structure are summarized.

\begin{theorem}[\protect{\cite[Theorem 4.1]{Filtser:2024aa}}, Rephrased] \label{thm:dynamic-emulator-Filster}
Let $(G,T)$ be a dynamic instance where $G$ is an $n$-vertex planar graph undergoing edge insertions and deletions with weights in $[1, n^{c}]$ for a constant $c$ at all times,\footnote{We note that \cite{Filtser:2024aa} assumes the weights lie in the range $[1, n^{4}]$ to ensure the data structure operates in the small-spread case of \cite{Chang:2022aa}. The same small-spread argument applies to weights in $[1,n^c]$ for any fixed constant $c$. Thus, we can extend the result to $[1, n^{c}]$.} and $T\subseteq V(G)$ is a terminal set of at most two vertices at any time.
Let $1 \leq \tau \leq n / \log n$ be a target parameter and $0 < \varepsilon < 1$ be an error parameter. 

Then, there is a deterministic data structure that explicitly maintains a collection $\mathcal P$ of paths in $G$ that induces a $(1+ \varepsilon)$-approximate planar emulator $(\mathcal M[\mathcal P],T)$ of instance $(G, T)$, such that for each batch $B$ of updates in $(G,T)$ (whether edge insertions or deletions in $G$, or terminal additions or removals in $T)$, 
\begin{itemize}
\item \underline{Emulator Size}: 
    At any time, \(|\mathcal P|= \Otil(\sqrt{n\tau}/ \varepsilon^{O(1)})\), and
\item \underline{Emulator Recourse:}  
    The amortized number of paths added or deleted from $\mathcal{P}$ after after processing a batch update $B$ is $\Otil(|B|/\varepsilon^{O(1)})$, and  
\item \underline{Update Time:} 
    The amortized update time is $\Otil\Big(n|B|/(\varepsilon^{O(1)} \tau)\Big)$.
\end{itemize}
    The data structure has an initialization time of $\Otil(n/\varepsilon^{O(1)})$.
    Moreover, $(\mathcal M[\mathcal P],T)$ is aligned with $(G,T)$.\footnote{We note that the alignment is not explicitly discussed in \cite{Filtser:2024aa}, as their primary focus was solely on maintaining all-pairs shortest paths. However, the alignment of their construction with $(G, T)$ directly follows, since \cite{Filtser:2024aa} dynamically adapts the construction of \cite{Chang:2022aa}, using the same reasoning as in the proof of \Cref{thm:constant-hole}.}
\end{theorem}

\begin{proof}[Proof of \Cref{thm:dynamic-pattern}]
We use the same data structure and the same update procedure as in
\Cref{thm:dynamic-emulator-Filster}.  Thus it remains only to verify that
the emulator maintained by the data structure is not merely a planar
terminal emulator, but a planar pattern emulator.

Let $\Tcal$ be the current decomposition tree.  For a node
$x=(R,S)\in V(\Tcal)$, let $\Mcal_x$ denote the emulator maintained for the
subinstance $(R,S)$ by the bottom-up construction. 
We prove that the graph $\Mcal_x$ is a planar pattern
emulator for $(R,S)$, with respect to the embedding and the terminal order
induced by the construction.

The proof is by induction on the height of $x$ in $\Tcal$.  If $x$ is a leaf,
then $x$ is a \one instance.  For such an instance we use the exact aligned
planar emulator given by \Cref{thm:emulator-exact}.  In particular, the
embedding of the emulator agrees with the embedding of the corresponding
instance on the terminals, and therefore the emulator is a planar pattern
emulator for the leaf instance.

Now consider an internal node $x=(R,S)$, and assume that we maintain emulators for all
children $x_1=(R_1,S_1),\ldots,x_k=(R_k,S_k)$ of $x$.  By the induction
hypothesis, each $\Mcal_{x_i}$ is a planar pattern emulator for
$(R_i,S_i)$.  The child instances are glued together along exactly the
terminals, portals, copied portals, and separator vertices introduced by the
decomposition step that produced them.  Moreover, these vertices occur in the
same cyclic order as in the original embedding: in the $r$-division step this
is inherited from the embedded pieces, in the slicing step it is enforced by
the copied path and copied portals, and in the terminal-reduction step it is
enforced by the separator path and its portals.  Hence the hypotheses of
\Cref{lem:composability-pattern-emulators} apply.  Therefore, replacing each
child instance $(R_i,S_i)$ by its pattern emulator $\Mcal_{x_i}$ and gluing
the resulting graphs according to the same boundary identifications gives a
planar pattern emulator for $(R,S)$.

If the bottom-up construction performs this replacement through an
intermediate emulator, then the preceding paragraph shows that the
intermediate graph is a pattern emulator for $(R,S)$, while the maintained
graph $\Mcal_x$ is a pattern emulator for that intermediate graph.  By
\Cref{lem:transitivity-pattern-emulator}, $\Mcal_x$ is again a pattern
emulator for $(R,S)$.  This completes the induction step.

Applying the invariant to the root of $\Tcal$ shows that the maintained graph
is a planar pattern emulator for the current instance $(G,T)$.  The size,
recourse, and update-time bounds are unchanged from
\Cref{thm:dynamic-emulator-Filster}, since the decomposition tree, the
rebuilding policy, and the set of updated child instances are exactly the
same.  The only additional requirement is that the local emulators used at
\one leaves are chosen to be the aligned planar emulators of
\Cref{thm:emulator-exact}, which have the same asymptotic size and can be
rebuilt within the same update-time bound.  The success probability is also
inherited directly from \Cref{thm:dynamic-emulator-Filster}.
\end{proof}

\subsection{Pattern Emulator Hierarchy: Proof of Theorem~\ref{thm:Intromain}} \label{subsec:all-pairs}

In this subsection, we explain how to obtain \Cref{thm:Intromain}.

\Intromain*

To achieve a dynamic algorithm with subpolynomial update and query time, we adopt a similar approach as outlined in \cite[Section 4]{Filtser:2024aa}. 
Specifically, we use \Cref{thm:dynamic-pattern} from the previous subsection to construct an emulator hierarchy.

\dynamicpattern*

Intuitively, we employ the same idea we used in \Cref{subsec:dynamic-pattern-emulator}: the dynamic pattern emualtor of \Cref{thm:dynamic-pattern} is significantly larger than the target size we are aiming for asking queries on.
We thus use \Cref{thm:dynamic-pattern} recursively until we obtain a pattern emulator with diseired size.
More precisely, we form an $L$-sequence $(1+\overbar\eps)$-approximate pattern emulator, for suitble values of $L$ and $\overbar\eps$, and intialize the data structure of \Cref{thm:dynamic-pattern} on each pattern emulator in the sequence.
We formalize this in the definition followed after restating \Cref{def:sequence}.

\sequence*

\begin{definition}[Emulator Hierarchy of $(G,T)$ with Parameter $\eps$] \label{def:emulator-hierarchy}
    Let $G$ be an $n$-vertex planar graph, $T \subseteq V(G)$ be a terminal set, $0 < \eps < 1$ be an error parameter. 
    Let $L = \log ^{2/3} n$ be the number of levels, $\overbar\eps = \eps/(2L)$ be the approximation parameter, and $\delta = 2^L$ be the size reduction parameter. 
    Let $(G_D,T_D)$ be the dual instance of $(G,T)$ and let $(G_D^*,T_D^*)$ be the dyanmic instance based on $(G_D,T_D)$ as defined in \Cref{subsec:dual}.
    The emulator hierarchy of $(G,T)$ is recursively constructed on an $L$-sequence $(1+\overbar\eps)$-approximate pattern emulators $(H_1^*,T_D^*), \dots, (H_L^*,T_D^*)$ of $(G_D^*,T_D^*)$, as defined in \Cref{def:sequence}, as follows.
    \begin{enumerate}
        \item 
            Let $(\mathcal{M}_{G_D^*}[\mathcal{P}],T^*_D)$ be the $(1+\overbar\eps)$-approximate pattern emulator of $(G_D^*,T_D^*)$ maintianed by initializing \Cref{thm:dynamic-pattern} on $(G_D^*,T_D^*)$ with $\tau_0 = n_0/\delta$, where $n_0 = |G_D^*|$.
            The planar pattern emualor $(H_1^*,T_D^*)$ of $(G_D^*,T^*_D)$ is obtained from $(\mathcal{M}_{H_0^*}[\mathcal{P}],T^*_D)$ by replacing every simple path in $\mathcal{M}_{H_0^*}[\mathcal{P}]$ with an edge of weight equal to the length of its corresponding path.

        \item 
            For $1\leq i\leq L-1$, let $(\mathcal{M}_{H_i^*}[\mathcal{P}],T^*_D)$ be the $(1+\overbar\eps)$-approximate pattern emulator of $(H_i^*,T_D^*)$ maintained by initializing \Cref{thm:dynamic-pattern} on $(H_i^*,T_D^*)$ with $\tau_i = n_i/\delta$, where $n_i = |H_i^*|$.
            The planar pattern emualor $(H_{i+1}^*,T_D^*)$ of $(H_{i}^*,T_D^*)$ is obtained from $(\mathcal{M}_{H_i^*}[\mathcal{P}],T^*_D)$ by replacing every simple path in $(\mathcal{M}_{H_i^*}[\mathcal{P}],T^*_D)$ with an edge of weight equal to the length of its corresponding path.
    \end{enumerate}    
\end{definition}

Note that instead of forming the hierarchy directly on $(G, T)$ as in \cite{Filtser:2024aa}, we construct the hierarchy on the dynamic instance $(G_D^*, T_D^*)$ of $(G, T)$ as defined in \Cref{subsec:dual}, and maintain a planar pattern emulator hierarchy rather than a standard emulator hierarchy.

The remainder of this subsection is structured as follows.
First, we initialize this hierarchy, then discuss how updates are handled, and finally prove the guarantees of \Cref{thm:Intromain} at the end of the subsection.

\paragraph{Initialization.}
We intitialize \Cref{def:emulator-hierarchy}.
This initializes a sequence $(H_1,T_D^*) \dots, (H_L^*,T_D^*)$ of $(1+\overbar\eps)$-approximate pattern emulators, where $L = \log^{2/3} n$ and $\overbar\eps = \eps/(2L)$.
Note that $2^L = \omega(\log n)$, and so for every $1 \leq i \leq L-1$, $\tau_i \leq n_i / \log n_i$, as required by \Cref{thm:dynamic-pattern}.

\paragraph{Handling the updates.}
The updates are propagated through the hierarchy naturally in a top-down approach.
More precisely, upon an update in $G$, we first pass the update to the data structure of the first level, thereby maintaining $(\mathcal{M}_{H_0}[\mathcal{P}],T)$.
We then update $H_1$ based on the new $\mathcal{M}_{H_0}[\mathcal{P}]$, and propagate these updates to the data structure of the second level, continuing this process until we update $(H_L,T)$.

It is important to note that the updates on $G$ consist solely of edge insertions and deletions (and not terminal additions or removals).
Terminal additions and removals are handled dynamically during query processing for $s, t \in V(G)$, as explained below.

\paragraph{Answering the Queries.}
Upon a query on $s, t \in V(G)$, we identify $f_s, f_t$ as the dual vertices in $G_D^*$ corresponding to $s$ and $t$, respectively, and add $f_s, f_t$ to $T_D^*$. 
Subsequently, we update the data structures by passing the addition of $f_s, f_t$ as terminals in a top-down manner, similar to what explained above.

After maintaining $(H_L^*, T_D^*)$ as the pattern emulator of the dynamic instance $(G_D^*,T_D^*)$, we compute the primal $(H_L, T)$ of $(H_L^*, T_D^*)$ using a static algorithm.
Since $T_D^* = \{f_s, f_t\}$, the primal $(H_L, T)$ inherently contains $s$ and $t$.
To compute the minimum $s$-$t$ cut in $(H_L, T)$, we utilize the following theorem.

\begin{theorem}[\cite{Kaplan:2011aa}, Rephrased] \label{thm:static-flow}
Given an $n$-vertex planar graph $G$ and vertices $s, t \in V(G)$, there is an algorithm that computes the $s$-$t$ min-cut in $\Otil(n)$ time.
\end{theorem}

After answering the $s$-$t$ query, we reverse all the changes in the data structure to restore the state of the emulator hierarchy to what it was before the query, ensuring that $T_D^* = \emptyset$.

\bigbreak

We are now ready to prove \Cref{thm:Intromain}.

\begin{proof}[Proof of Theorem~\ref{thm:Intromain}]
    We first derive bounds on the size of the emulators and the size of the update batch they receive as a result of applying the update batch $B$ to $(G_D^*,T_D^*)$.

    Using straightforward induction, it follows that the batch $B_i$ of updates applied to $(H_i^*,T_D^*)$ is of size
    \begin{equation} \label{eqn:hierarchy-recourse}
        |B_i| \leq \frac{|B|}{\overbar{\varepsilon}^{ci}},
    \end{equation}
    for $1 \leq i \leq L$, where $c$ is a sufficiently large constant in the exponent of $\overbar{\varepsilon}$ in \Cref{thm:dynamic-pattern}.

    We now show by induction that, for $0 \leq i \leq L$, the size of $(H_i^*,T_D^*)$ is
    \begin{equation} \label{eqn:hierarchy-size}
        n_i \leq n \cdot \Bigg( \frac{1}{\overbar{\varepsilon}^c \sqrt{\delta}} \Bigg)^i.
    \end{equation}

    For $i = 0$, the size trivially follows.

    Suppose that the size of $(H_i^*,T_D^*)$ is $n_i$ as defined in \Cref{def:emulator-hierarchy}.
    By the construction of $(H_{i+1}^*,T_D^*)$ and \Cref{thm:dynamic-pattern}, it follows that $n_{i+1} \leq \sqrt{n_i \tau_i} / \overbar{\varepsilon}^{c}$.
    Substituting the bound on $n_i$ into this expression results in
    \begin{align*}
        n_{i+1} 
        & \leq \sqrt{n_i \frac{n_i}{\delta}} \cdot \frac{1}{\overbar{\varepsilon}^c} 
        \leq n_i \cdot \Bigg( \frac{1}{\overbar{\varepsilon}^c \sqrt{\delta}} \Bigg) 
        \leq n \cdot \Bigg( \frac{1}{\overbar{\varepsilon}^c \sqrt{\delta}} \Bigg)^{i+1},
    \end{align*}
    as desired.

    We now prove each guarantee separately.
    
    \begin{itemize}
        \item \underline{Correctness:}
            The correctness of maintaining the hierarchy directly follows from \Cref{thm:dynamic-pattern}.
            The fact that $(H_L, T)$ contains the $s$-$t$ min-cut information follows from (1) the construction of $(H_L^*, T_D^*)$, which ensures that the shortest paths between the terminals in $T_D^*$ are approximately maintained in $H_L^*$, and (2) the correspondence of min-cuts in the primal $(H_L, T)$ is the union of paths between terminals in $(H_L^*, T_D^*)$ by \Cref{thm:cut-pattern-emulator}.

        \item \underline{Approximation Quality:}
            Using simple induction, it follows that $(H_i^*, T_D^*)$ approximates $(G^*_D, T^*_D)$ with quality $(1 + \overbar{\varepsilon})^i$ for $1 \leq i \leq L$.
            The quality for the primal $(H_L, T)$ is then given by 
            \[
                (1 + \overbar{\varepsilon})^L = (1 + \varepsilon / (2L))^L \leq e^{\varepsilon / (2L) \cdot L} \leq 1 + \varepsilon.
            \]

        \item \underline{Update and Query Time:}
            By \Cref{eqn:hierarchy-recourse,eqn:hierarchy-size}, the amortized update time for maintaining $H_i$ is 
            \begin{align*}
                \Otil\Bigg(|B_i| \frac{n_i}{\overbar{\varepsilon}^c \tau_i}\Bigg)
                = \Otil\Bigg(|B_i| \delta \Bigg)
                = \Otil\Bigg(\frac{|B|}{\overbar{\varepsilon}^{c(i+1)}} \delta \Bigg)
                = \log^{O(L)} n,
            \end{align*}
            where the last equality follows from the fact that $|B| = \Otil(1)$, $\overbar{\varepsilon} \geq 1 / \poly(\log n)$, and $\delta = 2^L$.

            The amortized update time across all levels then follows as $L = \log^{2/3} n$ and 
            \begin{equation*}
                \Otil(L \log^{O(L)} n) = 2^{\log^{2/3} n \cdot \log\log n} = n^{o(1)}.
            \end{equation*}

            For the query, we process the terminal addition of $s, t$ throughout the hierarchy, which takes $n^{o(1)}$ time as discussed above.
            To compute the primal $(H_L, T)$ of $(H_L^*, T^*_D)$, by \Cref{eqn:hierarchy-size}, we have
            \begin{equation*}
                n_L = n \cdot \Bigg( \frac{1}{\overbar{\varepsilon}^c \sqrt{\delta}} \Bigg)^{L-1} = n \cdot \log^{O(L)} n \cdot \Bigg( \frac{1}{2^{\log^{2/3} n}} \Bigg)^{\frac{\log^{2/3} n - 1}{2}} = n^{o(1)}.
            \end{equation*}
            Thus, computing the primal $(H_L, T)$ and executing \Cref{thm:static-flow} on it takes $n^{o(1)}$ time.

        \item \underline{Initialization Time:}
            By \Cref{thm:dynamic-pattern}, each $(H_i^*, T_D^*)$ is initialized in $\Otil(n_i / \overbar{\varepsilon}^c)$ time.
            Summing over all levels, we have
            \begin{equation*}
                \sum_{i = 1}^L n \cdot \Bigg( \frac{1}{\overbar{\varepsilon}^c \sqrt{\delta}} \Bigg)^i \cdot \frac{1}{\overbar{\varepsilon}^c} = \Otil(L n) = \Otil(n).
            \qedhere
            \end{equation*}
    \end{itemize}
\end{proof}

\section*{Acknowledgements}
\addcontentsline{toc}{section}{Acknowledgements}
ChatGPT 4.4 was used to improve grammar and exposition in parts of the manuscript.
This did not alter the scientific content, results, or conclusions. 
All edits were reviewed by the authors, who take full responsibility for the content.

\section*{References}
\addcontentsline{toc}{section}{References}
\printbibliography[heading=none]

\appendix
\crefalias{section}{appendix}

\section{Proof of Lemma~\ref{lem:one-hole-cut} Using Cut Properties} \label{sec:one-hole-case}

In this section, we outline a more intuitive proof of \Cref{lem:one-hole-cut} for constructing a $(1+\eps)$-approximate terminal cut sparsifier of a one-hole instance $(G,T)$. The proof is based on the cut properties of $(G,T)$, as discussed in \cite[Section 3.2]{Chen:2025aa}, and differs from our proof in \Cref{subsec:one-hole}.

\oneholecut*

The main ingredient in the proof is the following lemma.
Intuitively, the lemma describes how minimum cuts in the instance $(G, T)$ translate to shortest paths in the dual instance $(G_D, T_D)$.

\begin{restatable}[\protect{\cite[Lemma 9]{Chen:2025aa}}, Rephrased]{lemma}{onehole} \label{lem:one-hole-case}
    Let $G$ be a planar graph and $T \subseteq V(G)$ be a set of terminals all lying on a single hole. 
    Let $(G_D, T_D)$ be the dual instance of $(G, T)$. 
    Then, any minimum cut $\widehat E$ in $(G, T)$ that separates terminals $\emptyset \subset R \subset T$ from $T \setminus R$ corresponds to a union of edge-disjoint shortest paths between some dual terminals in $(G_D, T_D)$.
\end{restatable}

Before outlining the proof of \Cref{lem:one-hole-case}, we show how it can be used to construct the desired terminal cut sparsifier of \Cref{lem:one-hole-cut}.
For some error parameter $0 < \varepsilon < 1$, the construction of the $(1+\varepsilon)$-approximate terminal cut sparsifier $(H, T)$ of $(G, T)$ proceeds as follows. 
\begin{enumerate}
    \item Compute the dual instance $(G_D, T_D)$ of $(G,T)$.
    \item Compute a $(1+\varepsilon)$-approximate emulator $(H_D, T_D)$ of $(G_D, T_D)$ using \Cref{thm:emulator}.
    \item Declare the terminal cut sparsifier $(H, T)$ to be the primal instance of $(H_D, T_D)$.
\end{enumerate}
 
The following claim establishes the guarantees of this procedure, thus proving \Cref{lem:one-hole-cut}.

\begin{claim} \label{clm:one-hole-emulator} 
    The primal instance $(H, T)$ obtained from the procedure above is a $(1+\varepsilon)$-approximate terminal cut sparsifier of $(G, T)$. 
    Moreover, $(H, T)$ can be computed in time $\Otil(n / \varepsilon^{O(1)})$ and has size $|H| = \widetilde{O}(|T| / \varepsilon^{O(1)})$.
\end{claim}

\begin{proof}
    We address each guarantee separately.

    \underline{Construction Time:}  
    Since $|G_D| = O(n)$, the time required to compute $(H_D, T_D)$ is $\Otil(n / \varepsilon^{O(1)})$ by \Cref{thm:emulator}. 
    As dual graphs can be computed in $O(n)$ time, the overall construction time for $(H, T)$ follows.

    \underline{Sparsifier Size:}  
    The bound on the size of $H$ follows directly from \Cref{thm:emulator} and the fact that there is a one-to-one correspondence between the terminals in $T$ and $T_D$ due to the construction of the dual instance.

    \underline{Correctness:}  
    Let $\widehat{E}$ be a min-cut in $G$ separating $\emptyset \subset R \subset T$ from $T \setminus R$. 
    By \Cref{lem:one-hole-case}, $\widehat{E}$ corresponds to the union $\widehat{E}_D$ of edges from edge-disjoint shortest paths between some dual terminals in $G_D$. 
    Let $P_D$ denote the set of pairs $(t_1, t_2)$ that define $\widehat{E}_D$, 
    i.e., $\widehat{E}_D$ consists of all edges on the chosen shortest paths in $G_D$ for every $(t_1, t_2) \in P_D$. 

    By the construction of the emulator, for every $(t_1, t_2) \in P_D$, there is a $t_1$-$t_2$ shortest path in $H_D$ whose length is within a $(1 + \varepsilon)$ factor of $\dist_{G_D}(t_1, t_2)$. 
    Let $\overbar{E}_D$ denote the set of edges that belong to the $t_1$-$t_2$ shortest paths in $H_D$ for every $(t_1, t_2) \in P_D$. 

    Since $(t_1, t_2) \in P_D$, $\overbar{E}_D$ consists of edges of the $t_1$-$t_2$ shortest path $P_{12}$ in $H_D$.
    Since $(H_D, T_D)$ is aligned with $(G_D, T_D)$ by \Cref{thm:emulator}, the primal edges in $H$ corresponding to the dual edges of $P_{12}$ separate $R$ from $T \setminus R$ in $H$ (note that the shortest paths constructing $\overbar{E}_D$ in $H_D$ may overlap).  
    Furthermore, the weight of this cut is within a $(1 + \varepsilon)$ factor of the weight of the min-cut separating $R$ from $T \setminus R$ in $G$, as $P_{12}$ has length within a $(1 + \varepsilon)$ factor of the $t_1$-$t_2$ shortest path in $G_D$. 

    The reverse direction (i.e., showing that every min-cut in $H$ is within a $(1 + \varepsilon)$ factor of the corresponding min-cut in $G$) follows in a similar way.
\end{proof}

\subsection{Proof Outline for Lemma~\ref{lem:one-hole-case}}

Let $\widehat E$ be the minimum cut in $G$ that separates $\emptyset \subset R \subset T$ from $T \setminus R$.
Suppose $G \setminus \widehat E$ consists of some connected components $C_1, \dots, C_k$.
For each $1 \leq i \leq k$, we let $T_i = T \cap V(C_i)$ be the set of terminals in $C_i$.
Clearly, $T_i\cap T_j = \emptyset$ for $i\neq j$.

Since $\widehat E$ separates $R$ from $T \setminus R$, it follows that for each $1 \leq i \leq k$, either $T_i \subseteq R$ or $T_i \subseteq T \setminus R$; otherwise, $C_i$ contains terminals from both $R$ and $T\setminus R$, thus contradicting the fact that $\widehat E$ separates $R$ from $T \setminus R$.
Let $t_1, \dots , t_{k'}$ be all the terminals in $T$ ordered clockwise on the hole.\footnote{Since all terminals lie on a single hole, without loss of generality, we assume that the hole is the outer face of $G$.}
We interpret indices modulo $k'$.

For two terminals $t_a,t_b\in T$, let $[t_a,t_b]_\circ$ denote the cyclic
interval of terminals encountered when walking clockwise from $t_a$ to $t_b$,
including both endpoints. Similarly, let $(t_a,t_b)_\circ$ denote the same
interval but excluding the endpoints. We allow singleton intervals, and we
also regard $T$ itself as a cyclic interval.

For a nonempty set $S\subseteq T$, let $\hull(S)$ be a cyclic interval of
minimum cardinality containing $S$.
Note that the terminal set $T_i$ of $C_i$ is the union of (at most $|T_i|$) intervals, but all are covered by $\hull(S)$.
On the other hand, in the following claim, we show that there is a connected component $C_l$ whose terminal set $T_l$ consists of only a single interval.

\begin{claim} \label{clm:interval}
    There exists an index $l\in[k]$ such that $T_l$ is a cyclic interval of $T$.
\end{claim}

\begin{proof}

    We first state the only planar fact we need. Let $C_i$ and $C_j$ be two
    distinct connected components of $G\setminus \widehat E$. Then there do
    not exist terminals $a,b\in T_i$ and $x,y\in T_j$ that appear in the
    cyclic order
    \(
        a,x,b,y
    \)
    on the boundary of the hole.
    Indeed, since $C_i$ is connected, it contains an $a$-$b$ path. Similarly,
    since $C_j$ is connected, it contains an $x$-$y$ path. These two paths are
    vertex-disjoint, because $C_i$ and $C_j$ are distinct connected components
    of $G\setminus \widehat E$. But two vertex-disjoint paths whose endpoints
    alternate on the boundary of a face cannot both be drawn in the disk
    bounded by that face without crossing, contradicting planarity.

    Now choose any connected component $C_l$. If $T_l$
    is a cyclic interval, we are done. Otherwise, $T_l$ is not a cyclic interval.

    Since $T_l$ is not a cyclic interval, inside $\hull(T_l)$ there are two
    consecutive maximal cyclic subintervals of $T_l$, say
    $[t_a,t_b]_\circ$ and $[t_c,t_d]_\circ$       
    such that the open gap
    \(
        J := (t_b,t_c)_\circ
    \)
    is nonempty and contains no terminal of $T_l$. Thus $t_b,t_c\in T_l$, and
    every terminal in $J$ belongs to some component different from $C_l$.

    Pick any terminal $x\in J$, and let $C_{l'}$ be the connected component of
    $G\setminus \widehat E$ containing $x$. We claim that
    \(
        T_{l'}\subseteq J
    \).
    Suppose not. Then there is a terminal $y\in T_{l'}\setminus J$. Since
    $x\in J=(t_b,t_c)_\circ$ and $y\notin J$, the four terminals
    \(
        t_b,\ x,\ t_c,\ y
    \)
    appear in this cyclic order on the boundary of the hole. But
    $t_b,t_c\in T_l$ and $x,y\in T_{l'}$, where $C_l$ and $C_{l'}$ are distinct
    connected components. This contradicts the planar non-crossing fact proved
    above. Hence $T_{l'}\subseteq J$.

    Therefore
    \(
        |\hull(T_{l'})| \le |J| < |\hull(T_l)|
    \).
    Thus, whenever we start with a component whose terminal set is not a cyclic
    interval, we can move to another component with nonempty terminal set whose
    cyclic hull is strictly smaller.
    Repeating this argument must terminate, because the quantity
    $|\hull(T_i)|$ is a positive integer. The terminal component at which the
    process terminates cannot have a terminal set that is not a cyclic interval;
    otherwise the same argument would produce a strictly smaller hull. Hence
    for some index $l$, the set $T_l$ is a cyclic interval.
\end{proof}

For any connected component $C_i$, let $E_i \coloneq E\Big(V(C_i), V(G) \setminus V(C_i)\Big)$ be the edge set separating $C_i$ from the rest of the graph.
By definition, the min-cut separating $R$ from $T\setminus R$ is $\widehat E = \cup_{i=1}^k E_i$.
In the following claim, we show that the edge set $E_l$ of the specific connected component $C_l$ from \Cref{clm:interval} corresponds to a shortest path in the dual instance $(G_D,T_D)$.
We assume that $\widehat E$ is a minimal minimum cut separating $R$ from $(T\setminus R)$; this can be achieved by deleting redundant edges from an arbitrary minimum cut, thus merging the connected component with no terminals into other connected components.

\begin{claim} \label{clm:one-interval}
    Let $C_l$ be the connected component from \Cref{clm:interval} with the terminal set $T_l$ as a cyclic interval of $T$. Let
    \(
        E_l = E\bigl(V(C_l), V(G)\setminus V(C_l)\bigr)
    \).
    Then the dual edge set $E_l^D$ in the dual instance
    $(G_D,T_D)$ corresponding to $E_l$ in $(G,T)$ induces a shortest path  between the two dual terminals that separates the dual interval $T_l$ from $T\setminus T_l$.
\end{claim}

\begin{proof}
    Since $\widehat E$ separates $R$ from $T\setminus R$, every terminal set
    $T_i$ is contained either in $R$ or in $T\setminus R$. Thus either
    $T_l\subseteq R$ or $T_l\subseteq T\setminus R$. By symmetry, assume
    without loss of generality that
    \(
        T_l\subseteq R
    \).

    We first show that $E_l$ is a minimum cut separating $T_l$ from
    $T\setminus T_l$.
    Clearly, $E_l$ separates $T_l$ from $T\setminus T_l$, because $C_l$ is a
    connected component of $G\setminus \widehat E$ and
    $T_l=T\cap V(C_l)$.
    Let $F$ be a minimum cut in $G$ separating $T_l$ from $T\setminus T_l$.
    We claim that
    \(
        (\widehat E\setminus E_l)\cup F
    \)
    is a cut separating $R$ from $ T\setminus R$.
    To see this, first observe that every edge of $E_l$ has one endpoint in
    $C_l$ and the other endpoint in a component of $G\setminus \widehat E$
    whose terminals lie in $T\setminus R$. Indeed, all vertices of $C_l$ lie in the
    same component as the terminals $T_l\subseteq R$. If some edge of $E_l$
    had its other endpoint in a component containing terminals of $R$, then
    deleting this edge from $\widehat E$ would still leave $R$ separated from
    $T\setminus R$, contradicting the minimality of $\widehat E$.
    The same contradiction holds if the other component contains no terminal.
    Hence $\widehat E\setminus E_l$ separates $R\setminus T_l$ from
    $T\setminus R\cup T_l$. On the other hand, $F$ separates $T_l$ from
    $T\setminus T_l$, and in particular separates $T_l$ from $T\setminus R$.
    Therefore the union
    \(
        (\widehat E\setminus E_l)\cup F
    \)
    separates all terminals of $R=(R\setminus T_l)\cup T_l$ from all terminals
    of $T\setminus R$.

    Since $\widehat E$ is a minimum cut separating $R$ from $T\setminus R$, we have
    \[
        w(\widehat E)
        \le
        w\bigl((\widehat E\setminus E_l)\cup F\bigr)
        \le
        w(\widehat E\setminus E_l)+w(F).
    \]
    Because $E_l\subseteq \widehat E$, this gives
    \(
        w(E_l)\le w(F).
    \)
    But $F$ was chosen as a minimum cut separating $T_l$ from
    $T\setminus T_l$, and $E_l$ itself is such a cut. Therefore $E_l$ is a
    minimum cut separating $T_l$ from $T\setminus T_l$.

    Now write the cyclic interval $T_l$ as
    \(
        T_l=[t_a,t_b]_\circ 
    \).
    Let $p,q\in T_D$ be the two dual terminals corresponding to the two gaps
    on the boundary of the hole that separate this interval; namely, the gap
    between $t_{a-1}$ and $t_a$, and the gap between $t_b$ and $t_{b+1}$,
    where indices are interpreted cyclically.

    By planar cut/path duality for one-face instances, an edge set in $G$
    separates the cyclic interval $[t_a,t_b]_\circ$ from its complement if and
    only if the corresponding dual edge set contains a $p$-$q$ path in $G_D$.
    Moreover, such a minimal cut corresponds exactly to a simple
    $p$-$q$ path.

    Since $E_l$ is a minimum cut separating $T_l$ from $(T\setminus T_l)$, the dual edge set
    $E_l^D$ is therefore a $p$-$q$ path. It remains to show that this path is
    shortest. Suppose, for contradiction, that there is a $p$-$q$ path $P$ in
    $G_D$ with
    \(
        w_D(P) < w_D(E_l^D)
    \).
    Let $P^\star$ be the set of primal edges corresponding to the dual edges of
    $P$. Again by planar cut/path duality, $P^\star$ separates
    $T_l=[t_a,t_b]_\circ$ from $T\setminus T_l$ in $G$. Hence
    \[
        w(P^\star)=w_D(P)<w_D(E_l^D)=w(E_l),
    \]
    contradicting the fact that $E_l$ is a minimum cut separating $T_l$ from
    $T\setminus T_l$.

    Thus $E_l^D$ is a shortest $p$-$q$ path in $(G_D,T_D)$.
\end{proof}

We are now ready to prove \Cref{lem:one-hole-case}.

\begin{proof}[Proof of \Cref{lem:one-hole-case}]

    We prove the statement by induction on the
    number of edges in the cut: if $\widehat E$ is a minimum
    cut in $G$ separating $R$ from $T\setminus R$, then the dual edge set $\widehat E_D$ is a union of edge-disjoint shortest
    paths between dual terminals in $(G_D,T_D)$.

    The base case is $\widehat E=\emptyset$, in which case there is nothing to prove.
    Hence assume $\widehat E\neq\emptyset$, and assume the statement holds for every
    minimum terminal cut with fewer than $|\widehat E|$ edges.

    Let $C_1,\ldots,C_k$ be the connected components of $G\setminus \widehat E$, and
    for each $i\in[k]$, define
    \(
        T_i \coloneq T\cap V(C_i)
    \).
    Since $\widehat E$ separates $R$ from $T\setminus R$, each nonempty set $T_i$ is contained
    entirely in $R$ or entirely in $T\setminus R$.

    We first discuss a simple consequence of the minimality of
    $\widehat E$. Every edge of $\widehat E$ has its endpoints in two connected components of
    $G\setminus \widehat E$ whose terminal sets lie on opposite sides of the partition
    $(R,T\setminus R)$. Indeed, if an edge $e\in \widehat E$ joined two components both containing
    terminals only from $R$, or both containing terminals only from $T\setminus R$, then
    adding $e$ back could not create a path between $R$ and $T\setminus R$. Hence $\widehat E\setminus\{e\}$
    would still separate $R$ from $T\setminus R$, contradicting the
    minimality of $\widehat E$. The same argument rules out an edge of $\widehat E$ incident to
    a component containing no terminal.

    By \Cref{clm:interval}, there exists a connected component $C_l$
    such that $T_l$ is a cyclic interval of $T$.
    Define
    \(
        E_l 
        = E\bigl(V(C_l), V(G)\setminus V(C_l)\bigr)
    \).
    Since $C_l$ is a connected component of $G\setminus \widehat E$, every edge leaving
    $C_l$ belongs to $\widehat E$, and therefore
    \(
        E_l\subseteq \widehat E
    \).

    By \Cref{clm:one-interval}, the dual edge set $(E_l)_D$ induces a shortest
    path in $(G_D,T_D)$ between the two dual terminals that separate the cyclic
    interval $T_l$.

    It remains to show that the remaining edge set
    \(
        \widehat E' \coloneq \widehat E\setminus E_l
    \)
    can be handled recursively.
    Since $T_l$ is contained in exactly one of $R$ or $T\setminus R$, we
    distinguish two symmetric cases. Assume first that
    \(
        T_l\subseteq R
    \).
    Define the residual terminal partition
    \[
        R' \coloneq R\setminus T_l
        \quad \text{and} \quad
        Q' \coloneq T\setminus R \cup T_l.
    \]
    If $R'=\emptyset$, then $\widehat E'=\emptyset$. Indeed, by the minimality
    observation above, every edge of $\widehat E$ is incident to a component whose
    terminal set lies in $R$ and a component whose terminal set lies in $T\setminus R$.
    Since $C_l$ contains all terminals of $R$, every edge of $\widehat E$ is incident
    to $C_l$, and hence belongs to $E_l$. Therefore $\widehat E=E_l$ and there is no
    residual edge set.
    Thus assume $R'\neq\emptyset$. We claim that $\widehat E'$ is a
     minimum cut separating $R'$ from $Q'$.

    First, $\widehat E'$ separates $R'$ from $Q'$. To see this, consider any component
    of $G\setminus \widehat E$ whose terminal set lies in $R'$. Every edge leaving this
    component belongs to $\widehat E$. Moreover, none of these edges belongs to $E_l$:
    an edge of $E_l$ is incident to $C_l$, and by the minimality observation
    every edge of $E_l$ has its other endpoint in a component whose terminals
    lie in $T\setminus R$. Hence all edges leaving every $R'$-component remain deleted in
    $\widehat E'$. Therefore no terminal of $R'$ can be connected in $G\setminus \widehat E'$ to
    any terminal of $Q'=T\setminus R\cup T_l$.

    Next, we prove minimality of the weight of $\widehat E'$. Let $Y$ be a minimum
     cut separating $R'$ from $Q'$ in $G$. We claim that
    \(
        E_l\cup Y
    \)
    separates $R$ from $T\setminus R$. Indeed, any path from a terminal in $T_l$ to a
    terminal in $B$ must leave the component $C_l$, and hence must use an edge
    of $E_l$. On the other hand, any path from a terminal in
    $R'=R\setminus T_l$ to a terminal in $Q' = T\setminus R$ must cross $Y$ since
    $T\setminus R\subseteq Q'$ and $Y$ separates $R'$ from $Q'$.
    Thus $E_l\cup Y$ is a cut separating $R$ from $T\setminus R$. Since $\widehat E$ is a minimum cut separaring $R$ from $T\setminus R$,
    \[
        w(\widehat E)
        \le
        w(E_l\cup Y)
        \le
        w(E_l)+w(Y).
    \]
    As $\widehat E=E_l\cup \widehat E'$, this implies that
    \(
        w(F')\le w(Y).
    \)
    But $\widehat E'$ itself is a cut separating $R'$ from $Q'$, so $\widehat E'$ is a minimum
     cut separating $R'$ from $Q'$.

    Therefore $\widehat E'$ is a minimal minimum cut separating $R'$ from $Q'$ with
    fewer edges than $\widehat E$. By the induction hypothesis, the dual edge set
    $\widehat E'_D$ is a union of edge-disjoint shortest paths between dual terminals.
    Since $E_l$ and $\widehat E'$ are disjoint subsets of $\widehat E$, their dual edge sets are
    also disjoint. Hence
    \(
        \widehat E_D = (E_l)_D \cup \widehat E'_D
    \)
    is a union of edge-disjoint shortest paths between dual terminals.

    The case where $T_l\subseteq T\setminus R$ is handled in a symmetric way.
\end{proof}

\end{document}